# Preliminary design of a RESPER probe prototype, configured in a multi dipole-dipole array.

Alessandro Settimi[(1)], Giuseppe Tutone[(1)], James A. Baskaradas[(1)], Cesidio Bianchi[(1)], Achille E. Zirizzotti[(1)], Giovanni Santarato[(2)]

[(1)]*Istituto Nazionale di Geofisica e Vulcanologia (INGV), Geomagnetismo, Aeronomia e Geofisica Ambientale (Sezione Roma 2)*

[(2)]*Dipartimento di Scienze della Terra, Università degli Studi di Ferrara - Via Saragat 1, 44100 Ferrara, Italy.*

Corresponding author: Dr. Alessandro Settimi
Istituto Nazionale di Geofisica e Vulcanologia (INGV)
Via di Vigna Murata 605
I-00143 Rome, Italy
Tel: +39-0651860719
Fax: +39-0651860397
Email: alessandro.settimi@ingv.it

## Index

## Preface

*Electrical spectroscopy*

Electrical resistivity and relative dielectric permittivity are two independent physical properties which characterize the behaviour of bodies when these are excited by an electromagnetic field. The measurement of these properties provides crucial information regarding the practical use of bodies (for example, materials that conduct electricity), as well as numerous other purposes.

Some studies have shown that the electrical resistivity and dielectric permittivity of a body can be obtained by measuring the transfer impedance using a system with four electrodes, although these electrodes do not require resistive contact with the investigated body [Grard, 1990a, b; Grard and Tabbagh, 1991; Tabbagh et al., 1993; Vannaroni et al., 2004; Del Vento and Vannaroni, 2005]. In this case, the current is made to circulate in the body by electric coupling, supplying the electrodes with an alternating electrical signal of low or middle frequency (LF-MF). In this type of investigation, the range of optimal frequencies for electrical resistivity values of the more common materials is between ≈*10 kHz* and ≈*1 MHz*.

The lower limit is effectively imposed by two factors: a) firstly, the Maxwell-Wagner effect, which limits probe accuracy [Frolich, 1990], is the most important limitation and occurs because of interface polarization effects that are stronger at low frequencies, for example below *10 kHz* depending on medium resistivity; b) secondly, the need to maintain the amplitude of the current at measurable levels because, given the capacitive coupling between electrodes and soil, the current magnitude is proportional to frequency.

Conversely, the upper limit is fixed so as to permit analysis of the system under a regime of quasi static approximation, ignoring the factor of the velocity of the cables used for the electrode harness, which degrades the accuracy of the impedance phase measurements. It is therefore possible to make use of an analysis of the system in the LF and MF bands where the electrostatic term is significant. A general electromagnetic calculation produces lower values than a static one, and high resistivity reduces this differences. Consequently, above *1 MHz* a general electromagnetic calculation must be preferred, while below *500 kHz* a static calculation would be preferred, and between *500 kHz* and *1 MHz* both these methods could be applied [Tabbagh et al., 1993].

*Geoelectrical measurements. A comparison of two different electrode arrays*

The geoelectrical technique is based on the analysis of the underground electric fields generated by a current flow injected from the surface [Loke, 1999]. This resistivity method is based on the electric conduction in the ground, and it is governed by Ohm's law. From the current source $I$ and potential difference $\Delta V$ values, an apparent resistivity value $\rho_a$ can be calculated as $\rho_a = k(\Delta V/I)$, where k is a geometric factor that depends on the arrangement of the four electrodes. A pair of electrodes ($T_1$, $T_2$) is used for the current injection, while potential difference measurements are made using a second pair of electrodes ($R_1$, $R_2$). The potential is then converted into apparent resistivity, and then by inversion to the true resistivity, which depends on several factors: mainly the lithology of the soil, and its porosity, and the saturation and resistivity of its water pores.

The measurements can be carried out using different array configurations: for example, Wenner's or dipole-dipole [Loke, 1999]. The Wenner's array is an attractive choice for surveys carried out in areas with a lot of background noise (due to its high signal strength), and also when good vertical resolution is required. The dipole-dipole array might be a more suitable choice if good horizontal resolution and data coverage is important (assuming the resistivity meter is sufficiently sensitive and there is good ground contact).

In general, the Wenner's array is good for the resolving of vertical changes (i.e. horizontal structures), while it is relatively poor for the detection of horizontal changes (i.e. narrow vertical structures). Compared to the other arrays, the Wenner's array has a moderate depth of investigation. Among the common arrays, the Wenner's array has the strongest signal strength. This can be an important factor when a survey is carried in areas with high background noise. One disadvantage of this array for two-dimensional surveys is the relatively poor horizontal coverage as the electrode spacing is increased. This can be a problem if the system used has a relatively small number of electrodes.

The dipole-dipole array has been, and still is, widely used in resistivity and induced-polarization surveys, because of its low electromagnetic (EM) coupling between the current and potential circuits. The dipole-dipole array is very sensitive to horizontal changes in resistivity, although relatively insensitive to vertical changes in resistivity. This means that it is good for the mapping of vertical structures, such as dykes and cavities, but relatively poor for the mapping of horizontal structures, such as sills or sedimentary layers.

In general, this array has a shallower depth of investigation compared to the Wenner's array, although for two-dimensional surveys, this array has better horizontal data coverage than the Wenner's array. This can be an important advantage when the number of nodes available with the multi-electrode system is small. One possible disadvantage of this array is the very small signal strength.

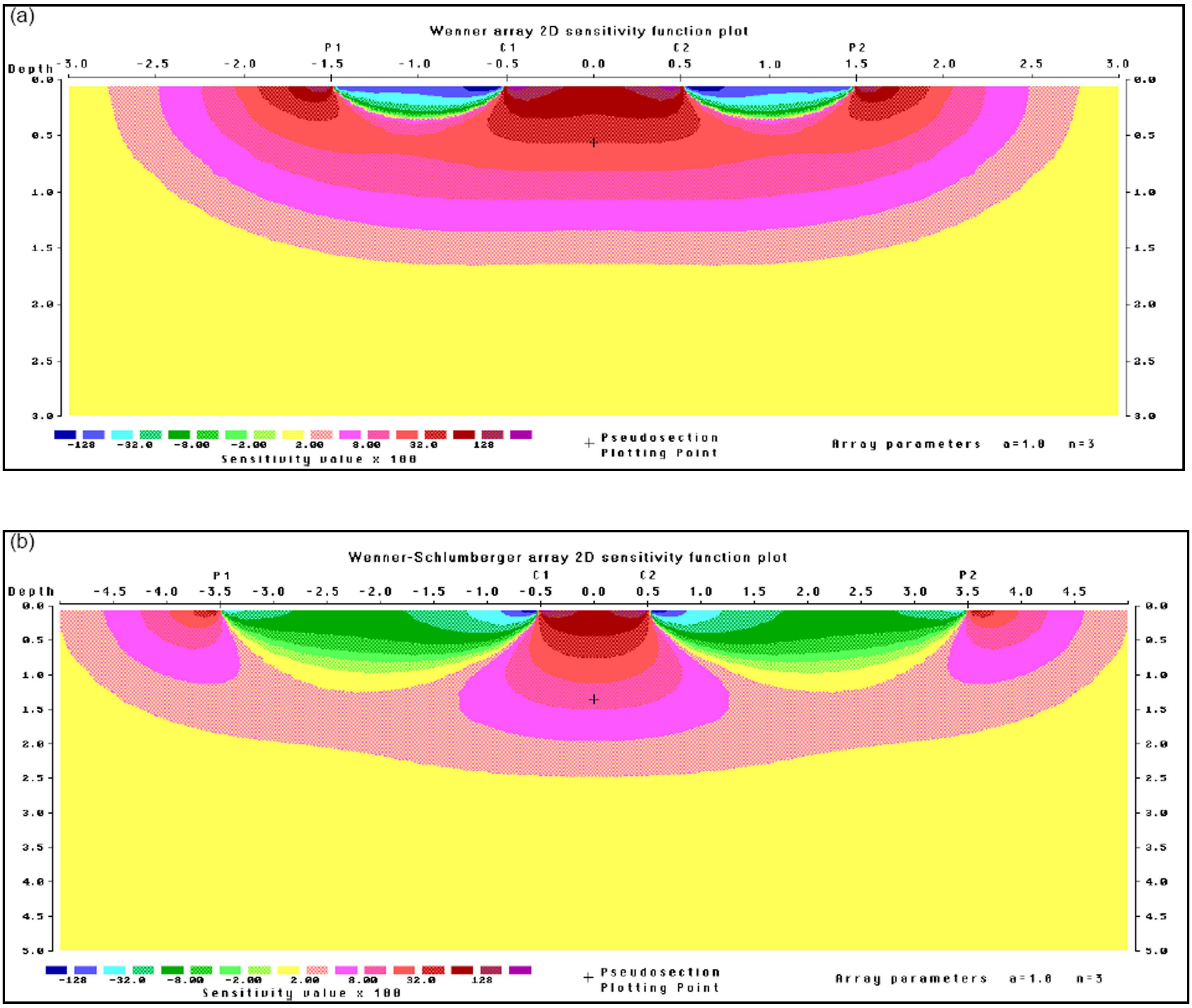

**Figure 1.** The sensitivity patterns for the (a) Wenner and (b) dipole-dipole arrays.

.

## 1. Introduction

In previous studies [Settimi et al., 2010a-c; Settimi, 2010d], the authors proposed a discussion of theoretical modelling and a move towards the practical implementation of an induction probe that can acquire transfer impedance in the field.

A RESPER probe enables measurement of electrical RESistivity and relative dielectric PERmittivity using alternating current at LFs ($30kHz<f<300kHz$) up to MFs ($300kHz<f<3MHz$) bands. Measurements are taken using four electrodes laid on the surface to be analyzed, and through measurement of transfer impedance, the electrical resistivity and dielectric permittivity of the material can be established. Furthermore, by increasing the distance between the electrodes, the electrical properties of the sub-surface structures can be investigated to greater depths. The main advantage of the RESPER is that measurements of electrical parameters can be conducted in a non-destructive manner, thereby enabling characterization of precious and unique materials. Also, in appropriate arrangements, measurements can be taken with the electrodes raised slightly above the surface, providing totally non-invasive analysis, although accompanied by a greater error. The probe can perform measurements on materials with high resistivity and permittivity in an immediate way, without the need for later stages of data post-analysis.

An initial paper [Settimi et al., 2010a] discussed the theoretical modelling of an induction probe which performs simultaneous non-invasive measurements of electrical RESistivity $1/\sigma$ and dielectric PERmittivity $\varepsilon_r$ of non-saturated media (RESPER probe). A mathematical-physical model was applied on the propagation of errors in the measurement of resistivity and permittivity, based on the sensitivity functions tool [Murray-Smith, 1987]. The findings were also compared with the results of the classical method of analysis in the frequency domain, which is useful for determining the behaviour of zero and pole frequencies in the linear time invariant circuit (LTI) of the RESPER probe. The paper underlined that the average values of electrical resistivity and dielectric permittivity can be used to estimate the transfer impedance over various terrestrial soil [Edwards, 1998] and concrete [Polder et al., 2000; Laurents et al., 2005] types, especially when they are characterized by low volumetric water content [Knight and Nur, 1987] and analyzed within a frequency bandwidth within the LFs [Al-Qadi et al., 1995; Myounghak et al., 2007]. To meet the design specifications required to ensure satisfactory performance of the RESPER, the forecasts of the sensitivity-functions approach are more reliable than the results foreseen by the transfer-functions method. In simpler terms, once the measurement inaccuracy is within an acceptable limit (*10%)*, the sensitivity approach provides more realistic values with respect to those provided by the transfer method. These numeric values concern both the band of frequency $f$ for the probe and the measurable range of resistivity $1/\sigma$ or permittivity $\varepsilon_r$ for the soils and concretes (the order of magnitude of these values is reported in the relevant literature; see [Settimi et al., 2010a] and references therein).

A second paper [Settimi et al., 2010b] moved towards a practical implementation of electrical spectroscopy. In order to design a RESPER probe to perform measurements of $1/\sigma$ and $\varepsilon_r$ on a subsurface with inaccuracies below a prefixed limit (*10%*) in a bandwidth of MFs, the RESPER should be connected to an appropriate analogical digital converter (ADC) that can sample in uniform mode [Razavi, 1963], or in phase and quadrature (IQ) [Jankovic and Öhman, 2001]. If the probe is characterized by a galvanic contact with the surface, then the inaccuracies in the measurement of resistivity and permittivity, due to the uniform or IQ sampling ADC, can be analytically expressed. A large number of numerical simulations have shown that performance depends on the selected sampler, and that the IQ is better, compared to the uniform mode, under the same operating conditions, i.e. number of bits and medium (see references therein [Settimi et al., 2010b]).

A third paper [Settimi et al., 2010c] developed just a suitable number of numerical simulations, using Mathcad program, which provide the working frequencies, the electrode-electrode distance and the optimization of the height above ground minimizing the inaccuracies of the RESPER, in galvanic or capacitive contact with terrestrial soils or concretes of low or high resistivity. As findings of simulations, the paper underlined that the performances of a lock-in amplifier [Scofield, 1994] are preferable even when compared to an IQ sampling ADC with high bit number, under the same operating conditions. As consequences in the practical applications: if the probe is connected to a data acquisition system (DAS) as an uniform or an IQ sampler, then it could be commercialized for companies of building and road paving, being employable for analyzing “in situ” only concretes; otherwise, if the DAS is a lock-in amplifier, the marketing would be for companies of geophysical prospecting, involved to analyze “in situ” even soils.

The report [Settimi, 2010d] proposed to discuss the Fourier domain analysis performances of a RESPER probe. A uniform ADC, which is characterized by a sensible phase inaccuracy depending on frequency, is connected to a Fast Fourier Transform (FFT) processor, that is especially affected by a round-off amplitude noise linked to both the FFT register length and samples number [Oppenheim et al., 1999]. If the register length is equal to *32* bits, then the round-off noise is entirely negligible, else, once bits are reduced to *16*, a technique of compensation must occur. In fact, oversampling can be employed within a short time window, reaching a compromise between the needs of limiting the phase inaccuracy due to ADC and not raising too much the number of averaged FFT values sufficient to bound the round-off.

As claimed by the applied patent [Zirizzotti et al., 2010], the device exploits the in-phase and quadrature sampling technique which, together with numerical operations performed by a microcontroller, allows the device to attain a required performance. It is possible to execute a number of numerical integrations which, combined with some circuit solutions, can reduce the amplitude and phase errors of the acquired signal. The device can operate at variable frequency, maintaining a suitable under-sampling frequency to fully exploit the analogical-digital acquisition performance both in velocity and dynamic range. In fact, assuming that the electric current injected in materials and so the voltage measured by probe are quasi-monochromatic signals, i.e. with a very narrow frequency band, an IQ down-sampling process can be employed [Andren and Fakatselis, 1995]. Besides the quantization error of IQ ADC, which can be assumed small both in amplitude and phase, as decreasing exponentially with the bit number, the electric signals are affected by two additional noises. The amplitude term noise, due to external environment, is modelled by the signal to noise ratio which can be reduced performing averages over a thousand of repeated measurements. The phase term noise, due to a phase-splitter detector, even if increasing linearly with the frequency, can be minimized by digital electronics providing a rise time of few nanoseconds.

## 2. RESPER probe

When using an induction probe (Fig. 2), the response depends on both geometrical parameters, like the height of each electrode above the ground surface and the separation of the electrodes, and physical parameters, including frequency, electrical resistivity and relative dielectric permittivity. When a medium is assumed to be linear and its response linearly dependent on the electrical charges of the two exciting electrodes, the simplest approach is a static calculation [Tabbagh et al., 1993], especially using a low operating frequency. If the electrodes have small dimensions relative to their separation, then they can be considered as points. Moreover, if the current wavelength is much larger than all of the dimensions under consideration, then the quasi-static approximation applies [Grard, 1990, a, b].

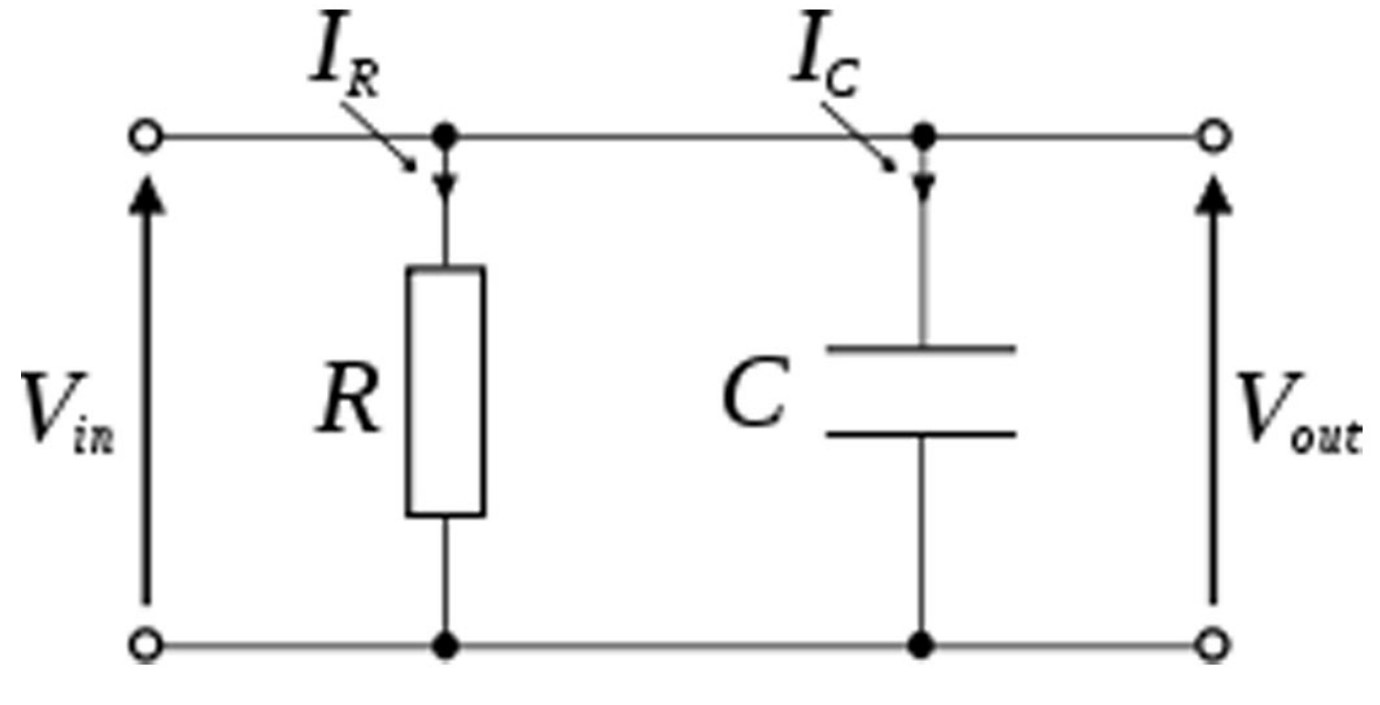

**Figure 2.** Equivalent circuit of the RESPER probe.

The RESPER probe measures a capacitance in a vacuum $C_0(L)$ that is directly proportional to its characteristic geometrical dimension, i.e. the electrode-electrode distance $L$ [see Appendix A]. When the RESPER shows a galvanic contact with the subjacent medium of electrical conductivity $\sigma$ and dielectric permittivity $\varepsilon_r$, it measures a complex impedance $Z_N(f,L,\sigma,\varepsilon_r)$ that consists of resistive $R_N(L,\sigma)$ and capacitive $C_N(L,\varepsilon_r)$ parallel components. The resistance $R_N(L,\sigma)$ depends only on $L$ and $\sigma$ [Grard and Tabbagh, 1991] (Tab.1),

$$R_N(L,\sigma) = 2\frac{\varepsilon_0/\sigma}{C_0(L)}, \tag{2.1}$$

while $C_N(L,\varepsilon_r)$ depends only on $L$ and $\varepsilon_r$ [Grard and Tabbagh, 1991] (Tab.1):

$$C_N(L,\varepsilon_r) = \frac{1}{2}C_0(L)\cdot(\varepsilon_r+1). \tag{2.2}$$

As a consequence, as well as grazing the medium, if the probe measures the conductivity $\sigma$ and permittivity $\varepsilon_r$ working in a frequency $f$ much lower than the cut-off frequency $f_T=f_T(\sigma,\varepsilon_r)= \sigma/(2\pi\varepsilon_0(\varepsilon_r+1))$, the complex impedance $Z_N(f,L,\sigma,\varepsilon_r)$ is characterized by the phase $\Phi_N(f,\sigma,\varepsilon_r)$ and modulus $|Z|_N(L,\sigma)$. The phase $\Phi_N(f,\sigma,\varepsilon_r)$ depends linearly on $f$, with a maximum value of $\pi/4$, and it is directly proportional to the ratio $(\varepsilon_r+1)/\sigma$; while $|Z|_N(L,\sigma)$ does not depend on $f$, and is inversely proportional to both $L$ and $\sigma$. Indeed, if $Z_N(f,L,\sigma,\varepsilon_r)$ consists of the parallel components of $R_N(L,\sigma)$ (see Eq. (2.1)) and $C_N(L,\varepsilon_r)$ (see Eq.(2.2)), then it is fully characterized by the high frequency (HF) pole $f_T=f_T(\sigma,\varepsilon_r)$, which cancels its denominator: the complex impedance acts as a LF-MF band-pass filter with cut-off $f_T=f_T(\sigma,\varepsilon_r)$; in simpler terms, the frequency equalizing Joule and displacement current. Under the operating conditions defined by Settimi et al. [2010a], average values of $\sigma$ can be used over the band ranging from LF to MF; therefore, $|Z|_N(L,\sigma)$ is not a function of frequency below $f_T$.

Instead, when the RESPER probe has capacitive contact with the subjacent medium and the geometry of the probe is characterized by the ratio $x$ between the height above ground $h$ and the electrode-electrode distance $L$,

$$x = \frac{h}{L}, \tag{2.3}$$

its configurations can be entirely defined by a suitable geometrical factor $K(x)$, which depends even on the height/ dimension ratio $x$. Actually, Grard and Tabbagh [1991] preferred to introduce the complementary $\delta(x)$ of the geometrical factor $K(x)$, i.e.:

$$\delta(x) = 1 - K(x), \tag{2.4}$$

where $K(x=0)=1$ and $\delta(x=0)=0$ [Appendix A].

So, if the RESPER works in the pulse frequency $\omega=2\pi f$, which can be normalized with respect to the cut-off $\omega_T=2\pi f_T$ [Grard and Tabbagh, 1991],

$$\Omega = \frac{\omega}{\omega_T} = \omega R_N C_N = \omega\frac{\varepsilon_0(\varepsilon_r+1)}{\sigma}, \tag{2.5}$$

then the probe measures a complex impedance $Z(\Omega,x,\sigma,\varepsilon_r)$ which consists of the resistance $R(\Omega,x,\sigma,\varepsilon_r)$ and capacitance $C(\Omega,x,\sigma,\varepsilon_r)$ components [Grard and Tabbagh, 1991] (Fig. 3):

$$R(\Omega,x,\sigma,\varepsilon_r) = R_N(L,\sigma)\frac{[1+\delta(x)\frac{\varepsilon_r-1}{2}]^2+[\frac{\delta(x)}{\Omega}\frac{\varepsilon_r+1}{2}]^2}{1-\delta(x)}, \tag{2.6}$$

$$C(\Omega,x,\sigma,\varepsilon_r) = C_N(L,\varepsilon_r)\frac{1+\delta(x)(\frac{\varepsilon_r-1}{2}+\frac{1}{\Omega^2}\frac{\varepsilon_r+1}{2})}{[1+\delta(x)\frac{\varepsilon_r-1}{2}]^2+[\frac{\delta(x)}{\Omega}\frac{\varepsilon_r+1}{2}]^2}. \tag{2.7}$$

Inverting Eqs. (2.6) and (2.7), $\sigma$ and $\varepsilon_r$ can be expressed as functions of $R$ and $C$, i.e. [Settimi et al., 2010a]:

$$\sigma(\omega,x,R,C)=\frac{2[1-\delta(x)]\varepsilon_0\omega^2 RC_0}{\delta^2(x)+\omega^2R^2[C_0-\delta(x)C]^2}, \qquad (2.8)$$

$$\varepsilon_r(\omega,x,R,C)=\frac{\delta(x)[\delta(x)-2]-\omega^2R^2[C_0-\delta(x)C]\{C_0+[\delta(x)-2]C\}}{\delta^2(x)+\omega^2R^2[C_0-\delta(x)C]^2}. \qquad (2.9)$$

In our opinion, once the degrees of freedom of the $(f, x)$ pair are fixed, it is not suitable to choose $(R,C)$ as independent variables and then $(\sigma, \varepsilon_r)$ as dependent variables (Eqs. (2.8) and (2.9)). Instead, it is more appropriate to consider $(\sigma, \varepsilon_r)$ as quantities of physical interest, and consequently Eqs. (2.6) and (2.7) as the starting points for the theoretical development. Indeed, even if the physics does not forbid the choice of $(R,C)$ as independent variables, applying the function $(R,C) \rightarrow (\sigma, \varepsilon_r)$, the procedures of the design should anyway choose $(\sigma, \varepsilon_r)$ as independent variables, preferentially applying the inverse function $(\sigma, \varepsilon_r) \rightarrow (R,C)$. There are the following two practical approaches: (a) $(\sigma, \varepsilon_r)$ as independent variables, in order to establish the class of media with electrical conductivity and dielectric permittivity $(\sigma, \varepsilon_r)$ that can be investigated by the RESPER working in a fixed band $B$ and specified by a known geometry $x$; (b) preferential way $(\sigma, \varepsilon_r) \rightarrow (R,C)$, since once a subjacent medium with conductivity $\sigma$ and permittivity $\varepsilon_r$ is selected, the probe specifications $R$ and $C$ can be projected both in frequency $f$ and in height/ dimension ratio $x$.

| ***GALVANIC CONTACT: Wenner, $L_0$ = 10 cm*** | **SOIL** | **CONCRETE** |
|---|---|---|
| *Low Resistivity* | *$1/\sigma$ = 130 Ω·m → R =206.901 Ω*<br>*$\varepsilon_r$ = 13 → C = 0.078 nF* | *$1/\sigma$ = 4000 Ω·m → R = 6.366 kΩ*<br>*$\varepsilon_r$ = 9 → C = 55.633 pF* |
| *High Resistivity* | *$1/\sigma$ = 3000 Ω·m → R = 4.775 kΩ*<br>*$\varepsilon_r$ = 4 → C = 27.816 pF* | *$1/\sigma$ = 10000 Ω·m → R = 15.915 kΩ*<br>*$\varepsilon_r$ = 4 → C = 27.816 pF* |

**Table 1.** Resistive and capacitive parallel components of the complex impedance, corresponding to the electrical resistivity and dielectric permittivity of non-saturated terrestrial soils and concretes, characterized by a low or high resistivity, and in galvanic contact with the RESPER, configured in a Wenner's array with characteristic geometrical dimension $L_0$ = *10 cm*.

**(a)**

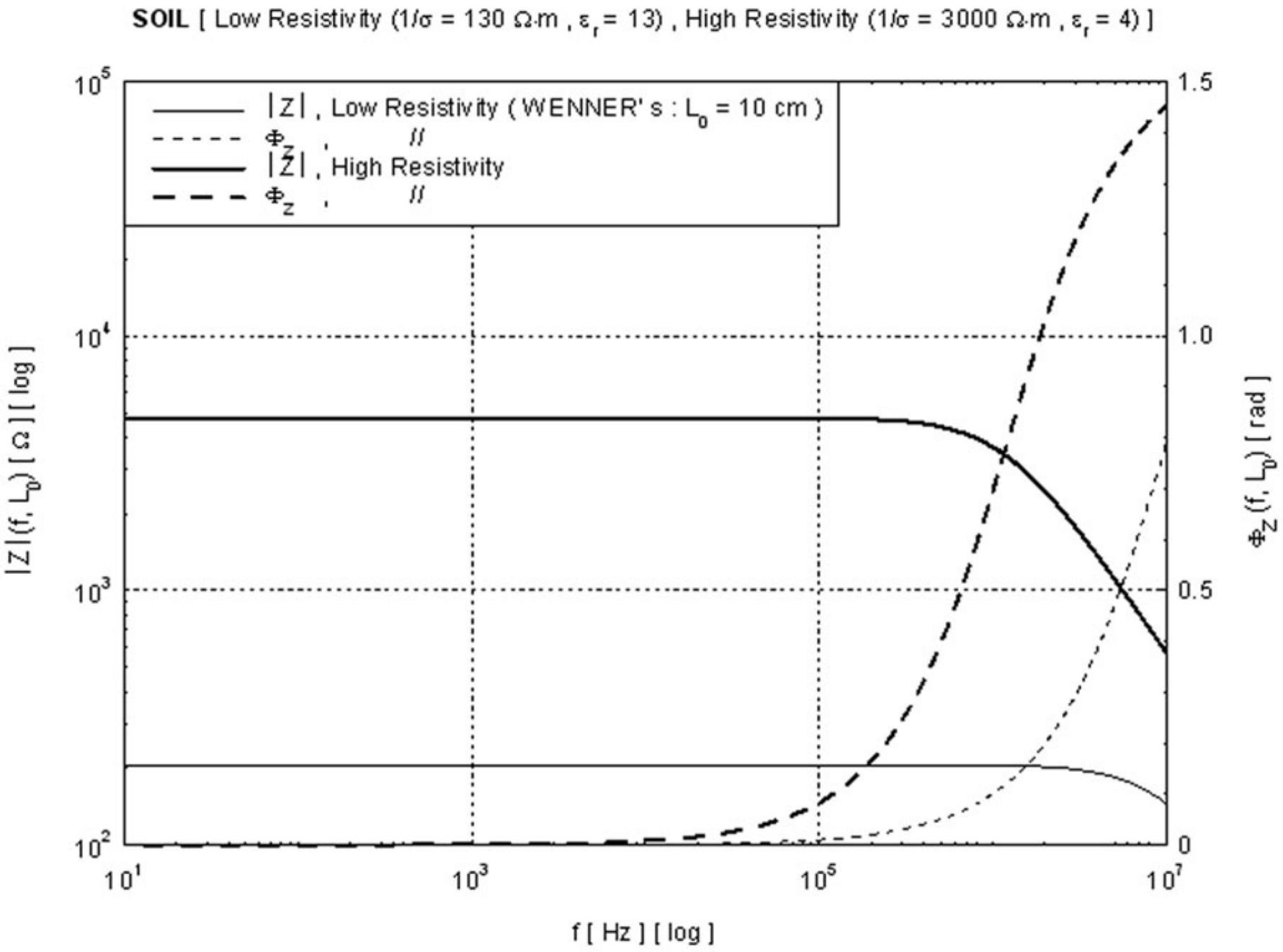

**(b)**

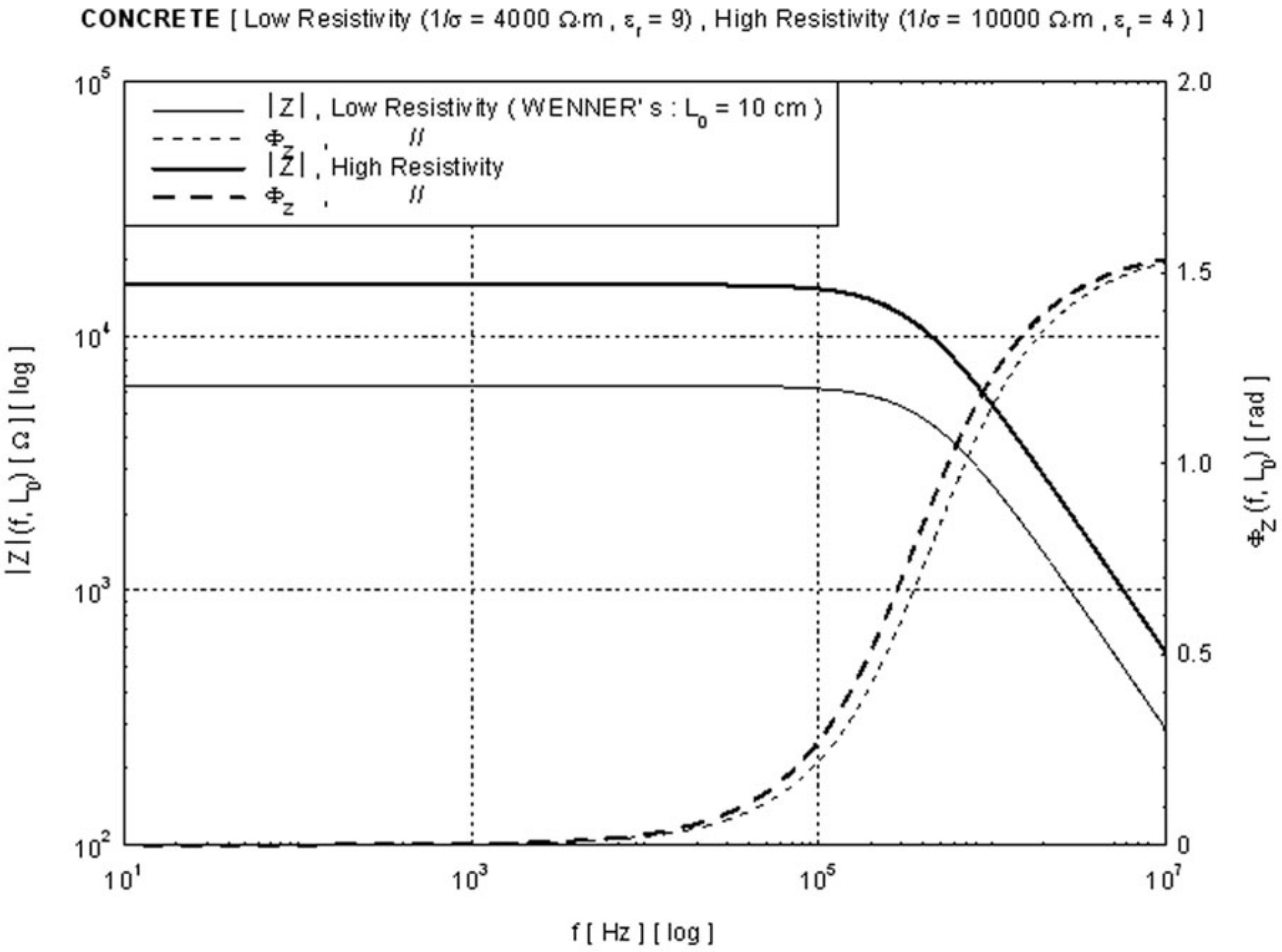

**Figure 3.** Refer to the caption of Tab. 1. A probe is configured in the Wenner's array, according to an electrode-electrode distance $L_0=10\ cm$, and it can show a galvanic contact (height above ground $h=0$) both on the terrestrial soils, characterized by a low ($1/\sigma=130\Omega\cdot m$, $\varepsilon_r=13$) or high ($1/\sigma=3000\Omega\cdot m$, $\varepsilon_r=4$) electrical resistivity, and on the concretes, with a low ($1/\sigma=4000\Omega\cdot m$, $\varepsilon_r=9$) or high ($1/\sigma=10000\Omega\cdot m$, $\varepsilon_r=4$) resistivity. Bode's diagrams showing the complex impedance both in modulus $|Z|(f,L_0)$ and phase $\Phi_Z(f,L_0)$, plotted as functions of the frequency $f$, in the band $f \in [0, f_{\mathrm{lim}}]$ with $f_{lim}=10\ MHz$, for both the soils (a) and concretes (b) analyses.

**2.1 IQ Down-Sampling in noisy conditions**

In signal processing, down-sampling (or "sub-sampling") is the process of reducing the sampling rate of a signal. This is usually done to reduce the data rate or the size of the data [Andren and Fakatselis, 1995].

The down-sampling factor, commonly denoted by *M*, is usually an integer or a rational fraction greater than unity. If the RESPER probe injects electric current into materials at a radio frequency (RF) frequency *f*, then the ADC samples at a rate $f_s$ fixed by:

$$f_S = \frac{f}{M}\,, \tag{2.10}$$

being *M* preferably, but not necessary, a power of *2* to facilitate the digital circuitry ( $M = 2^m,\ m \in \mathbb{N}$ ).

The employment of a down sampler allows operating at various frequencies *f* and the only constraint is due to Eq. (2.10). Since the sounding frequency is known, the signal is assumed to show a very narrow band such that the IQ sampling process, associated to an average process, contributes to reject off the unwanted carrier frequencies *f*. The down sampler acts as a phase sensitive detector and integrator. In fact, the sampling process detects the amplitude of signal. Moreover, the phase relation is maintained, as it is sensitive only to the carrier frequency, coincident with the sampling or down sampling frequency. Finally, the average operation on digital value acts like an analog integrator.

A practical scheme to select the sampling rate is to launch two time sequences as in Fig. 4. Now there is a problem due to timing. If the RESPER would work at a fixed frequency *f*, then the proper relationship between the rate $f_s$ and the down-sampling factor *M* could be easily found. Instead, if the probe is performing a sort of electrical spectroscopy, then an enable signal for the sampling and holding circuit (S&H) must be generated. The time frame should be such that the sequences $n{\cdot}M{\cdot}T$ for the sample I and $n{\cdot}M{\cdot}T+T/4$ for the sample Q could be obtained, corresponding to the period *T* of the maximum working frequency. So the rate *fs* would be ensured as a *M* factor sub-multiple of frequency *f*. A possible conceptual scheme of this implementation is shown in Fig. 5 [Zirizzotti et al.,2010].

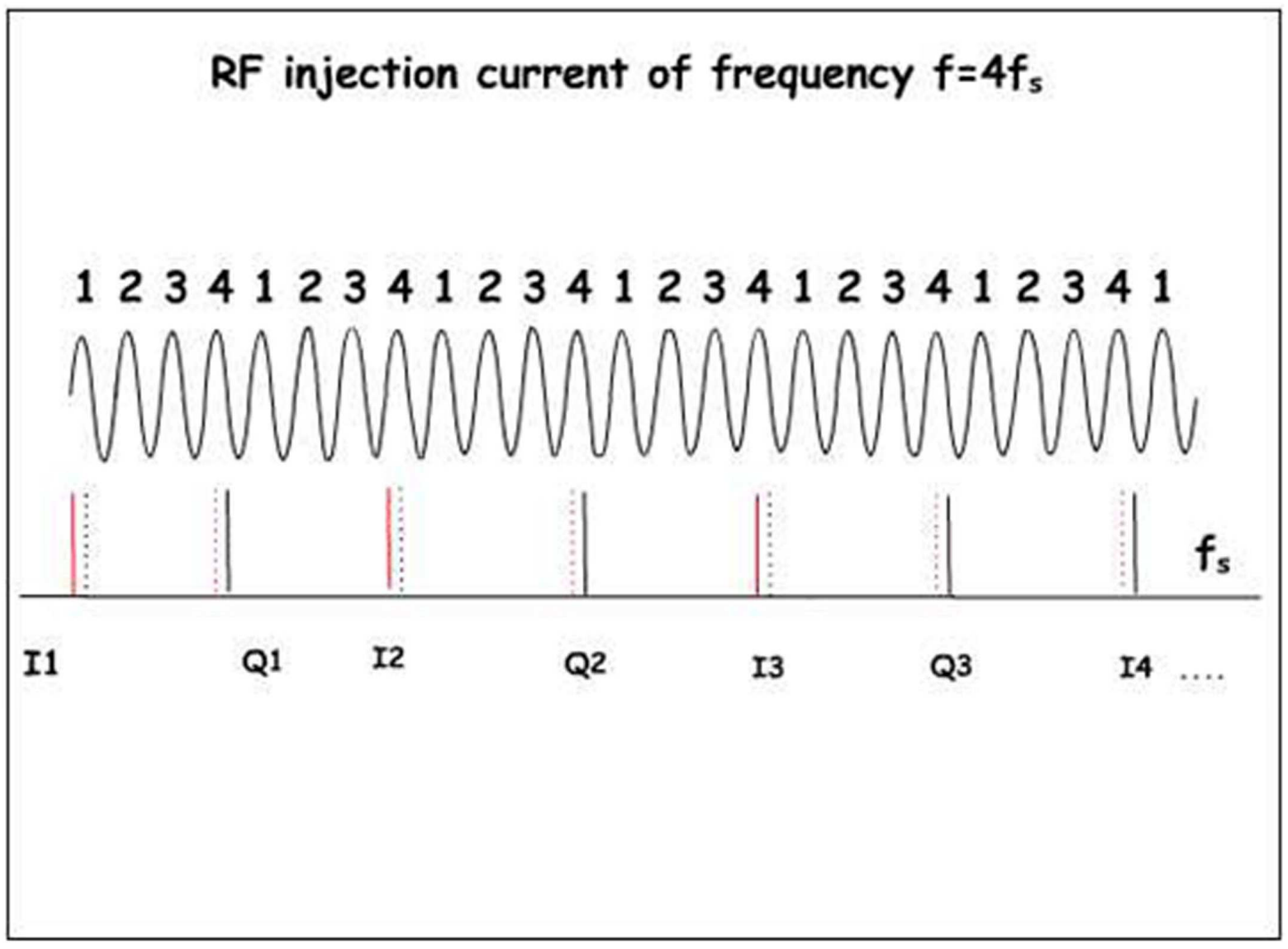

**Figure 4.** Practical scheme of an in phase and quadrature (IQ) down-sampling process. Samples I and Q can be enabled depending on whether the discrete time *n* is even or odd (*n* is a natural number). The sample in phase I is picked up at time $n{\cdot}M{\cdot}T$ and the sample in quadrature Q at time $n{\cdot}M{\cdot}T + T/4$ (*T* is the signal period and *M* the down-sampling factor). Obviously one can also choose a different values of *M* shown by the figure (*M* = *4*).

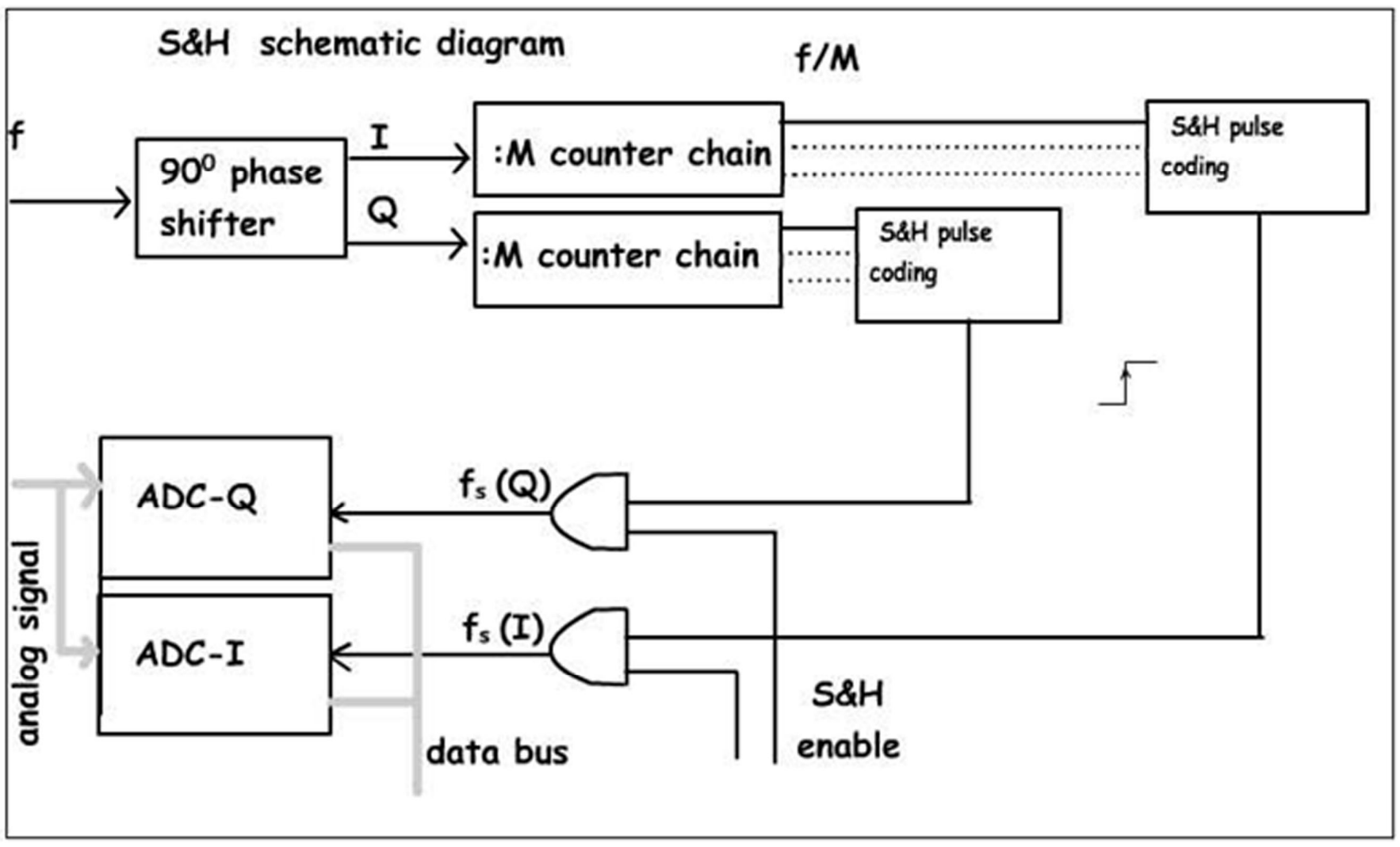

**Figure 5.** Refer to the caption of Fig. 4. Logical scheme of a sampling and holding circuit (S&H) which employs two IQ analogical digital converters (ADC). The frequency *f* of input signal is forwarded to a $90^{\circ}$ degree phase-shifter. Two chains of identical programmable counters operate a division for the down-sampling factor *M*. The rate is precisely $f_S = f/M$.

Besides the quantization error of IQ ADC, which can be assumed small both in amplitude and phase, as decreasing exponentially with the bit number $n_{bit}$, the electric signals are affected by two additional noises [Settimi et al. , 2010b]. The amplitude term noise, due to external environment, is modelled by the signal to noise ratio $SNR = 30dB$ which can be reduced performing averages over one thousand of repeated measurements ($A = 10^3$). The phase term noise, due to a phase-splitter detector, which, even if increasing linearly with the frequency *f*, can be minimized by digital electronics providing a rise time of few nanoseconds ($\tau = 1ns$). In analytical terms:

$$\frac{\Delta|Z|}{|Z|}=\left.\frac{\Delta|Z|}{|Z|}\right|_{IQ}+\frac{1}{\sqrt{A}}\left.\frac{\Delta|Z|}{|Z|}\right|_{Enviroment}=\frac{1}{2^{n_{bit}}}+\frac{1}{\sqrt{A}}\frac{2}{SNR}, \qquad (2.11)$$

$$\frac{\Delta\Phi_Z}{\Phi_Z}=\left.\frac{\Delta\Phi_Z}{\Phi_Z}\right|_{IQ}+\left.\frac{\Delta\Phi_Z}{\Phi_Z}\right|_{Phase-Shifter}=\frac{1}{2^{n_{bit}}}+\tau\cdot f\ . \qquad (2.12)$$

With respect to the ideal case involving only a quantization error, the additional noise both in amplitude, due to external environment, and especially in phase, due to phase-shifter detector, produce two effects: firstly, both the curves of inaccuracy $\Delta R/R(f)$ and $\Delta C/C(f)$ and in measurement of the resistive *R* and capacitive *C* parallel components for complex impedance are shifted upwards, to values larger of almost half a magnitude order, at most; and, secondly, the inaccuracy curve $\Delta C/C(f)$ of capacitance *C* is narrowed, even of almost half a MF decade. So, both the optimal value of frequency $f_{opt}$, which minimizes the inaccuracy $\Delta C/C(f)$ of *C*, and the maximum frequency $f_{max}$, allowing an inaccuracy $\Delta C/C(f)$ below the prefixed limit $\Delta C/C|_{fixed}$ (*10%*), are left shifted towards lower frequencies, even of half a MF decade. Instead, the phase-splitter is affected by a noise directly proportional to frequency, which is significant just from MFs; so, the minimum frequency $f_{min}$, allowing $\Delta C/C(f)$ below $\Delta C/C|_{fixed}$ (*10%*), remains almost invariant at LFs (Fig. 6) [Settimi et al., 2010a-d].

Therefore, the profit of employing the down-sampling method is obvious. This method would allow running a real electric spectroscopy because, theoretically, measurements could be performed at any frequency. An advantage is that there are virtually no limitations due to the sampling rate of ADCs and associated S&H circuitry.

**(a)**

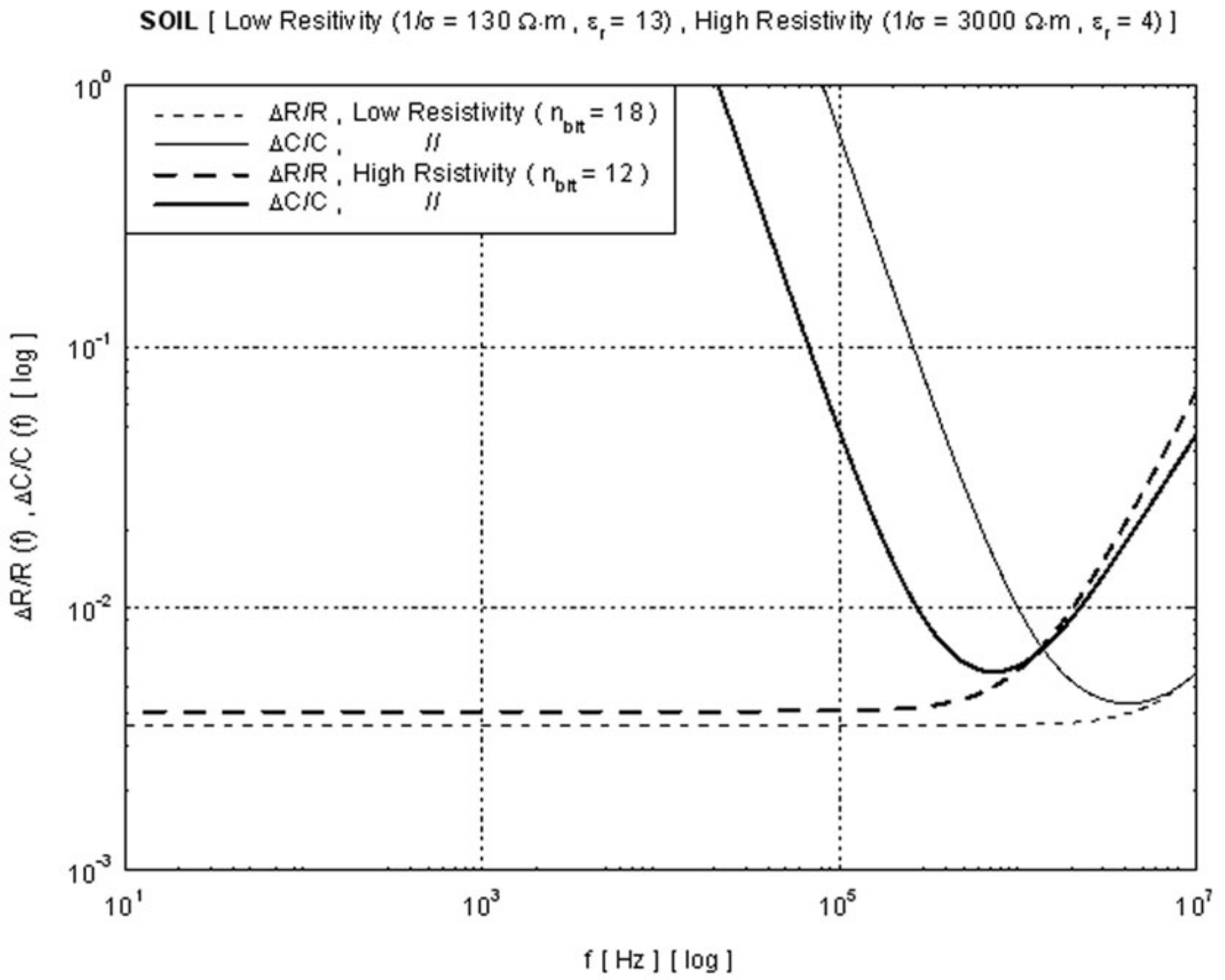

**(b)**

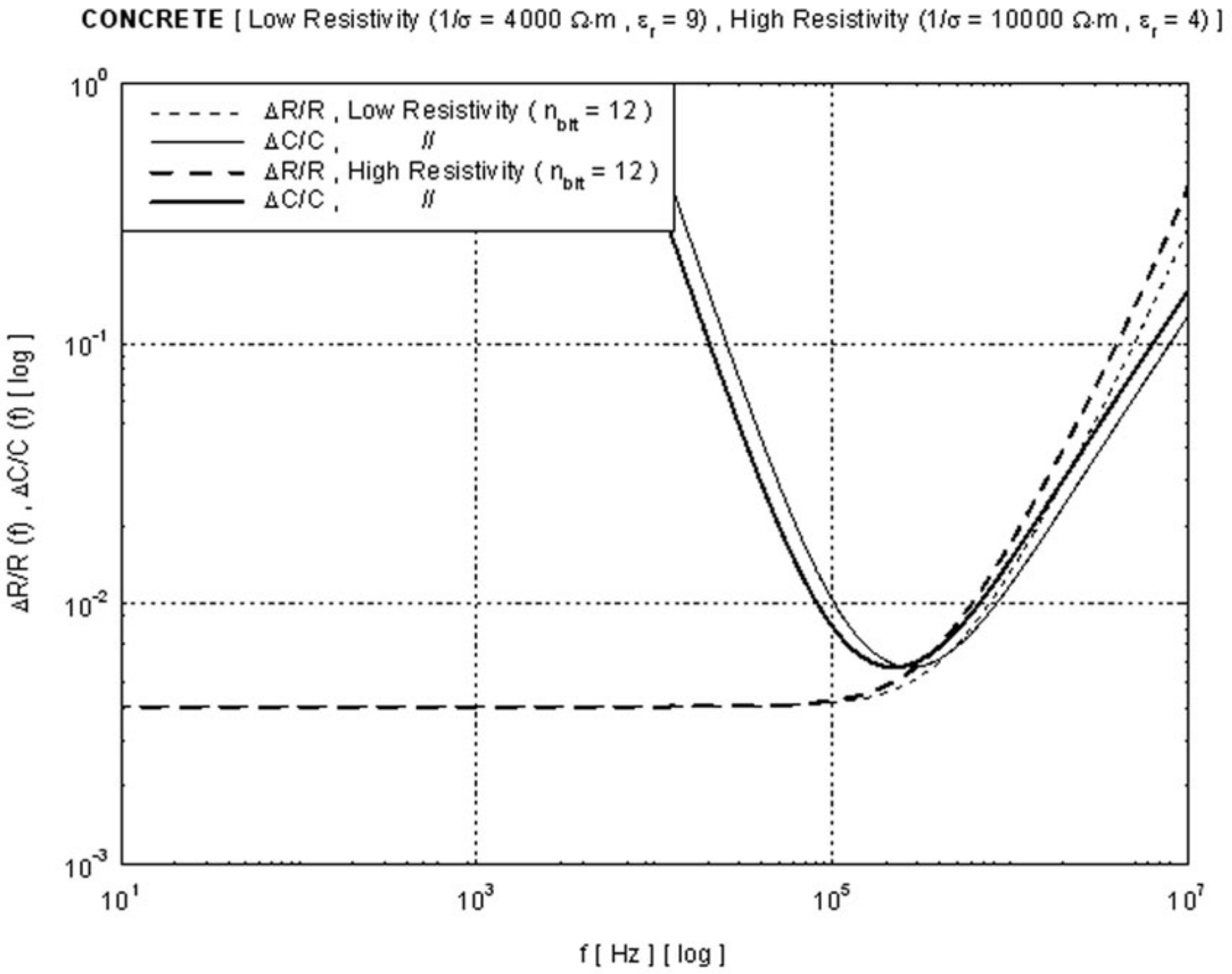

**Figure. 6.** Refer to the caption of Fig. 3. The RESPER probe is connected to an IQ ADC when the internal quartz is oscillating at its lowest merit factor $Q \approx 10^4$, so in the worst case. The bit number $n_{bit} = 18$ is used for soils characterized by a low electrical resistivity, while $n_{bit}=12$ for soils with high resistivity and for all the concretes. The RESPER is affected by an additional noise both in amplitude, due to the external environment (signal to noise ratio $SNR = 30\ dB$, averaged terms $A = 10^3$), and in phase, due to a phase-splitter detector specified by (phase inaccuracy, $\Delta\Phi_Z/\Phi_Z = 0.2°$). Bode's diagrams showing the inaccuracies $\Delta R/R(f)$ and $\Delta C/C(f)$ in the measurement of resistance $R$ and capacitance $C$, plotted as functions of the frequency $f$, for both the soils (a) and concretes (b) analyses (Tab. 2) [Settimi et al., 2010a-d].

If the RESPER grazes the subjacent medium [see Eq. (2.4): $x=0 \rightarrow \delta(x)=0$], then the electrical conductivity $\sigma$ and the dielectric permittivity $\varepsilon_r$ can be simply linked to the resistance $R$ and the capacitance $C$ components of complex impedance, [see Eqs. (2.8) and (2.9)]:

$$\sigma(R) = \frac{2\varepsilon_0}{RC_0}, \tag{2.13}$$

$$\varepsilon_r(C) = 2\frac{C}{C_0} - 1. \tag{2.14}$$

Only in this case, the inaccuracies $\Delta\sigma/\sigma$ and $\Delta\varepsilon_r/\varepsilon_r$ in the measurement of the conductivity $\sigma$ and permittivity $\varepsilon_r$ characterizing a medium coincide respectively with the inaccuracies $\Delta R/R(f)$ and $\Delta C/C(f)$ of the resistance $R$ and capacitance $C$ constituting the parallel RC circuit equivalent to the medium. In fact, elaborating Eqs (2.13) and (2.14):

$$\frac{\Delta\sigma}{\sigma} = \frac{\Delta R}{R}, \tag{2.15}$$

$$\frac{\Delta\varepsilon_r}{\varepsilon_r} = (1 + \frac{1}{\varepsilon_r})\frac{\Delta C}{C} \overset{\varepsilon_r > 1}{\simeq} \frac{\Delta C}{C}. \tag{2.16}$$

If the probe is connected to an ideal IQ sampler specified by the number of bits $n_{bit}$, then the cut-off frequency of the complex impedance coincides with the optimal frequency $f_{opt}$ (Tab. 2),

$$\omega_T = 2\pi f_T = \frac{\sigma}{\varepsilon_0(\varepsilon_r + 1)}, \tag{2.17}$$

which minimizes the inaccuracies $\Delta\sigma/\sigma$ and $\Delta\varepsilon_r/\varepsilon_r$ in the measurement of $\sigma$ and $\varepsilon_r$:

$$\left.\frac{\Delta\sigma}{\sigma}\right|_{IQ,\min} \approx \left.\frac{\Delta\varepsilon_r}{\varepsilon_r}\right|_{IQ,\min} \approx \frac{1}{2^{n_{bit}-2}}, \tag{2.18}$$

| *IQ DOWN-SAMPLING + Electric Noise (SNR = 30 dB, A = $10^3$) + Phase Shifter ($\Delta\Phi_Z/\Phi_Z$ = 0.2°)* | SOIL | CONCRETE |
|---|---|---|
| *Low Resistivity* | $f_{opt}$ = 4.135 MHz | $f_{opt}$ = 733.816 kHz |
| | $f_{min}$ = 270.663 kHz | $f_{min}$ = 76.256 kHz |
| | $f_{max}$ = 155.553 MHz | $f_{max}$ = 13.293MHz |
| *High Resistivity* | $f_{opt}$ = 275.144 kHz | $f_{opt}$ = 220.109 kHz |
| | $f_{min}$ = 26.881 kHz | $f_{min}$ = 22.877 kHz |
| | $f_{max}$ = 4.985 MHz | $f_{max}$ = 3.988 MHz |

**Table 2.** Refer to the caption of Fig. 6. A probe is connected to an IQ ADC (bit number $n_{bit}$). Data are shown for the optimal working frequency, $f_{opt}$, which minimizes the inaccuracy in the measurement of capacitance, $\Delta C/C(f)$, minimum and maximum frequencies, $f_{min}$ and $f_{max}$, that limit the inaccuracies of resistance and capacitance below a prefixed limit, $\Delta R/R(f) \leq 0.1$ and $\Delta C/C(f) \leq 0.1$, for measurements performed on soils characterized by a low electrical resistivity ($n_{bit}$= 18), on soils with high resistivity and all the concretes ($n_{bit}$= 12).

## 3. Design of the capacitive contact for terrestrial soils or concretes

The measurements taken using the induction probe are affected by errors that mainly originate from uncertainties associated with transfer impedance, from dishomogeneities between the modelled and the actual stratigraphy, and from inaccuracies of the electrode array deployment above the surface [Vannaroni et al., 2004]. Errors in impedance result mainly from uncertainties in the electronic systems that perform the amplitude and phase measurements of the voltages and currents [Del Vento and Vannaroni, 2005]. These uncertainties are assumed to be constant throughout the whole frequency band, even though their effects that propagate through the transfer function will produce a frequency-dependent perturbation.

*Sensitivity functions approach*

Our previous study [Settimi et al., 2010a] proposed to develop explicitly the sensitivity functions approach that is implied in the theory of error propagation suggested by Vannaroni et al. (2004). Indeed, this section recalls a mathematical–physical model for the propagation of errors in the measurement of electrical conductivity $\sigma$ and relative dielectric permittivity $\varepsilon_r$, based on the sensitivity functions tool [Murray-Smith, 1987]. This is useful for expressing inaccuracies in the measurements of conductivity and permittivity as a linear combination of the inaccuracies for the transfer impedance, both in modulus |Z| and in phase $\Phi_Z$, where the weight functions are inversely proportional only to the sensitivity functions for |Z| and $\Phi_Z$ relative to $\sigma$ and $\varepsilon_r$. The inaccuracies of transfer impedance depend on the inaccuracies of electrical voltage and current which are assigned by the electronics used and, in particular, by the sampling methods.

Therefore, the inaccuracies $\Delta\sigma/\sigma$ in the measurement of the electrical conductivity $\sigma$, and $\Delta\varepsilon_r/\varepsilon_r$ in the dielectric permittivity $\varepsilon_r$, can be expressed as a linear combination of the inaccuracies $\Delta|Z|/|Z|$ and $\Delta\Phi_Z/\Phi_Z$ in the measurement of the transfer impedance, respectively in modulus |Z| and in phase $\Phi_Z$,

$$\frac{\Delta\sigma}{\sigma}=\left|S_{|Z|}^{\sigma}\right|\frac{\Delta|Z|}{|Z|}+\left|S_{\Phi_Z}^{\sigma}\right|\frac{\Delta\Phi_Z}{\Phi_Z}=\frac{1}{\left|S_{\sigma}^{|Z|}\right|}\frac{\Delta|Z|}{|Z|}+\frac{1}{\left|S_{\sigma}^{\Phi_Z}\right|}\frac{\Delta\Phi_Z}{\Phi_Z} \quad , \text{ for } \quad \varepsilon_r=const\,, \tag{3.1}$$

$$\frac{\Delta\varepsilon_r}{\varepsilon_r}=\left|S_{|Z|}^{\varepsilon_r}\right|\frac{\Delta|Z|}{|Z|}+\left|S_{\Phi_Z}^{\varepsilon_r}\right|\frac{\Delta\Phi_Z}{\Phi_Z}=\frac{1}{\left|S_{\varepsilon_r}^{|Z|}\right|}\frac{\Delta|Z|}{|Z|}+\frac{1}{\left|S_{\varepsilon_r}^{\Phi_Z}\right|}\frac{\Delta\Phi_Z}{\Phi_Z} \quad , \text{ for } \quad \sigma=const\,, \tag{3.2}$$

where ($S_{\sigma}^{|Z|}$, $S_{\sigma}^{\Phi_Z}$) and ($S_{\varepsilon_r}^{|Z|}$, $S_{\varepsilon_r}^{\Phi_Z}$) are the pairs of sensitivity functions for the transfer impedance, both in |Z| and $\Phi_Z$, relative to the conductivity $\sigma$ and permittivity $\varepsilon_r$, whose expressions are reported by Settimi et al. (2010a). The conditions $\sigma=const$ and $\varepsilon_r=const$ in Eqs. (3.1) and (3.2) underline not so much that constant values of electrical conductivity and dielectric permittivity are used to estimate the complex impedance over various terrains and concretes, but that the quantities $\sigma$ and $\varepsilon_r$ are not independent of each other, since the electrical displacement shows a phase-shift with respect to the electrical field [Frolich, 1990]. So, for the need to distinguish the inaccuracies in measurements of conductivity and permittivity, the inaccuracy $\Delta\sigma/\sigma$ can only be calculated assuming there is no uncertainty for $\varepsilon_r$ ($\Delta\varepsilon_r/\varepsilon_r=0 \Leftrightarrow \varepsilon_r=const$), and *vice versa*.

The interesting physical results obtained using this sensitivity functions approach are discussed by Settimi (2010a).

In particular, if the RESPER probe shows a galvanic contact with a subjacent medium, i.e. $h=0$, then the inaccuracies $\Delta\sigma/\sigma$ in the measurement of the electrical conductivity $\sigma$, and $\Delta\varepsilon_r/\varepsilon_r$ for the dielectric permittivity $\varepsilon_r$, are minimized in the frequency band $B$ of the RESPER, for all of its geometrical configurations and media, and even if $h\neq 0$, the design of the RESPER probe must still be optimized with respect to the minimum value of the inaccuracy $\Delta\varepsilon_r/\varepsilon_r$ for permittivity $\varepsilon_r$, which is always higher than the corresponding minimum value of the inaccuracy $\Delta\sigma/\sigma$ for conductivity $\sigma$ in the band $B$ of the probe, for all of its configurations and media [Tabbagh et al., 1993; Vannaroni et al., 2004].

Under the quasi static approximation [Tabbagh et al., 1993], consider a RESPER probe grazing a medium, i.e. $x=h/L=0$. Then, the sensitivities function $S_{\sigma}^{|Z|}$ and $S_{\sigma}^{\Phi_Z}$, relative to the electrical conductivity $\sigma$, and $S_{\varepsilon_r}^{|Z|}$ and $S_{\varepsilon_r}^{\Phi_Z}$, relative to the dielectric permittivity $\varepsilon_r$, for the transfer impedance, both in modulus |Z|

and in phase $\Phi_Z$, are independent of the characteristic geometrical dimension $L$ of the RESPER, i.e. $S_{\sigma,\varepsilon_r}^{|Z|,\Phi_Z}(f,L) = S_{\sigma,\varepsilon_r}^{|Z|,\Phi_Z}(f)$. In fact, the sensitivities functions are defined as normalized functions [Murray-Smith, 1987]. In simpler terms, our mathematical–physical model [Settimi et al., 2010a] predicts that, only if $x=0$, then the inaccuracies $\Delta\sigma/\sigma$, in the measurement of conductivity $\sigma$, and $\Delta\varepsilon_r/\varepsilon_r$, for permittivity $\varepsilon_r$, are invariant with the Wenner's or dipole-dipole arrays (Fig. 6). So, the probe is characterized by the same performances in the frequency band $B$ and in the measurable ranges of $\sigma$ and $\varepsilon_r$ (Tab. 2).

*Transfer functions method*

Our previous study [Settimi et al., 2010a] proposed to deepen the transfer functions method by analyzing the zero and pole behavior, which were implied in the frequency domain analysis suggested by Grard and Tabbagh (1991). Indeed, this section recalls the method of analysis in the frequency domain for determining the behavior of the zero and pole frequencies in the LTI circuit of the induction probe.

To satisfy the operative conditions of linearity for the measurements, if the RESPER probe shows a capacitive contact with the subjacent medium then the frequency $f$ of the RESPER should be imposed as included between the zero $z_M$ and the pole $p_M$ of the transfer impedance, and so its modulus is almost constant within the frequency band [Grard and Tabbagh, 1991],

$$z_M(x,\varepsilon_r,\sigma) \le f \le p_M(x,\varepsilon_r,\sigma) . \tag{3.3}$$

Based on the above conditions, an optimization equation is deduced for the probe that links the optimal ratio $x$ between its height $h$ above ground and its characteristic geometrical dimensions $L$ only to the dielectric permittivity $\varepsilon_r$ of the medium, so that:

$$\delta(x) \cong \frac{2}{15\varepsilon_r + 17} . \tag{3.4}$$

To satisfy the operative conditions of linearity for the measurements, if the RESPER probe shows a galvanic contact with the medium, then the working frequency $f$ of the RESPER should be imposed as lower than the cut-off frequency $f_T=f_T(\sigma,\varepsilon_r)$ of the transfer impedance, and so its modulus as constant below the cut-off frequency $f_T$. It is only under these conditions that it is optimal to design the electrode-electrode distance of the probe or to establish the measurable ranges of the conductivity $\sigma$ and permittivity $\varepsilon_r$ of the medium. Eqs. (3.3) and (3.4), derived by the classical transfer function method, have been demonstrated by Settimi et al. (2010a).

### 3.1 Wenner's and dipole-dipole arrays: geometrical factor

The transfer impedance of an induction probe can be evaluated for any arbitrary arrays. As a general rule, it is assumed that subsurface electrical sounding becomes scarcely effective at depths greater than the horizontal distance between the electrodes [Grard and Tabbagh, 1991; Vannaroni et al., 2004].

Our study considers two kinds of probes, i.e. with Wenner's and dipole-dipole arrays. The Wenner's (Fig. 7.a) consists of four electrodes along a straight horizontal line, separated by equal intervals, denoted *L*. Instead, the convention for the dipole-dipole (Fig. 7.b) is that current and voltage spacing is the same, *L*, and the spacing between them is an integer *n* multiple of *L*.

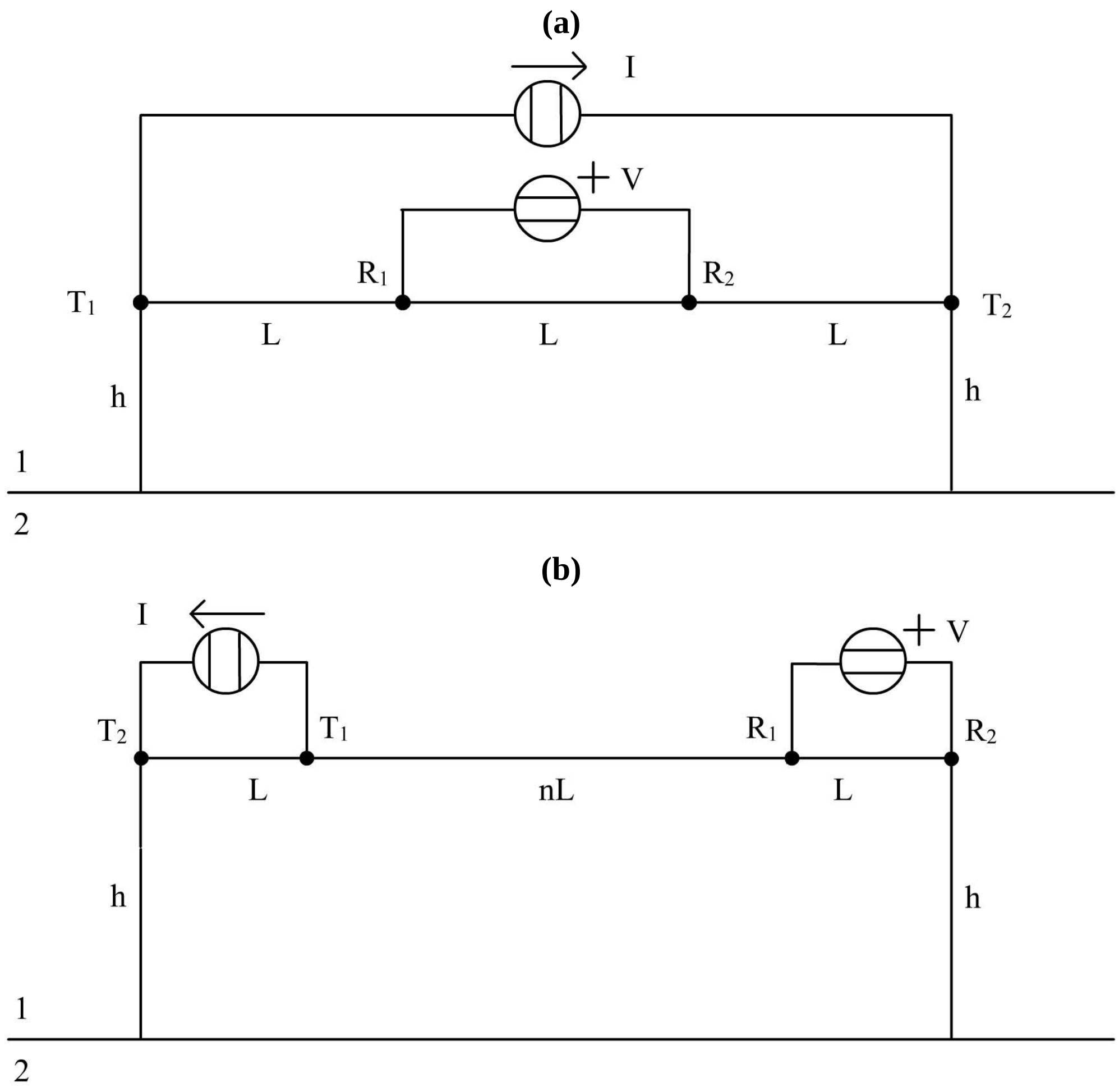

**Figure 7.** The geometry of the Wenner's (a) or dipole-dipole (b) arrays.

Once fixed the total length $L_{TOT}$ of the RESPER probe, the characteristic geometrical dimension is $L=L_{TOT}/3$ for the Wenner's (Fig. 7.a) and $L=L_{TOT}/(n+2)$ for the dipole-dipole (Fig. 7.b) array defined by the integer parameter $n \in \mathbb{N} = \{1, 2, 3, \dots\}$, i.e.:

$$L = \begin{cases} L_{TOT}/3 & \text{for Wenner's} \\ L_{TOT}/(n+2) \quad , \quad n \in \mathbb{N} & \text{for dipole-dipole} \end{cases} . \tag{3.5}$$

So, the Wenner's (Fig. 7.a) measures the capacitance in a vacuum

$$C_0^{(W)}(L) = 4\pi\varepsilon_0 L \tag{3.6}$$

and, once defined the ratio $x=h/L$ between the height above ground $h$ and the electrode-electrode distance $L$, Eq. (2.3), it can be entirely specified by the geometrical factor [see Appendix A]

$$K^{(W)}(x) = 2\left(\frac{1}{\sqrt{1+4x^2}} - \frac{1}{2\sqrt{1+x^2}}\right), \tag{3.7}$$

such that: $K^{(W)}(x=0)=1$.

Instead, the dipole-dipole (Fig. 7.b) measures a capacitance in a vacuum

$$C_0^{(DD)}(L,n) = 2\pi\varepsilon_0 Ln(n+1)(n+2) = C_0^{(W)}(L)\cdot\frac{n(n+1)(n+2)}{2} \tag{3.8}$$

and, once introduced the integer parameter $n \in \mathbb{N} = \{1,2,3,\ldots\}$, it can be entirely specified by the geometrical factor [Appendix A]

$$K^{(DD)}(x,n) = \frac{1}{2}n(n+1)(n+2)\left(\frac{1}{\sqrt{n^2+4x^2}} + \frac{1}{\sqrt{(n+2)^2+4x^2}} - \frac{2}{\sqrt{(n+1)^2+4x^2}}\right), \tag{3.9}$$

being: $K^{(DD)}(x=0,n)=1$, $\forall n \in \mathbb{N}$.

When the RESPER probe shows a capacitive contact with the subjacent medium, and so the height/dimension ratio $x=h/L$ is not zero, i.e. $0<x\leq 1$, then the RESPER is characterized by a geometrical factor $\delta(x)=1-K(x)$, increasing function of $x$, which in the Wenner's array slopes up less (more) steeply than in the dipole-dipole defined by $n=1$ ($n=6$), so assuming smaller (or larger) values especially for $1/2<x<1$ (Fig. 8). As a consequence, a RESPER probe, with a fixed characteristic geometrical dimension $L$, that performs measurements on a medium of dielectric permittivity $\varepsilon_r$ could be designed with an optimal ratio $x_0=h_0/L$, which in the Wenner's array is larger (smaller) than in the dipole-dipole with $n=1$ ($n=6$), because its factor $\delta(x)$ slopes up less (more) steeply, increasing the ratio $x$, so reaching the prefixed optimal value $\delta_0(\varepsilon_r)\approx 2/(15\varepsilon_r+17)$ at a smaller $x_0$. In simpler terms, if the probe is in capacitive contact with the medium, to perform optimal measurements of the permittivity, the Wenner's array needs to be raised above ground by more (less) than in the dipole-dipole with $n=1$ ($n=6$), if their electrode-electrode distances are equal (Tab. 3).

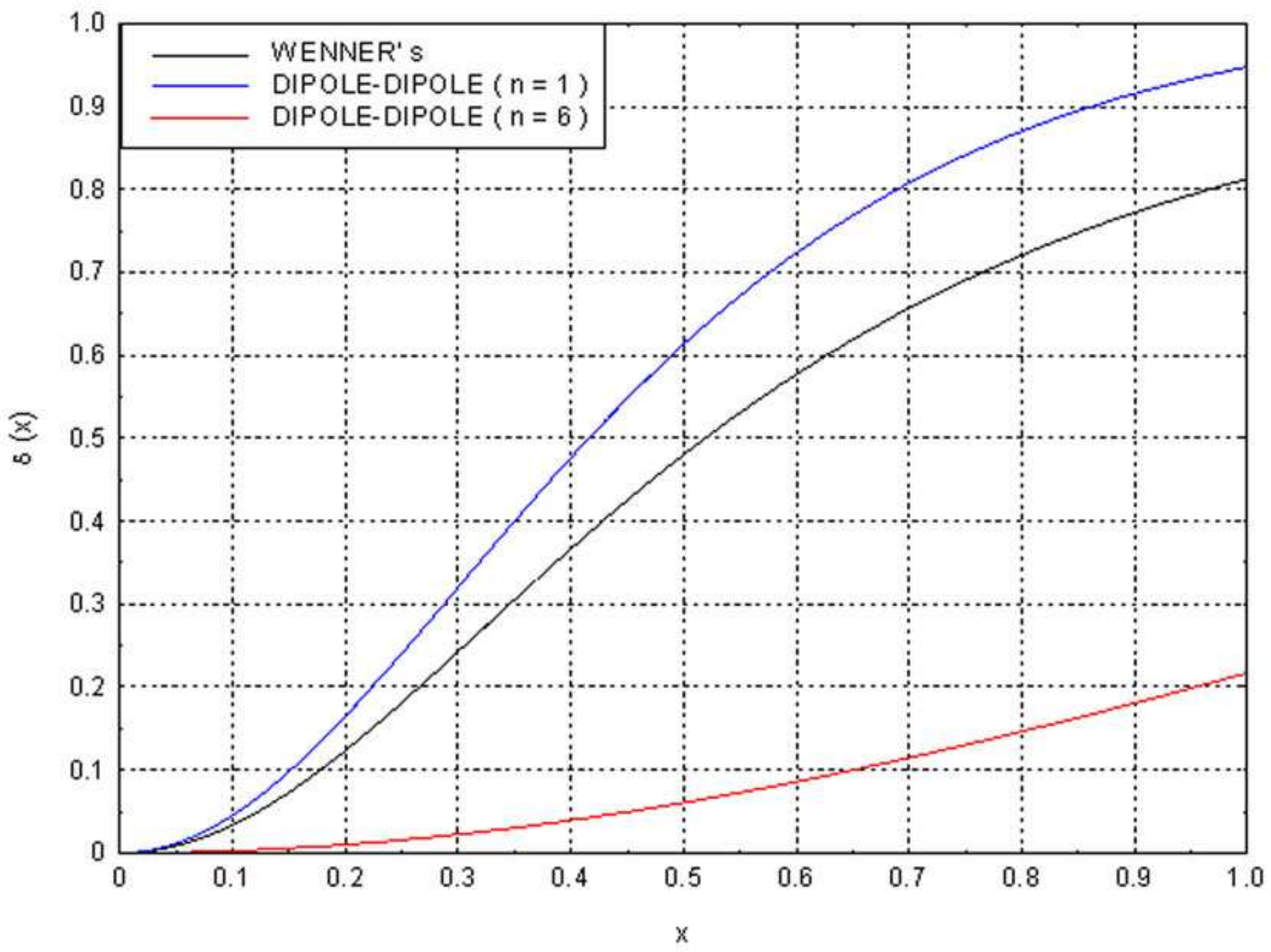

**Figure 8.** The RESPER probe shows a capacitive contact on a subjacent medium and is configured in the Wenner's or dipole-dipole ($n=1,6$) array. Plots showing the geometrical factor $\delta(x)$, as a function of the ratio $x=h/L$ between the height $h$ above ground and the characteristic geometrical dimension $L$, in the range $x \in (0,1]$.

Moreover, in the case of capacitive contact, if the RESPER probe with a characteristic geometrical dimension $L$ is designed by the optimal height/dimension ratio $x_0=h_0/L$ working in a frequency $f$, then the transfer impedance $Z(f,x_0)$, in units of $1/h$, calculated in $x_0$, is defined by suitable phase $\Phi(f,x_0)$ and modulus $|Z|(f,x_0)$. The phase $\Phi(f,x_0)$ does not depend on the Wenner's or dipole-dipole arrays (only choosing $n_0=1$). Instead, the modulus $|Z|(f,x_0)$, in units of $1/h$, in the Wenner's is shifted up compared to the dipole-dipole array ($n_0=1$) by a factor $n_0(n_0+1)(n_0+2)/2=3$ (Figs. 9,11); though it remains almost unvaried in both the afore mentioned arrays not only the shape of modulus $|Z|(f,x_0)$, but also the position of its zero $z_M(x_0)$ and pole $p_M(x_0)$ frequencies (Tab. 3).

The inaccuracies $\Delta\sigma/\sigma(f,x_0)$ in the measurements of the conductivity $\sigma$ and $\Delta\varepsilon_r/\varepsilon_r(f,x_0)$ for the permittivity $\varepsilon_r$, calculated in $x_0$, do not depend on the Wenner's or dipole-dipole arrays (only choosing $n_0=1$). If the upper limit $f_{up}$ of LF band is defined as the frequency where the inaccuracy $\Delta\sigma/\sigma(f,x_0)$ in measurement of $\sigma$ shows a point of minimum, i.e. $\partial_f\, \Delta\sigma/\sigma(f,x_0)|_{f=fup}=0$; and the lower limit $f_{low}$ of MF band as the frequency where the inaccuracy $\Delta\varepsilon_r/\varepsilon_r(f,x_0)$ in measurement of $\varepsilon_r$ shows a point of minimum, i.e. $\partial_f\, \Delta\varepsilon_r/\varepsilon_r(f,x_0)|_{f=flow}=0$ [Settimi et al., 2010c]: then the upper limit $f_{up}$ of LF band and the lower limit $f_{low}$ of MF band result invariant in both the arrays (Fig.10, 12; Tab. 3).

In simpler terms, to perform an optimal measurement of permittivity considering different height/dimension ratios, the design of both the two above arrays establishes (almost) invariant trends in frequency, both for their transfer impedances and measurement inaccuracies. In other words, once selected a subsurface to be analyzed, there exists one and only one "optimum"; anyway sized both the arrays, there is that height "ad hoc", which gives rise to that one "optimum"; the optimal inaccuracy depends on the surface but not on the arrays and is achieved by adjusting their heights.

Finally, compared to the dipole-dipole array defined by $n_0=1$, the dipole-dipole with $n=6$ is characterized especially by a zero $z_M(x_0)$ (and an upper limit $f_{up}$ of LF band) left shifted towards lower frequencies (Tab. 3a, b), like both the modulus $|Z|(f,x_0)$, in units of $1/h$ and the phase $\Phi(f,x_0)$ (or the inaccuracy $\Delta\sigma/\sigma(f,x_0)$ in the measurement of $\sigma$) (Figs. 11, 12). The modulus $|Z|(f,x_0)$ is shifted downwards until values smaller of almost two magnitude orders [by a factor $(3/2)n(n+1)(n+2)$], similarly to the phase $\Phi(f,x_0)$ (or the inaccuracy $\Delta\sigma/\sigma(f,x_0)$ just partially) which shows a squashed shape around its minimum (or the upper limit $f_{up}$). In simpler terms, the more the integer parameter $n$ increases, the more the working frequency can be modulated at LF band, reducing the measurement inaccuracies; in other words, this procedure allows the dipole-dipole array improving its performances.

**(a)**

| ***CAPACITIVE CONTACT: (W) Wenner' s (DD) Dipole-dipole (n =1)*** | **SOIL** | | | **CONCRETE** | | |
|---|---|---|---|---|---|---|
| *Low Resistivity* | ***(W)*** $x_0$ = 0.052 ***(DD)*** $x_0$ = 0.045 | $z_M(x_0)$ = 3.878MHz | $p_M(x_0)$ = 15.514MHz | ***(W)*** $x_0$ = 0.062 ***(DD)*** $x_0$ = 0.053 | $z_M(x_0)$ = 176.47kHz | $p_M(x_0)$ = 705.881kHz |
| | $f_{up}$ = 832.206kHz | $\Delta\sigma/\sigma(f_{up}, x_0)$ = $6.553\cdot10^{-3}$ | $\Delta\varepsilon_r/\varepsilon_r(f_{up}, x_0)$ = 0.053 | $f_{up}$ = 38.772kHz | $\Delta\sigma/\sigma(f_{up}, x_0)$ = $7.398\cdot10^{-3}$ | $\Delta\varepsilon_r/\varepsilon_r(f_{up}, x_0)$ = 0.063 |
| | $f_{low}$ = 6.348MHz | $\Delta\sigma/\sigma(f_{low}, x_0)$ = $6.62\cdot10^{-3}$ | $\Delta\varepsilon_r/\varepsilon_r(f_{low}, x_0)$ = $5.92\cdot10^{-3}$ | $f_{low}$ = 343.101kHz | $\Delta\sigma/\sigma(f_{low}, x_0)$ = $7.34\cdot10^{-3}$ | $\Delta\varepsilon_r/\varepsilon_r(f_{low}, x_0)$ = $7.503\cdot10^{-3}$ |
| *High Resistivity* | ***(W)*** $x_0$ = 0.087 ***(DD)*** $x_0$ = 0.075 | $z_M(x_0)$ = 470.587kHz | $p_M(x_0)$ = 1.882MHz | ***(W)*** $x_0$ = 0.087 ***(DD)*** $x_0$ = 0.075 | $z_M(x_0)$ = 141.176kHz | $p_M(x_0)$ = 564.705kHz |
| | $f_{up}$ = 103.637kHz | $\Delta\sigma/\sigma(f_{up}, x_0)$ = $7.485\cdot10^{-3}$ | $\Delta\varepsilon_r/\varepsilon_r(f_{up}, x_0)$ = 0.071 | $f_{up}$ = 31.091kHz | $\Delta\sigma/\sigma(f_{up}, x_0)$ = $7.485\cdot10^{-3}$ | $\Delta\varepsilon_r/\varepsilon_r(f_{up}, x_0)$ = 0.071 |
| | $f_{low}$ = 917.365kHz | $\Delta\sigma/\sigma(f_{low}, x_0)$ = $7.353\cdot10^{-3}$ | $\Delta\varepsilon_r/\varepsilon_r(f_{low}, x_0)$ = $8.446\cdot10^{-3}$ | $f_{low}$ = 275.209kHz | $\Delta\sigma/\sigma(f_{low}, x_0)$ = $7.535\cdot10^{-3}$ | $\Delta\varepsilon_r/\varepsilon_r(f_{low}, x_0)$ = $8.446\cdot10^{-3}$ |

**(b)**

| ***CAPACITIVE CONTACT: (DD) Dipole-dipole (n = 6)*** | **SOIL** | | | **CONCRETE** | | |
|---|---|---|---|---|---|---|
| *Low Resistivity* | $x_0$ = 0.045 | $z_M(x_0)$ = 220.07kHz | $p_M(x_0)$ = 3.695MHz | $x_0$ = 0.053 | $z_M(x_0)$ = 10.005kHz | $p_M(x_0)$ = 168.704kHz |
| | $f_{up}$ = 106.838kHz | $\Delta\sigma/\sigma(f_{up}, x_0)$ = $4.09\cdot10^{-3}$ | $\Delta\varepsilon_r/\varepsilon_r(f_{up}, x_0)$ = 0.15 | $f_{up}$ = 4.969kHz | $\Delta\sigma/\sigma(f_{up}, x_0)$ = $4.632\cdot10^{-3}$ | $\Delta\varepsilon_r/\varepsilon_r(f_{up}, x_0)$ = 0.243 |
| | $f_{low}$ = 4.362MHz | $\Delta\sigma/\sigma(f_{low}, x_0)$ = $4.178\cdot10^{-3}$ | $\Delta\varepsilon_r/\varepsilon_r(f_{low}, x_0)$ = $4.769\cdot10^{-3}$ | $f_{low}$ = 280.097kHz | $\Delta\sigma/\sigma(f_{low}, x_0)$ = $5.172\cdot10^{-3}$ | $\Delta\varepsilon_r/\varepsilon_r(f_{low}, x_0)$ = $6.442\cdot10^{-3}$ |
| *High Resistivity* | $x_0$ = 0.075 | $z_M(x_0)$ = 26.602kHz | $p_M(x_0)$ = 447.547kHz | $x_0$ = 0.075 | $z_M(x_0)$ = 7.981kHz | $p_M(x_0)$ = 134.264kHz |
| | $f_{up}$ = 13.232kHz | $\Delta\sigma/\sigma(f_{up}, x_0)$ = $4.623\cdot10^{-3}$ | $\Delta\varepsilon_r/\varepsilon_r(f_{up}, x_0)$ = 0.274 | $f_{up}$ = 3.97kHz | $\Delta\sigma/\sigma(f_{up}, x_0)$ = $4.633\cdot10^{-3}$ | $\Delta\varepsilon_r/\varepsilon_r(f_{up}, x_0)$ = 0.274 |
| | $f_{low}$ = 746.901kHz | $\Delta\sigma/\sigma(f_{low}, x_0)$ = $5.171\cdot10^{-3}$ | $\Delta\varepsilon_r/\varepsilon_r(f_{low}, x_0)$ = $7.347\cdot10^{-3}$ | $f_{low}$ = 224.07kHz | $\Delta\sigma/\sigma(f_{low}, x_0)$ = $5.171\cdot10^{-3}$ | $\Delta\varepsilon_r/\varepsilon_r(f_{low}, x_0)$ = $7.247\cdot10^{-3}$ |

**Table 3.** Refer to the captions of Figs. 3, 6 and 8. The RESPER can show a capacitive contact both on non-saturated terrestrial soils and concretes with low or high electrical resistivity, it is connected to an IQ ADC and is configured in the Wenner's (W) or dipole-dipole (DD) array defined by $n=1$ (a) and $n=6$ (b). Data are shown for: the height/dimension ratio $x=x_0$ optimally sized for W or DD array; both the low frequency (LF) zero $z_M(x_0)$ and the middle frequency (MF) pole $p_M(x_0)$ of complex impedance; both the upper limit $f_{up}$ of LF band and the lower limit $f_{low}$ of MF band; and all the inaccuracies $\Delta\sigma/\sigma(f_{up,low}, x_0)$ and $\Delta\varepsilon_r/\varepsilon_r(f_{up,low}, x_0)$ in measurement of the electrical conductivity and dielectric permittivity.

**(a)**

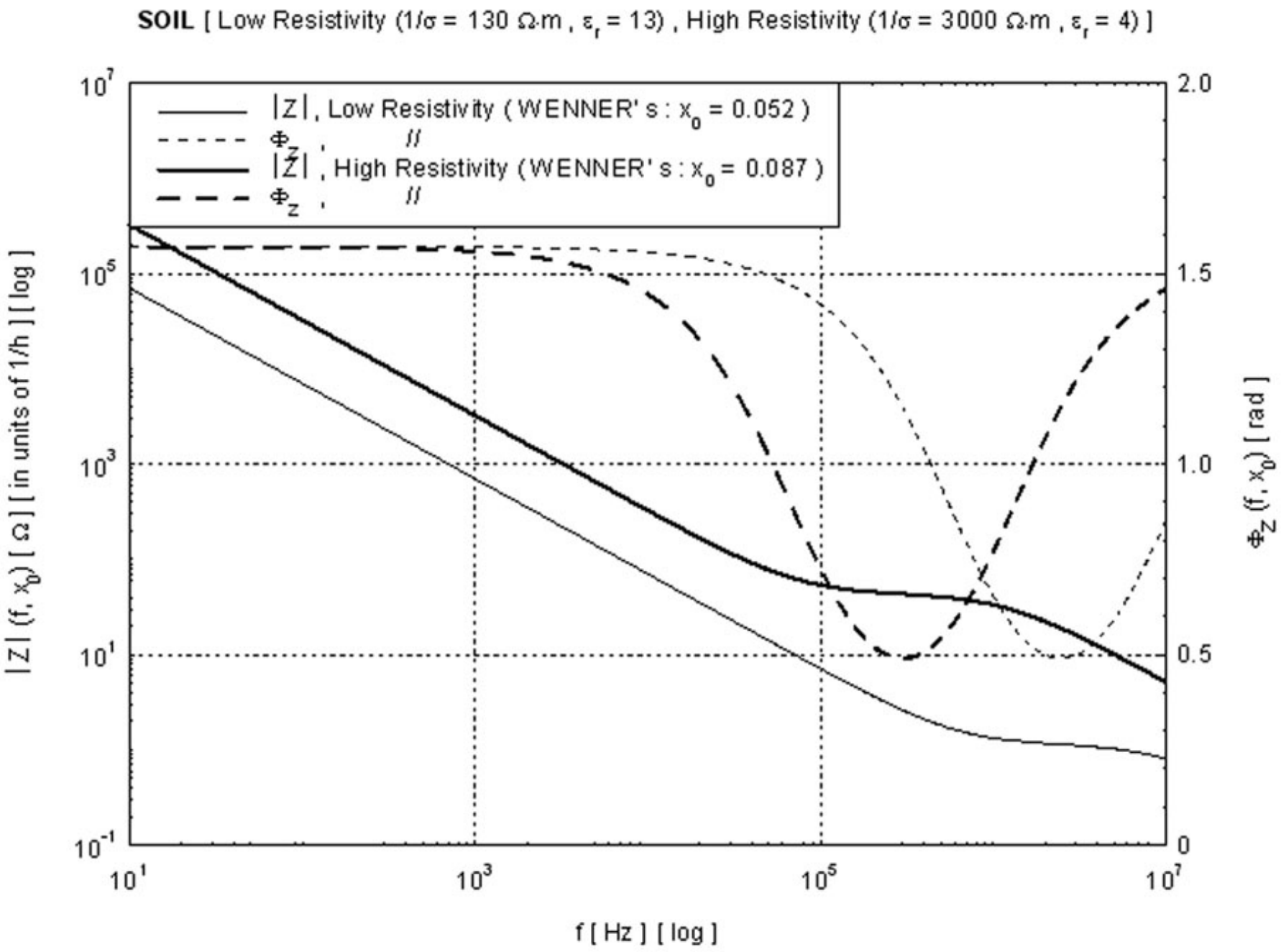

**(b)**

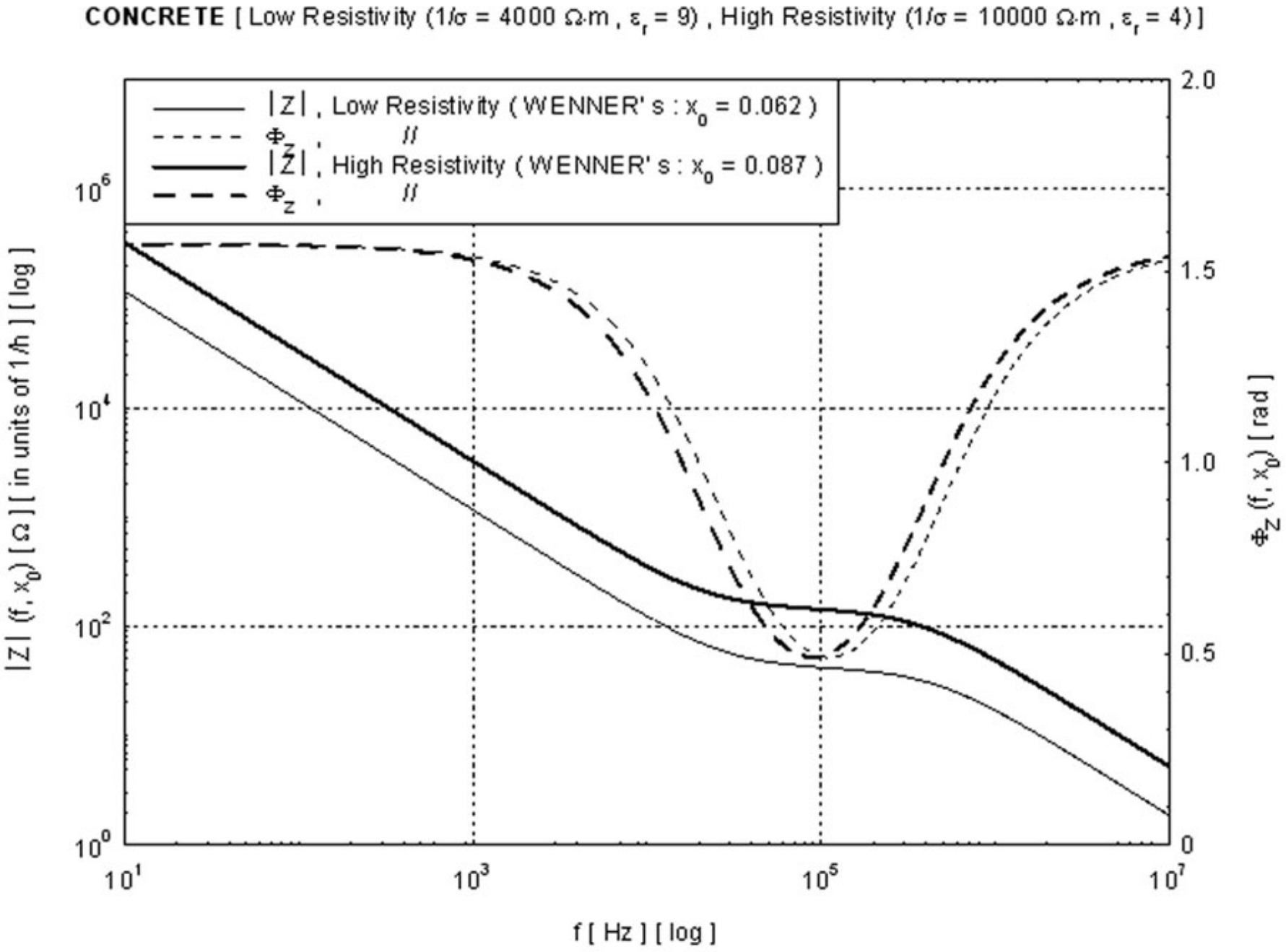

**Figure 9.** Refer to the caption of Tab. 3. The probe is configured in the Wenner's (W) array and it shows a capacitive contact on a subjacent medium. The height/dimension ratio $x=x_0$ is optimally sized for W array, according to the terrestrial soils and concretes. Bode's diagrams showing the complex impedance both in modulus $|Z|(f,x_0)$ [units of *1/h*] and phase $\Phi_Z(f,x_0)$, plotted as functions of the frequency *f*, in the band $f \in [0, f_{\lim}]$ with $f_{lim}=10$ *MHz*, for both the soils (a) and concretes (b) analyses.

**(a)**

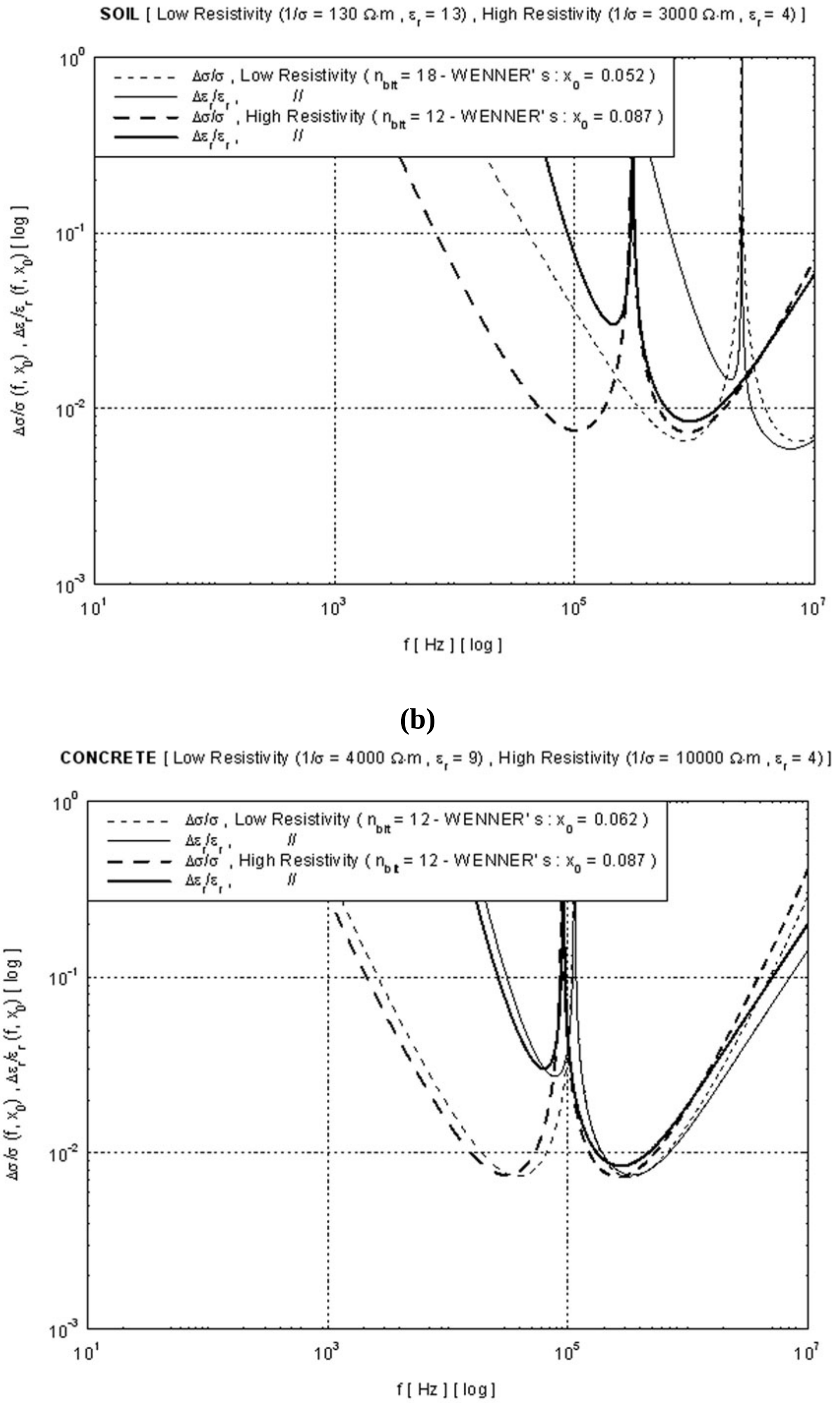

**Figure 10.** Refer to the captions of Tab. 3 and Fig. 9. The probe is configured in the Wenner's array and it shows a capacitive contact on a medium. Bode's diagrams showing the inaccuracies $\Delta\sigma/\sigma(f, x_0)$ and $\Delta\varepsilon_r/\varepsilon_r(f, x_0)$ in the measurement of conductivity $\sigma$ and permittivity $\varepsilon_r$, plotted as functions of the frequency $f$, for both the soils (a) and concretes (b) analyses [Settimi et al., 2010a-d].

**(a)**

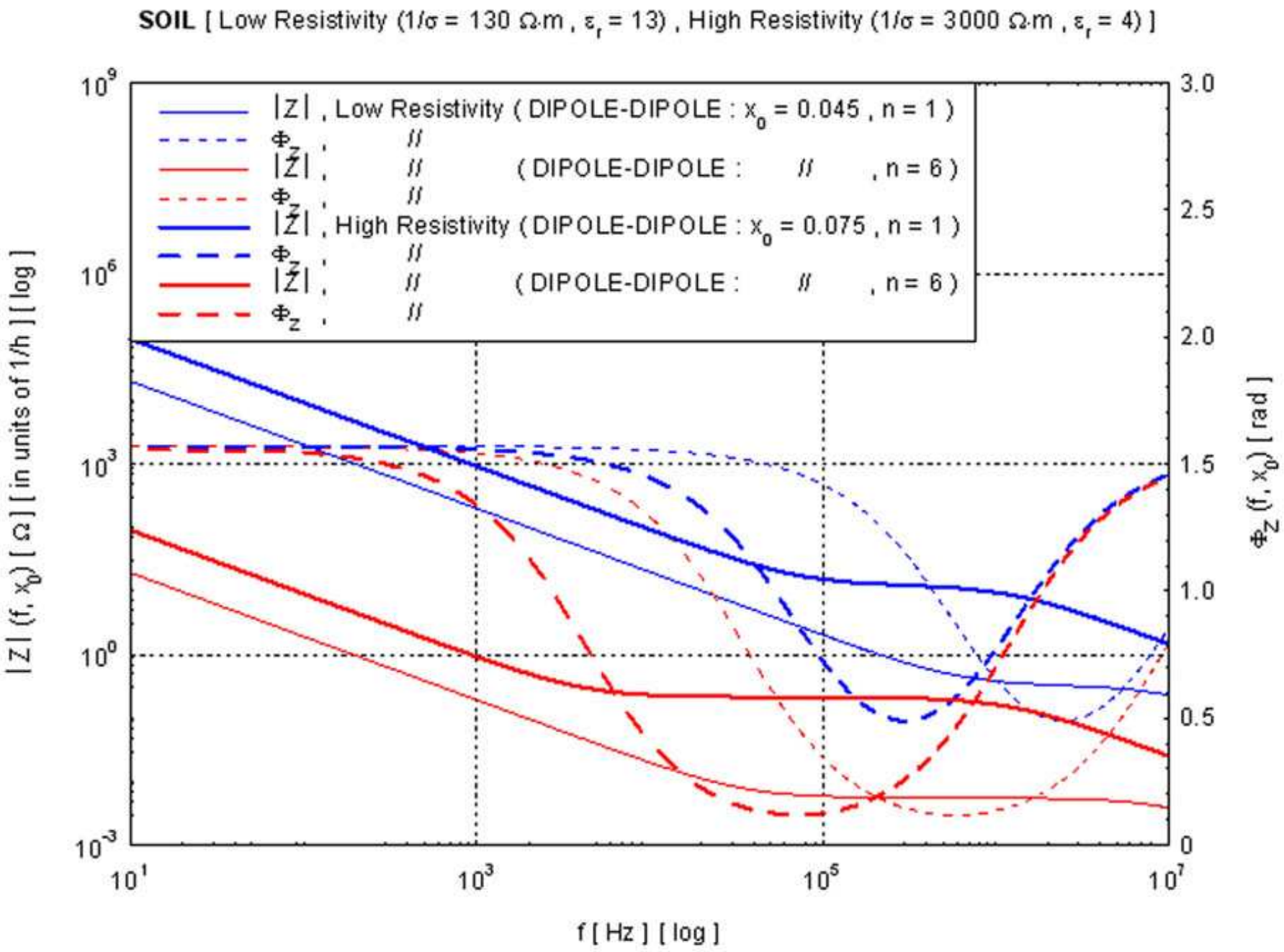

**(b)**

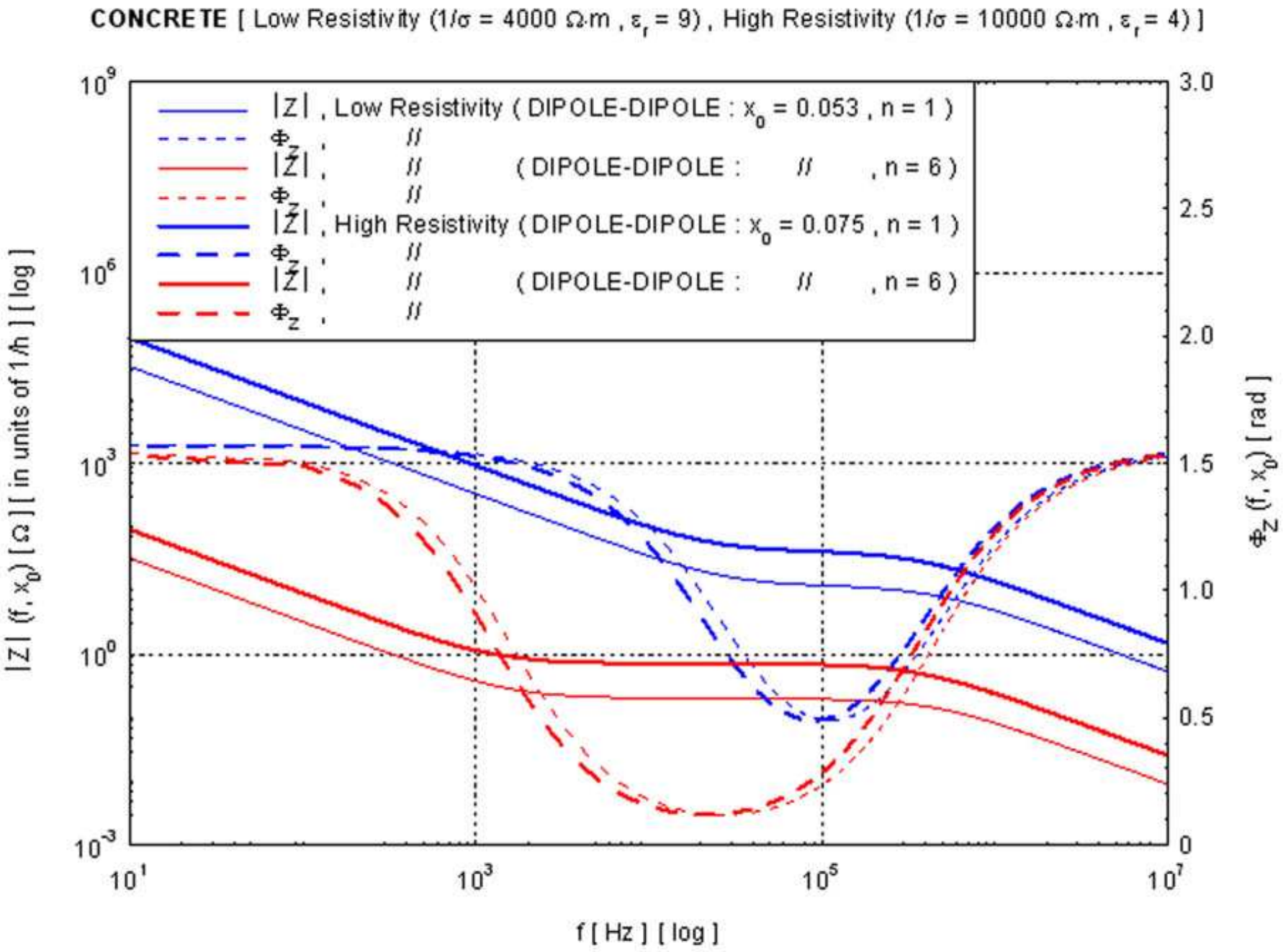

**Figure 11.** Refer to the caption of Tab. 3. The probe is configured in the dipole-dipole (DD) array (*n=1,6*) and it shows a capacitive contact on a subjacent medium. The height/ dimension ratio $x=x_0$ is optimally sized for DD array, according to the terrestrial soils and concretes. Bode's diagrams showing the complex impedance both in modulus $|Z|(f,x_0)$ [units of *1/h*] and phase $\Phi_Z(f,x_0)$, plotted as functions of the frequency *f*, in the band $f \in [0, f_{\lim}]$ with $f_{lim}=10$ *MHz*, for both the soils (a) and concretes (b) analyses

**(a)**

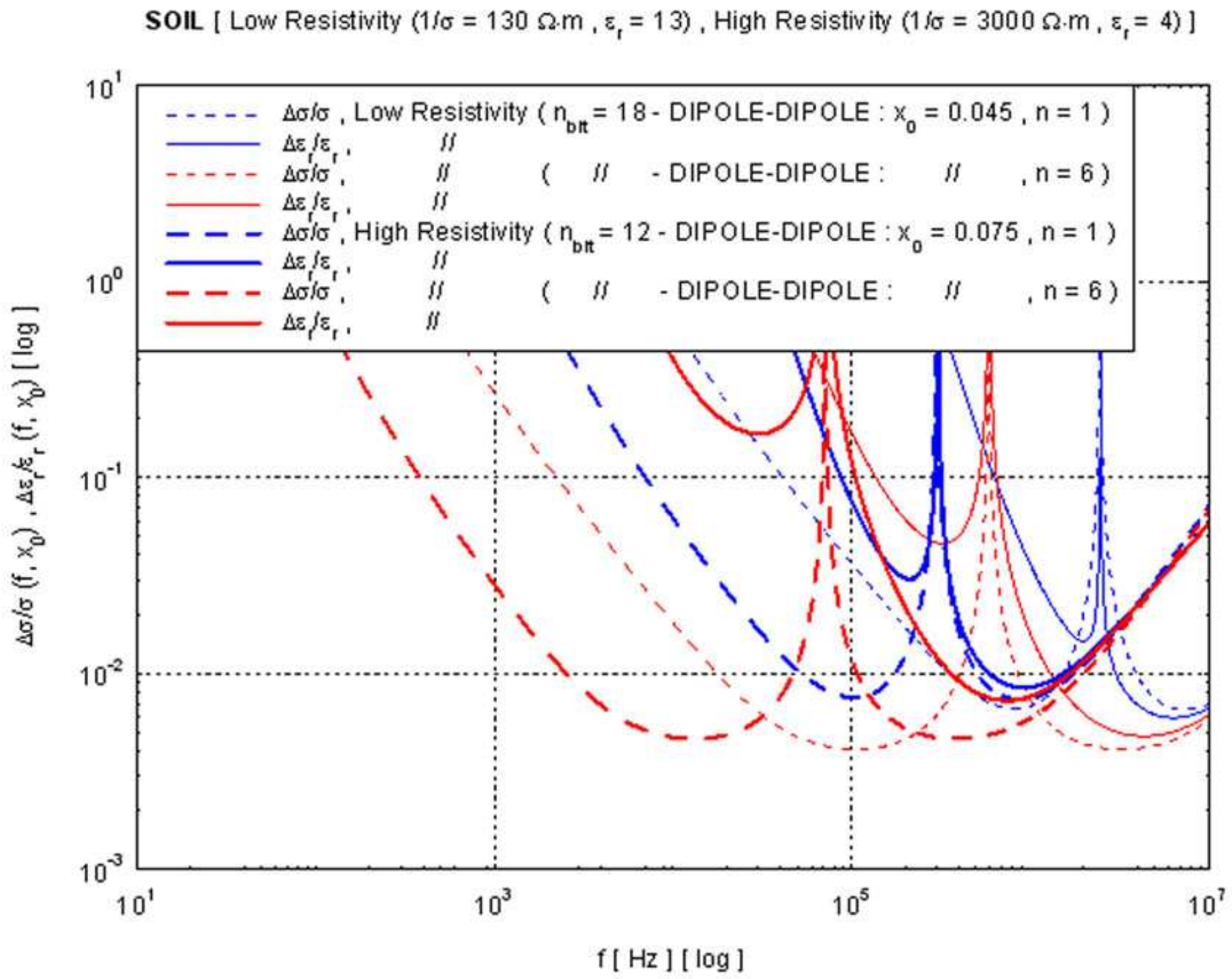

**(b)**

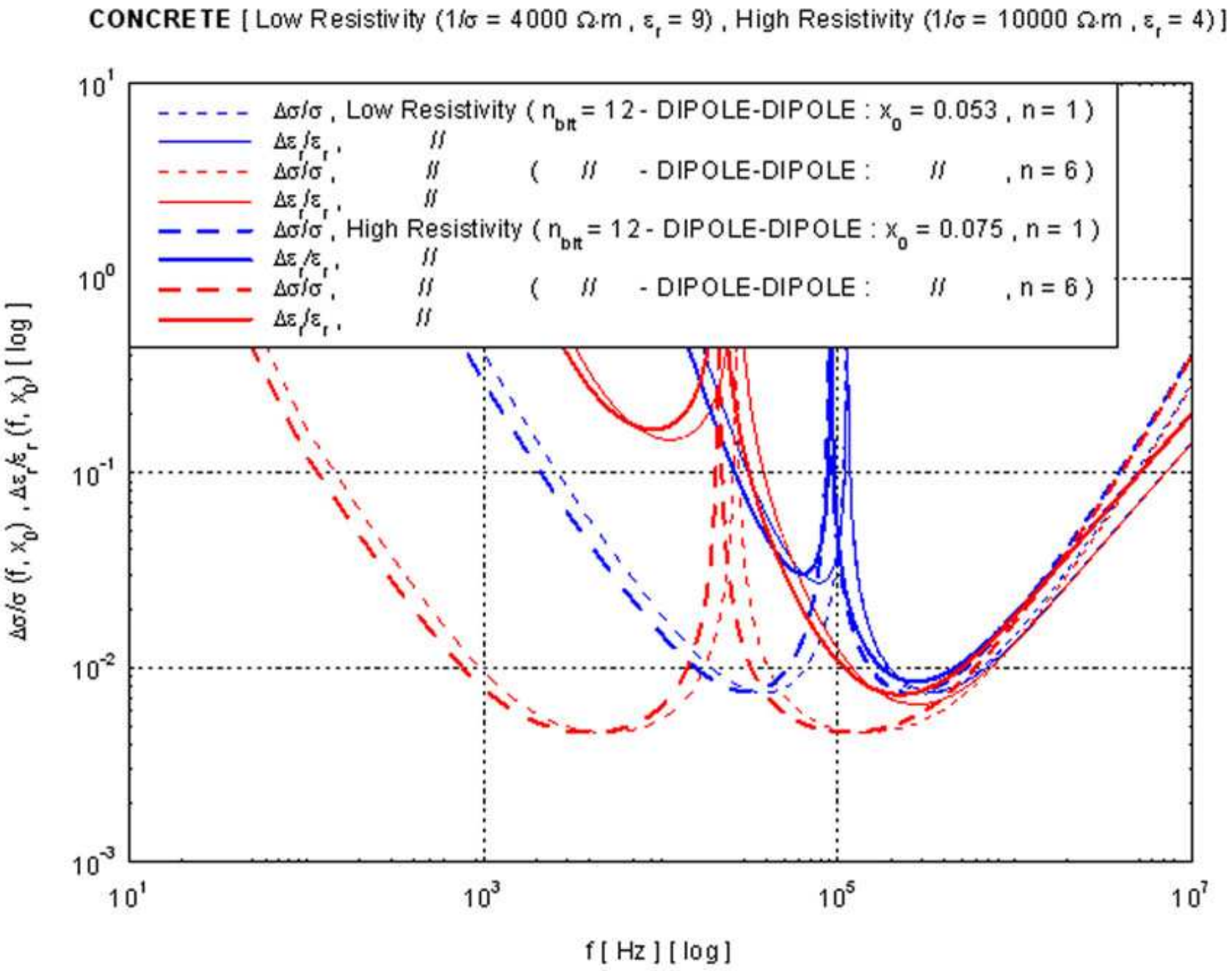

**Figure 12.** Refer to the captions of Tab. 3 and Fig. 11. The probe is configured in the dipole-dipole array (*n=1,6*) and it shows a capacitive contact on a medium. Bode's diagrams showing the inaccuracies $\Delta\sigma/\sigma(f, x_0)$ and $\Delta\varepsilon_r/\varepsilon_r(f, x_0)$ in the measurement of conductivity $\sigma$ and permittivity $\varepsilon_r$, plotted as functions of the frequency *f*, for both the soils (a) and concretes (b) analyses [Settimi et al., 2010a-d].

## 4. Characteristic geometrical dimensions

This section refers to Vannaroni et al. (2004), who discussed the dependence of the transmitting (TX) current and the reading (RX) voltage on the array and electrode dimensions. The dimensions of the induction probe terminals are not critical in the definition of the system, because they can be considered point electrodes with respect to their spacing distances. In this respect, the induction probes to consider are either the Wenner's (W) or the dipole-dipole (DD) array (Fig. 7a, b). The only aspect that can be of importance for the practical implementation of the system is the relationship between the electrode dimensions and the magnitude of the current injected into the subsuface. The current is a critical parameter of the transfer impedance probe, because in general, given the practical voltage levels applicable to the electrodes and the capacitive coupling with terrestrial soils or concretes, the current levels are expected to be quite low, with a resulting limit to the accuracy that can be achieved for the amplitude and phase measurements. Furthermore, low currents imply a reduction in the voltage signal read across the RX terminals and more stringent requirements for the reading amplifier (Fig. 13).

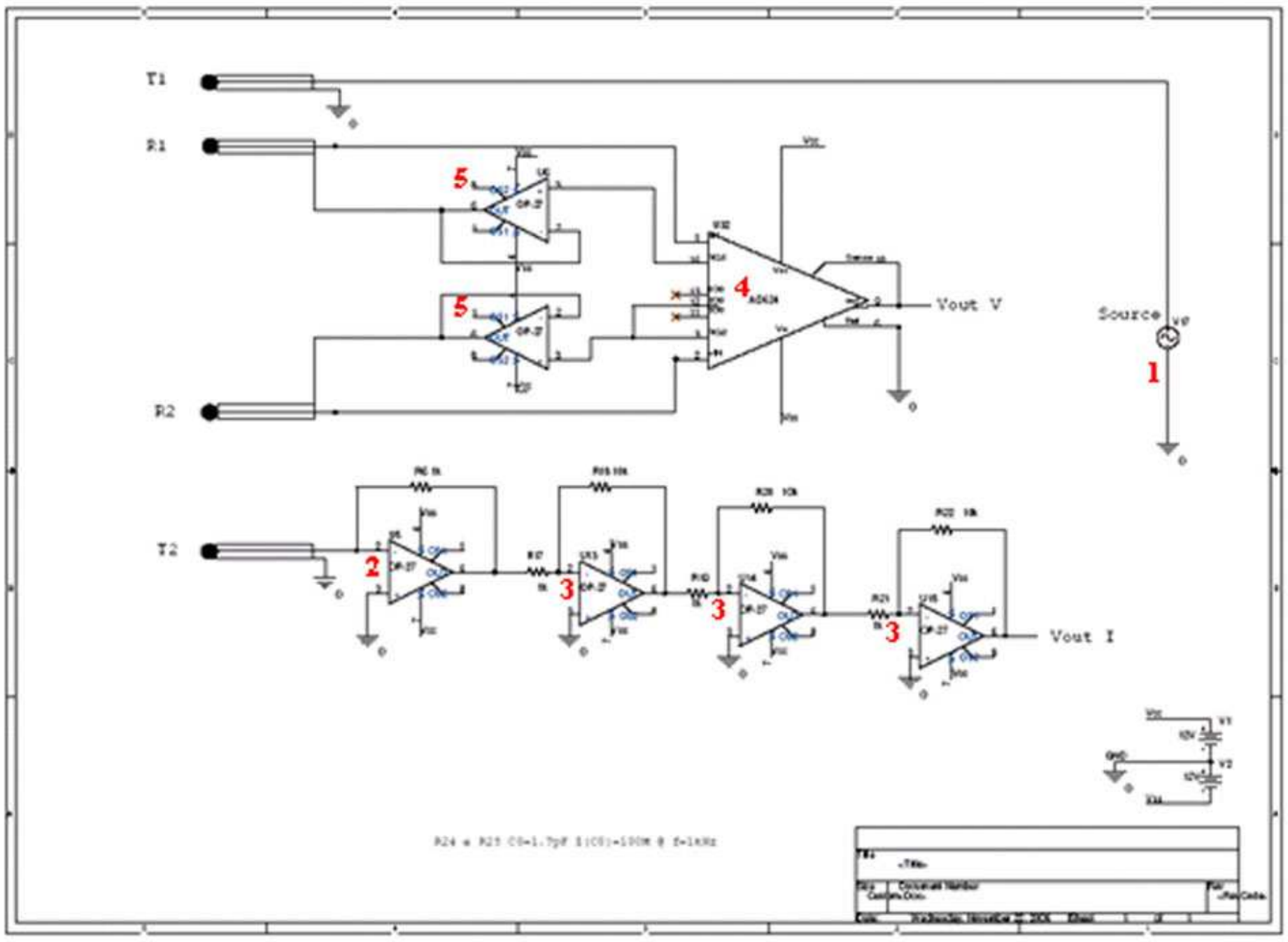

**Figure 13.** Electrical scheme for the analogical part of the measuring system: a signal generator (1), coupled to an amplifier stage, feeds one of two current electrodes (T1). The same current signal, picked to the other electrode (T2), is converted into voltage (2) and then amplified (3). The stage of differentiation for the voltage (4) is preceded by the feedback device to compensate the parasite capacities (5). The signal is sent to an ADC and transferred to a personal computer, where it can be properly processed. The electronic circuit is composed primarily from two stages. The first consists of a current-voltage converter followed by a cascade of amplifiers, to amplify the weak currents typical of high impedances and the second consists of a voltage amplifier with a retroactive chain of capacitive compensation. The circuit has been designed to work linearly at LF in the band from DC to LF-MF. The selected components have been developed specifically for electronic instruments of precision. The circuit techniques adopted for the compensation of the parasite capacities are innovative and allow performing measurements of high impedances. This analogical device is connected to an analogical digital conversion board which contains even a digital analogical converter (DAC) used as a signal generator that, properly projected, can generate a whole series of measurements in an automatic way even at different frequencies for a full analysis.

The Appendix B demonstrates that once the input resistance $R_{in}$ of voltage amplifier stage is fixed and the working frequency $f$ for RESPER probe is selected, which falls in a LF band starting from the minimum frequency $f_{min}$ for operating conditions of galvanic contact or in a MF band from the lower limit $f_{low}$ for capacitive contact, then the height $l(R_{in},f)$ of electrodes can be designed, as this height depends only on the input resistance $R_{in}$ and frequency $f$ , being a function inversely proportional to both $R_{in}$ and $f$. In Appendix B, a suitable proportionality coefficient $\alpha>>1$ is defined, so that the height $l_{W,DD}(R_{in},f)$ of a Wenner's or dipole-dipole array can be calculated as:

$$l_{W,DD} \cong \frac{\ln\alpha}{2\pi^2\varepsilon_0 R_{in} f} \quad , \quad f \geq f_{\min}, f_{low} \quad , \quad \text{for} \quad \alpha>>1 . \tag{3.10}$$

The functional trend of the height $l_{W,DD}(R_{in},f)$ with resistance $R_{in}$ is invariant whether the RESPER is configured in the W or DD array.
Note that the LF band $f \geq f_{min}$ for operating conditions of galvanic contact or the MF band $f \geq f_{low}$ for capacitive contact are defined by the performed measurement, being specified by the number of bits $n_{bit}$ for ADC and the subsurface, i.e. terrestrial soils or concretes.

Moreover, once the minimum bit number $n_{bit}$ for the IQ sampling ADC is known, which allows an inaccuracy $\Delta\varepsilon_r/\varepsilon_r$ for dielectric permittivity $\varepsilon_r$ below the limit of $10\%$, then the characteristic geometrical dimension $L(l,n_{bit})$ can be also defined, as this dimension depends only on the height $l(R_{in},f_{min})$ and bit number $n_{bit}$, being a function directly proportional to $l(R_{in},f_{min})$ and increasing as the power function $2^{nbit}$ of $n_{bit}$.
The electrode-electrode distance $L_W(l,n_{bit})$ of Wenner's array,

$$L_W \simeq l_W \frac{2^{n_{bit}-1}}{\ln\alpha} \quad , \quad \text{for} \quad \alpha>>1 , \tag{3.11}$$

can be compared to the characteristic geometrical dimension $L_{DD}(l,n_{bit})$ of dipole-dipole array, with integer parameter $n \in \mathbb{N} = \{1,2,3,\ldots\}$, according to the proportionality law:

$$\left.\frac{l}{L}\right|_{DD} = \frac{n(n+1)(n+2)}{2} \left.\frac{l}{L}\right|_{W} \quad , \quad \forall n \in \mathbb{N} . \tag{3.12}$$

The radius $r(L)$ must be smaller by a factor $\alpha>>1$ of the characteristic geometrical dimension $L$. The radius $r_{W,DD}(L)$ of W or DD array is:

$$r_{W,DD} = \frac{L_{W,DD}}{\alpha} \quad , \quad \text{for} \quad \alpha>>1 . \tag{3.13}$$

Once fixed the total length of W or DD array $L_{TOT}=3L_W(R_{in})=(n+1)L_{DD}(R_{in})=const$, then are determined the input resistance $R_{in}$ and so the characteristic geometrical dimensions $l_{W,DD}(R_{in})$, $r_{W,DD}(R_{in})$ [Tabs. 4]. Indeed, the dimensions $l_{W,DD}(R_{in})$, $L_{W,DD}(R_{in})$ and so $r_{W,DD}(R_{in})$ result as functions depending only on the resistance $R_{in}$ [Figs. 14, 15].

**a)**

| WENNER’s (W) — Galvanic contact / Capacitive contact | SOIL | | CONCRETE | |
|---|---|---|---|---|
| *Low Resistivity* | ***x = 0*** | ***$x_0$ = 0.052*** | ***x = 0*** | ***$x_0$ = 0.062*** |
| | $R_{in}$ = 138.5 GΩ | $R_{in}$ = 5.906 GΩ | $R_{in}$ = 21.79 GΩ | $R_{in}$ = 1.707 GΩ |
| | $l_W$ = 3.514 µm | $l_W$ = 3.514 µm | $l_W$ = 224.923 µm | $l_W$ = 224.948 µm |
| | $L_W$ = 10.003 cm | $L_W$ = 10.002 cm | $L_W$ = 10.003 cm | $L_W$ = 10.004 cm |
| | $r_W$ = 1 mm | $r_W$ = 1 mm | $r_W$ = 1 mm | $r_W$ = 1 mm |
| *High Resistivity* | ***x = 0*** | ***$x_0$ = 0.087*** | ***x = 0*** | ***$x_0$ = 0.087*** |
| | $R_{in}$ = 7.638 GΩ | $R_{in}$ = 638.6 MΩ | $R_{in}$ = 25.61 GΩ | $R_{in}$ = 2.128 GΩ |
| | $l_W$ = 224.87 µm | $l_W$ = 224.888 µm | $l_W$ = 224.868 µm | $l_W$ = 224.959 µm |
| | $L_W$ = 10 cm | $L_W$ = 10.001 cm | $L_W$ = 10 cm | $L_W$ = 10.004 cm |
| | $r_W$ = 1 mm | $r_W$ = 1 mm | $r_W$ = 1 mm | $r_W$ = 1 mm |

**(b)**

| DIPOLE-DIPOLE (DD) – n = 1 — Galvanic contact / Capacitive contact | SOIL | | CONCRETE | |
|---|---|---|---|---|
| *Low Resistivity* | ***x = 0*** | ***$x_0$ = 0.045*** | ***x = 0*** | ***$x_0$ = 0.053*** |
| | $R_{in}$ = 46.18 GΩ | $R_{in}$ = 1.969 GΩ | $R_{in}$ = 7.265 GΩ | $R_{in}$ = 569.2 MΩ |
| | $l_{DD}$ = 10.54 µm | $l_{DD}$ = 10.54 µm | $l_{DD}$ = 674.615 µm | $l_{DD}$ = 674.606 µm |
| | $L_{TOT}$ = 30 cm | $L_{TOT}$ = 30 cm | $L_{TOT}$ = 30.001 cm | $L_{TOT}$ = 30.001 cm |
| | $r_{DD}$ = 1 mm | $r_{DD}$ = 1 mm | $r_{DD}$ = 1 mm | $r_{DD}$ = 1 mm |
| *High Resistivity* | ***x = 0*** | ***$x_0$ = 0.075*** | ***x = 0*** | ***$x_0$ = 0.075*** |
| | $R_{in}$ = 2.561 GΩ | $R_{in}$ = 212.8 MΩ | $R_{in}$ = 8.537 GΩ | $R_{in}$ = 709.6 MΩ |
| | $l_{DD}$ = 674.611 µm | $l_{DD}$ = 674.875 µm | $l_{DD}$ = 674.579 µm | $l_{DD}$ = 674.623 µm |
| | $L_{TOT}$ = 30.001 cm | $L_{TOT}$ = 30.013 cm | $L_{TOT}$ = 30 cm | $L_{TOT}$ = 30.002 cm |
| | $r_{DD}$ = 1 mm | $r_{DD}$ = 1 mm | $r_{DD}$ = 1 mm | $r_{DD}$ = 1 mm |

**(c)**

| DIPOLE-DIPOLE (DD) – n = 6 — Galvanic contact / Capacitive contact | SOIL | | CONCRETE | |
|---|---|---|---|---|
| *Low Resistivity* | ***x = 0*** | ***$x_0$ = 0.045*** | ***x = 0*** | ***$x_0$ = 0.053*** |
| | $R_{in}$ = 2.199 GΩ | $R_{in}$ = 93.76 MΩ | $R_{in}$ = 345.9 MΩ | $R_{in}$ = 27.1 MΩ |
| | $l_{DD}$ = 221.352 µm | $l_{DD}$ = 221.352 µm | $l_{DD}$ = 1.4169 cm | $l_{DD}$ = 1.4169 cm |
| | $L_{TOT}$ = 30 cm | $L_{TOT}$ = 30 cm | $L_{TOT}$ = 30.006 cm | $L_{TOT}$ = 30.006 cm |
| | $r_{DD}$ = 375.006 µm | $r_{DD}$ = 375.006 µm | $r_{DD}$ = 375.073 µm | $r_{DD}$ = 375.077 µm |
| *High Resistivity* | ***x = 0*** | ***$x_0$ = 0.075*** | ***x = 0*** | ***$x_0$ = 0.075*** |
| | $R_{in}$ = 121.9 MΩ | $R_{in}$ = 10.13 MΩ | $R_{in}$ = 406.5 MΩ | $R_{in}$ = 33.79 MΩ |
| | $l_{DD}$ = 14.173 mm | $l_{DD}$ = 14.177 mm | $l_{DD}$ = 1.4167 cm | $l_{DD}$ = 1.4167 cm |
| | $L_{TOT}$ = 30.014 cm | $L_{TOT}$ = 30.023 cm | $L_{TOT}$ = 30.001 cm | $L_{TOT}$ = 30.002 cm |
| | $r_{DD}$ = 375.176 µm | $r_{DD}$ = 375.284 µm | $r_{DD}$ = 375.018 µm | $r_{DD}$ = 375.026 µm |

**Table 4.** Refer to the captions of Figs. 3, 6 and Tab. 3. The RESPER probe can show a galvanic or capacitive contact both on non-saturated terrestrial soils and concretes with low or high electrical resistivity, it is connected to an IQ ADC and is configured in the Wenner’s (W) or dipole-dipole (DD) array ($n=1,6$). The voltage signal of RESPER is amplified by a downstream stage of input resistance $R_{in}$. The probe is designed with a total length $L_{TOT} = 30$ *cm* and the Wenner’s (dipole-dipole) array is implemented by four cylindrical electrodes of height $l_W(R_{in})$ [$l_{DD}(R_{in})$] and radius $r_W(R_{in})$ [$r_{DD}(R_{in})$]. Data are shown for the input resistance $R_{in}$ of voltage amplifier stage, the height $l_{W,DD}(R_{in})$ and radius $r_{W,DD}(R_{in})$ of electrodes in the W (a) or DD array, defined by $n=1$ (b) and $n=6$ (c).

## 4.1 Wenner's and dipole-dipole arrays: cylindrical electrodes

Let us design a RESPER probe, configured in Wenner's (W) or dipole-dipole array (DD) ($n=1,6$), with a total length of $L_{TOT} = 30\ cm$. Indeed, as a feature of the project, the RESPER is by laboratory bench, in order to perform the measurements on samples of various terrestrial soils or concretes, drawn from the outside environment. Moreover, the total length $L_{TOT}$ of probe is chosen $30\ cm$ at least, since a smaller length would mean choosing an input resistance $R_{in}$ of the voltage amplifier stage much higher than some tens or even hundreds of $G\Omega$ (Figs. 14, 15). For understandable technical limitations, our Ambient Geophysics Laboratory (LGA) has managed in building the electrodes with an height $l_{W,DD}$ no smaller than $200\mu m$. Finally, the input resistance $R_{in}$ is chosen around easily available values, so below the almost unmarketable limit of $1G\Omega$.

Firstly, design the RESPER probe as configured in a Wenner's array (Tab. 4a). The RESPER configured in a W array is not fit to analyze the terrestrial soils characterized by a low electrical resistivity, both in galvanic and capacitive contacts (highlighted in red colour), because the probe should be built with electrodes of height $l_W = 3.514\ \mu m$ and the voltage amplifier stage would require an input resistance higher than $R_{in} = 5.906\ G\Omega$. In order to meet the project features, suitable Mathcad simulations have specified a probe with the electrodes of radius $r_W = 1\ mm$ and the height around $l_W \approx 225\ \mu m$. Implementing these characteristic geometrical dimensions, the probe is not fit to show a galvanic contact both on the soils with an high resistivity and the concretes (highlighted in red colour), because the amplifier stage would require a resistance higher than $R_{in} = 7.638\ G\Omega$. Moreover, the probe is partially fit to show a capacitive contact on the concretes of low and high resistivity (yellow colour). Indeed, the amplifier would require: $1.707\ G\Omega \leq R_{in} \leq 2.128\ G\Omega$. Finally, the probe is fully fit to show a capacitive contact on the soils with high resistivity (green). Indeed, $R_{in} = 638.6\ M\Omega$.

Secondly, design the RESPER probe as configured in the dipole-dipole array defined by $n_0=1$ (Tab. 4b). The RESPER configured in the DD array with $n_0=1$ is not fit to analyze the terrestrial soils characterized by a low electrical resistivity, both in galvanic and capacitive contacts (highlighted in red colour), because primarily the probe should be built with electrodes of height $l_{DD}(n_0=1) = 10.54\ \mu m$ and even the voltage amplifier stage would require an input resistance higher than $R_{in} = 1.969\ G\Omega$. In order to meet the project features, suitable Mathcad simulations have specified a probe with the electrodes of radius $r_{DD}(n_0=1) = 1\ mm$ and the height around $l_{DD}(n_0=1) \approx 675\ \mu m$. Implementing these characteristic geometrical dimensions, the probe is not fit to show a galvanic contact on the concretes with a low and high resistivity (highlighted in red colour), because the amplifier stage would require a resistance higher than $R_{in} = 7.265\ G\Omega$. Moreover, the probe is partially fit to show a galvanic contact on the soils of high resistivity (yellow colour). Indeed, the amplifier would require: $R_{in} = 2.561\ G\Omega$. Finally, the probe is fully fit to show a capacitive contact on the soils of high resistivity and all concretes (green). Indeed, $212.8\ M\Omega \leq R_{in} \leq 709.6\ M\Omega$.

Therefore, the dipole-dipole (DD) array defined by $n_0=1$ is better compared to the Wenner's (W) since it is fully fit to show not only a capacitive contact on the terrestrial soils characterized by an high electrical resistivity but also on all concretes. Farther, the DD array with $n_0=1$ is partially fit to show a galvanic contact on the soils with a high resistivity. As a physical explanation, the geometrical dimension ratio is larger in the dipole-dipole array $n_0=1$ [Eq. (3.12)] than in Wenner's array [Eq. (3.11)] by a factor $n_0(n_0+1)(n_0+2)/2=3$; consequently, the total length of $L_{TOT} = 30\ cm$ is matched for an input resistance $R_{in}$ which is lower in the DD array compared to W's array: necessarily, the height of electrodes is larger in the DD array than in W's array (Figs. 14, 15).

Thirdly, design the RESPER probe as configured in the dipole-dipole array defined by $n=6$ (Tab. 4c). The RESPER configured in the DD array with $n=6$ is fit to analyze the terrestrial soils characterized by a low electrical resistivity, in capacitive (highlighted in green colour) and partially in galvanic (yellow colour) contact; indeed, though the probe is built with electrodes of height slightly larger than $200\mu m$, i.e. $l_{DD}(n=6) = 221.352\ \mu m$, anyway the voltage amplifier stage would require an input resistance which can reach values slightly higher than $1G\Omega$, i.e. in the range $93.76\ M\Omega \leq R_{in} \leq 2.199\ G\Omega$. In order to meet the project features, suitable Mathcad simulations have specified a probe with the electrodes of radius around $r_{DD}(n=6) \approx 375\ \mu m$ and the height around $l_{DD}(n=6) \approx 1.4\ cm$. Implementing these characteristic geometrical dimensions, the probe is fully fit to show both galvanic and capacitive contacts on the soils with a low resistivity and all concretes (green), because the amplifier stage would require a resistance in the range $10.13\ M\Omega \leq R_{in} \leq 345.9\ M\Omega$.

Therefore, the dipole-dipole (DD) array defined by $n=6$ is better compared to $n_0=1$ since it is fully fit to show not only a capacitive contact but also a galvanic contact on the terrestrial soils characterized by a high electrical resistivity and all concretes. Farther, the DD array with $n=6$ is fit to analyze the terrestrial soils with a low resistivity, in capacitive and partially in galvanic contact. As a physical explanation, the geometrical dimension ratio is larger in the dipole-dipole array $n=6$ than in $n_0=1$ by a factor $(3/2)n(n+1)(n+2)$ [Eq. (3.12)]; consequently, the total length of $L_{TOT} = 30\ cm$ is matched for an input resistance $R_{in}$ which is lower in the DD array $n=6$ compared to $n_0=1$: necessarily, the height of electrodes is larger in the DD array $n=6$ than in $n_0=1$ (Figs. 14, 15). Note that if the coefficient $\alpha$ is chosen invariant for both the Wenner's and dipole-dipole arrays, i.e. $\alpha_W = \alpha_{DD} = 10^2$, then the radius of electrodes is generally larger in the W's than in DD arrays, i.e. $r_{DD}(n_0=1)=r_W$ and $r_{DD}(n)=3r_{DD}(n_0=1)/(n+2)$ [see Eq. (3.5) and Eq. (3.13)].

**(a)**

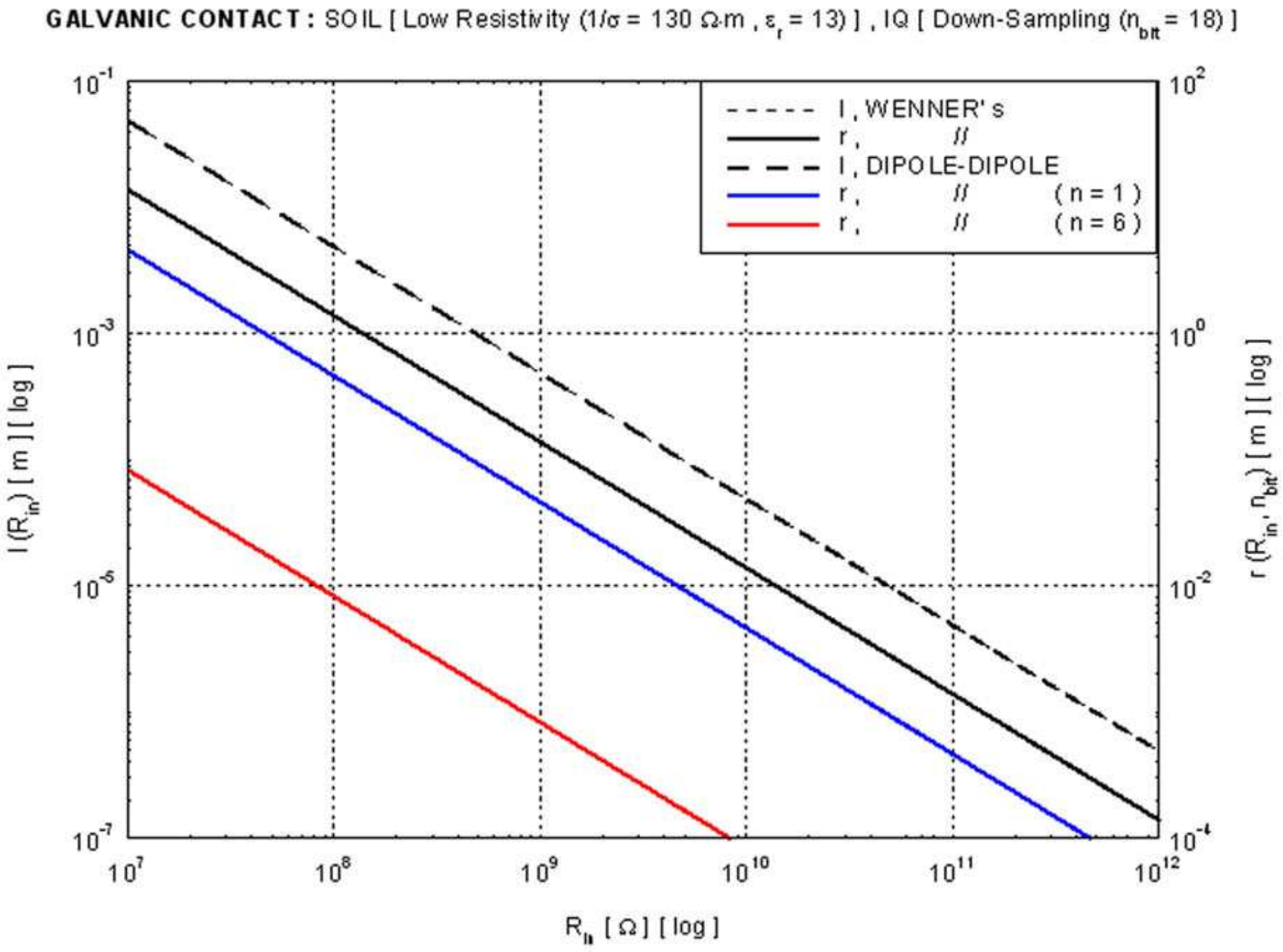

**(b)**

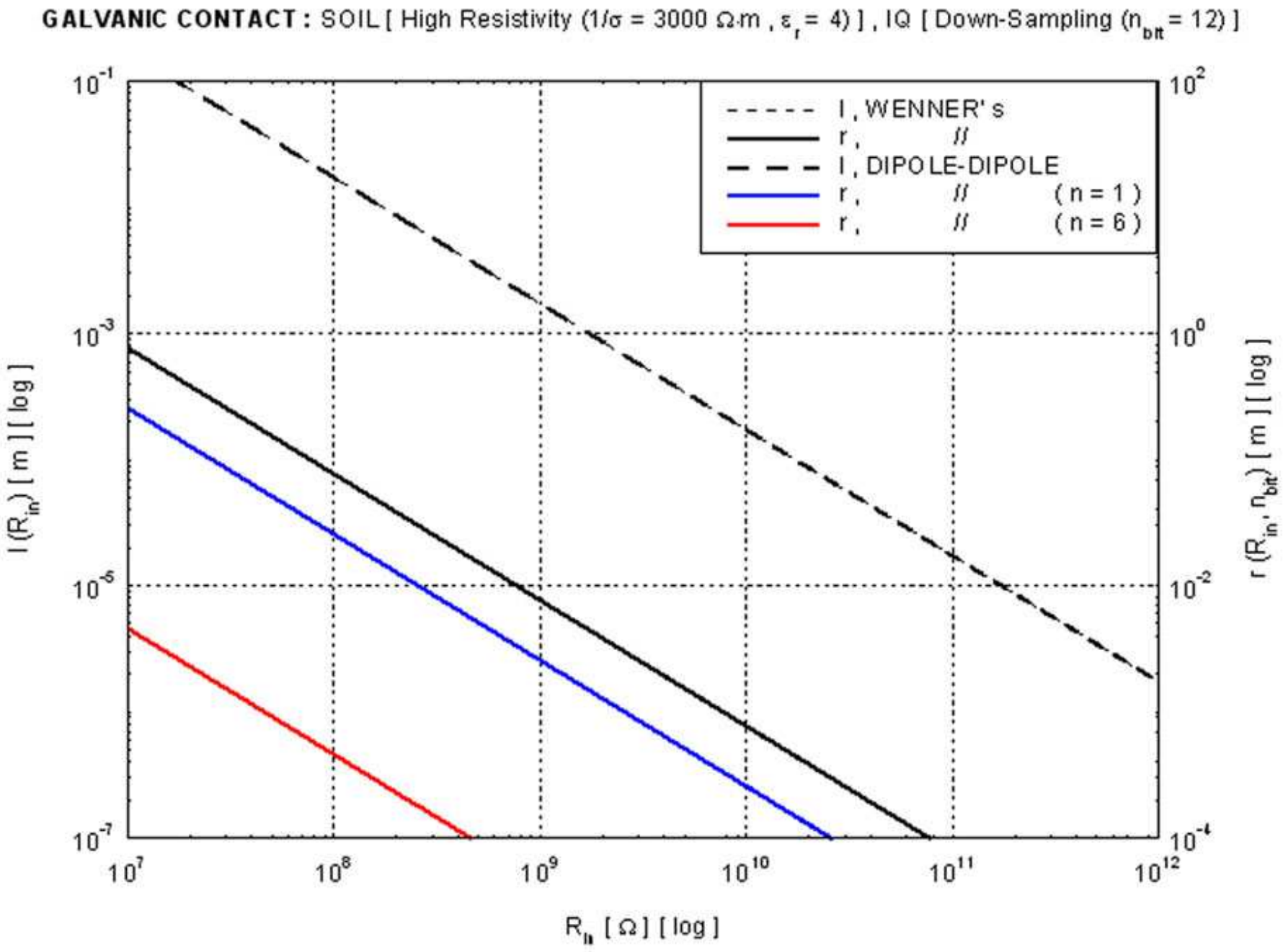

**(c)**

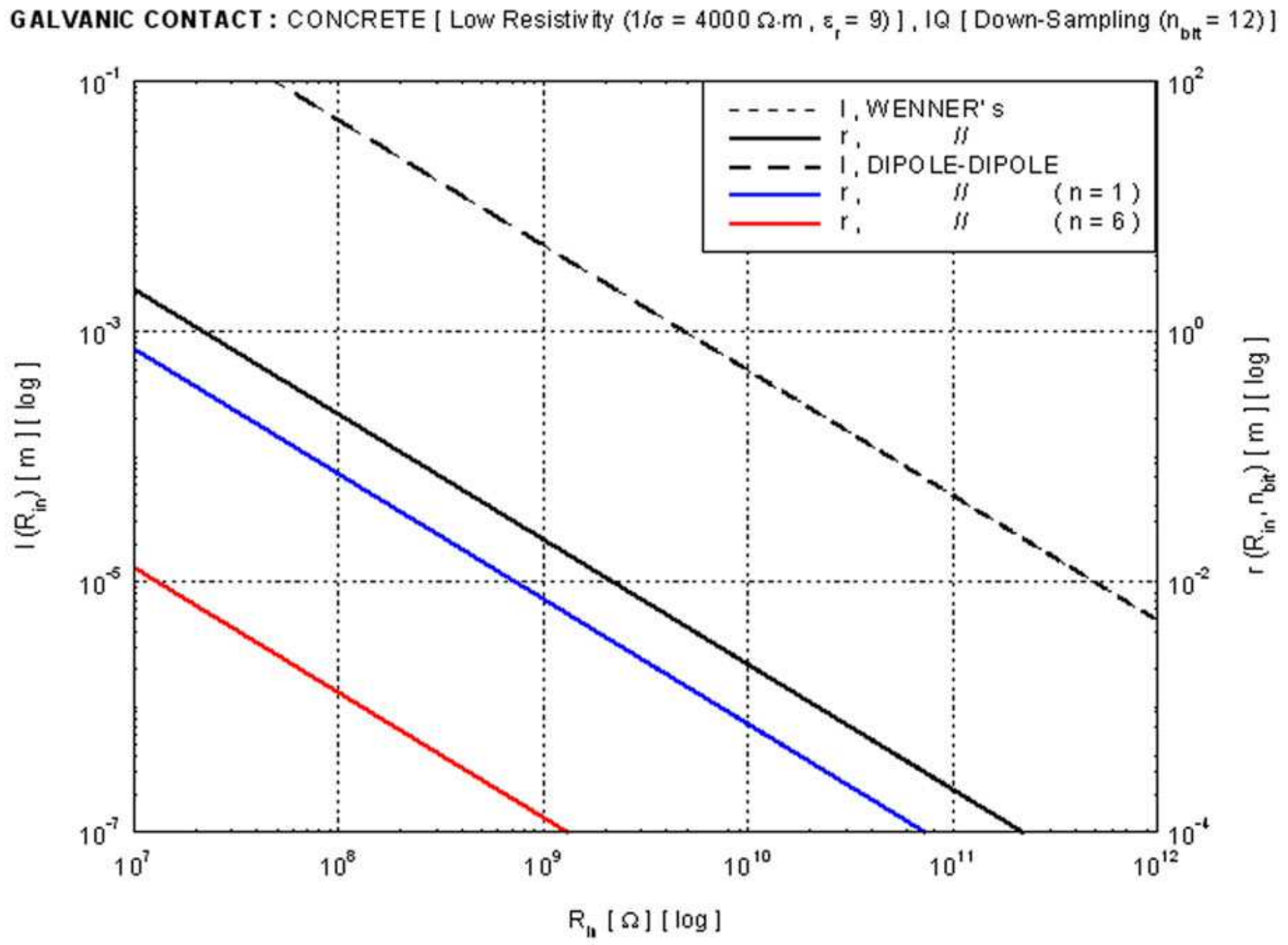

**(d)**

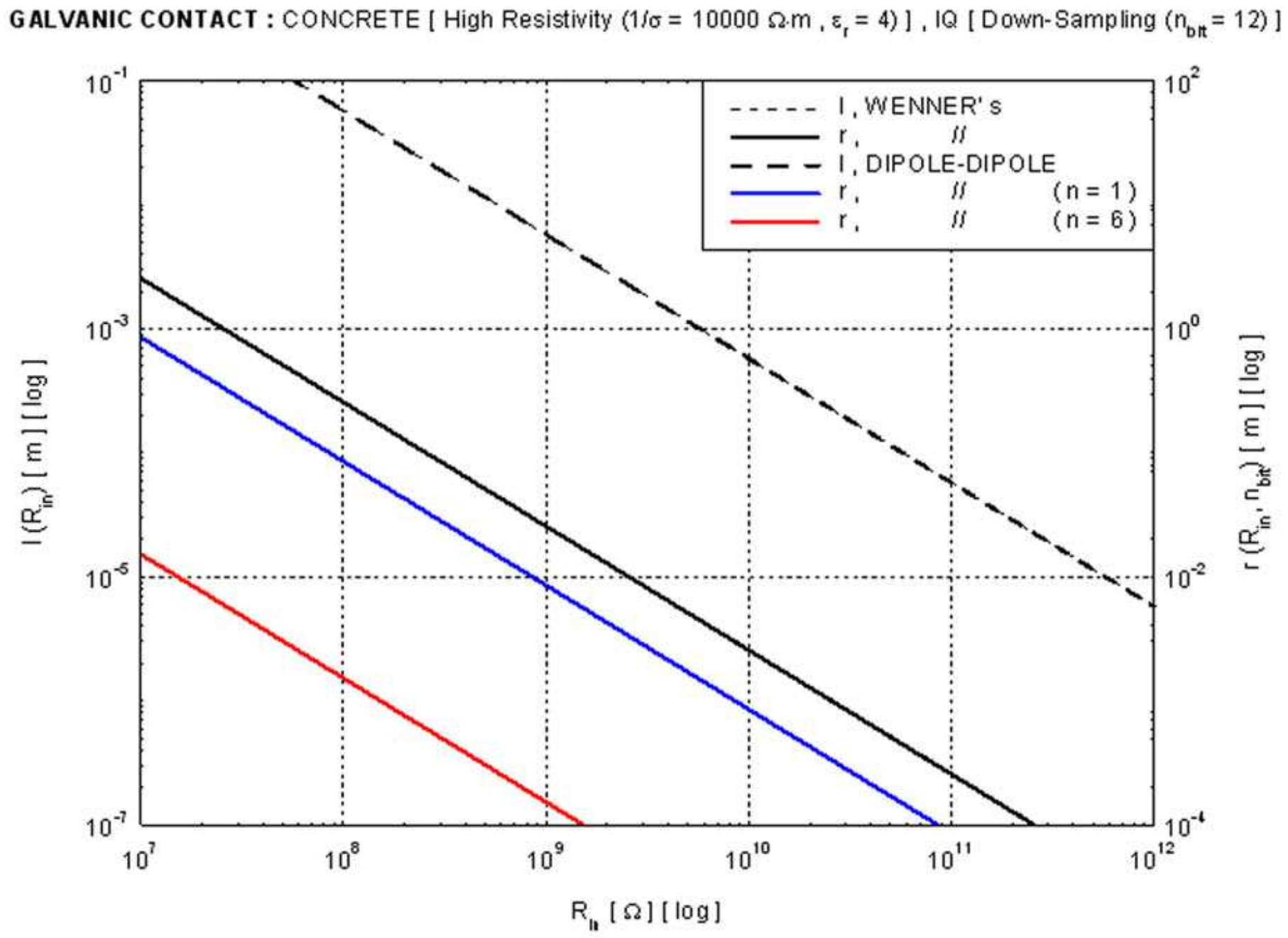

**Figure 14.** Refer to the captions of Tab. 4. The RESPER show only a galvanic contact both on both the terrestrial soils (a)-(b) and concretes (c)-(d) with low or high resistivity, it is connected to an IQ ADC and is configured in the Wenner's (W) or dipole-dipole (DD) array (*n=1,6*). Like-Bode's diagrams showing the height $l(R_{in})$ and radius $r(R_{in})$ of the cylindrical electrodes for W or DD array, with *n=1* and *n=6*, plotted as functions of the input resistance $R_{in}$ for amplifier stage, in the range $R_{in} \in [R_{in}^{(\min)}, R_{in}^{(\max)}]$ with $R_{in}^{(\min)} = 10M\Omega$ and $R_{in}^{(\max)} = 1000G\Omega$.

**(a)**

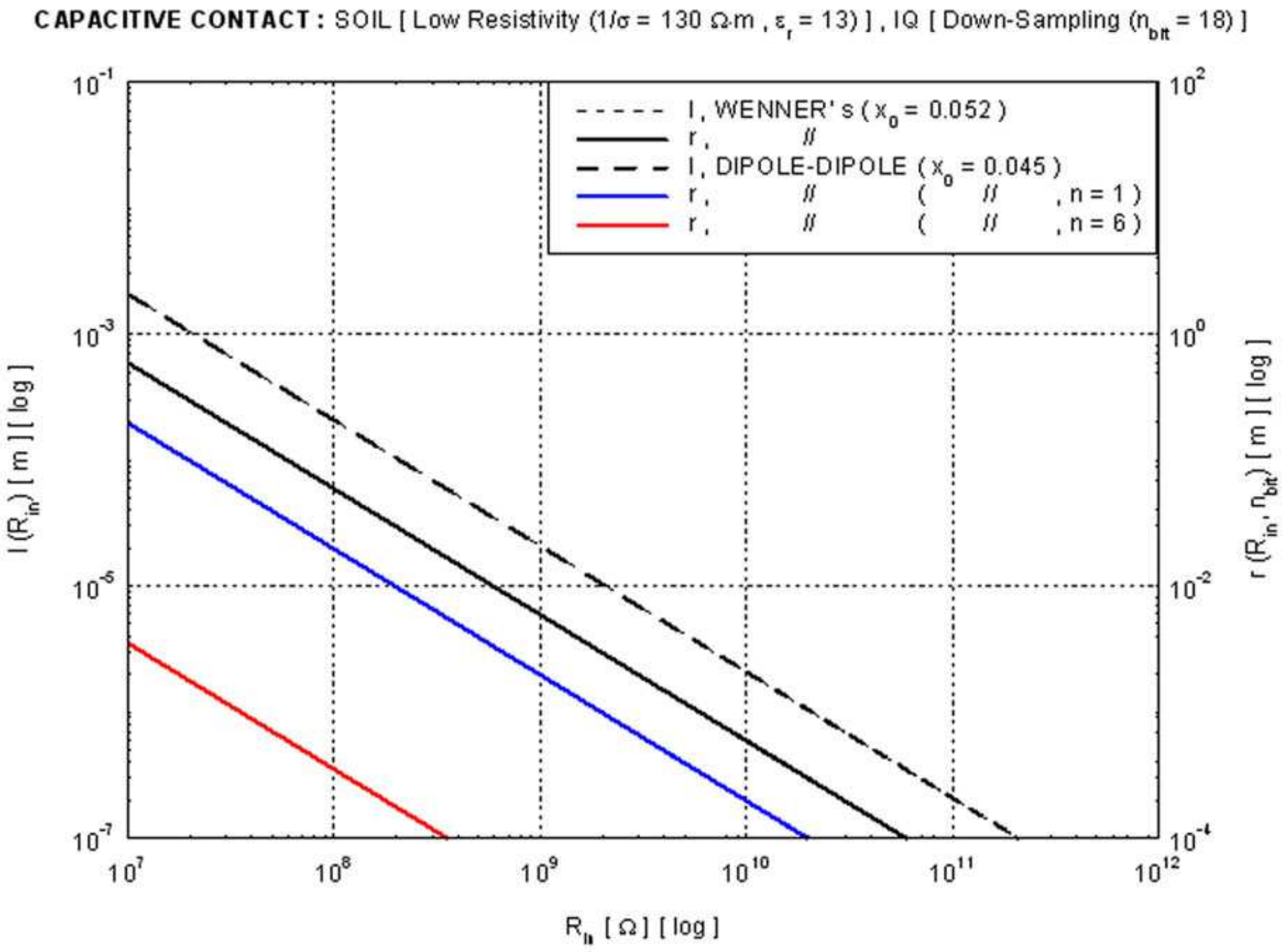
CAPACITIVE CONTACT : SOIL [ Low Resistivity (1/σ = 130 Ω·m , ε_r = 13) ] , IQ [ Down-Sampling (n_bit = 18) ]
l , WENNER' s ( x_0 = 0.052 )
r , //
l , DIPOLE-DIPOLE ( x_0 = 0.045 )
r , // ( // , n = 1 )
r , // ( // , n = 6 )
l (R_in) [ m ] [ log ]
r (R_in, n_bit) [ m ] [ log ]
R_in [ Ω ] [ log ]

**(b)**

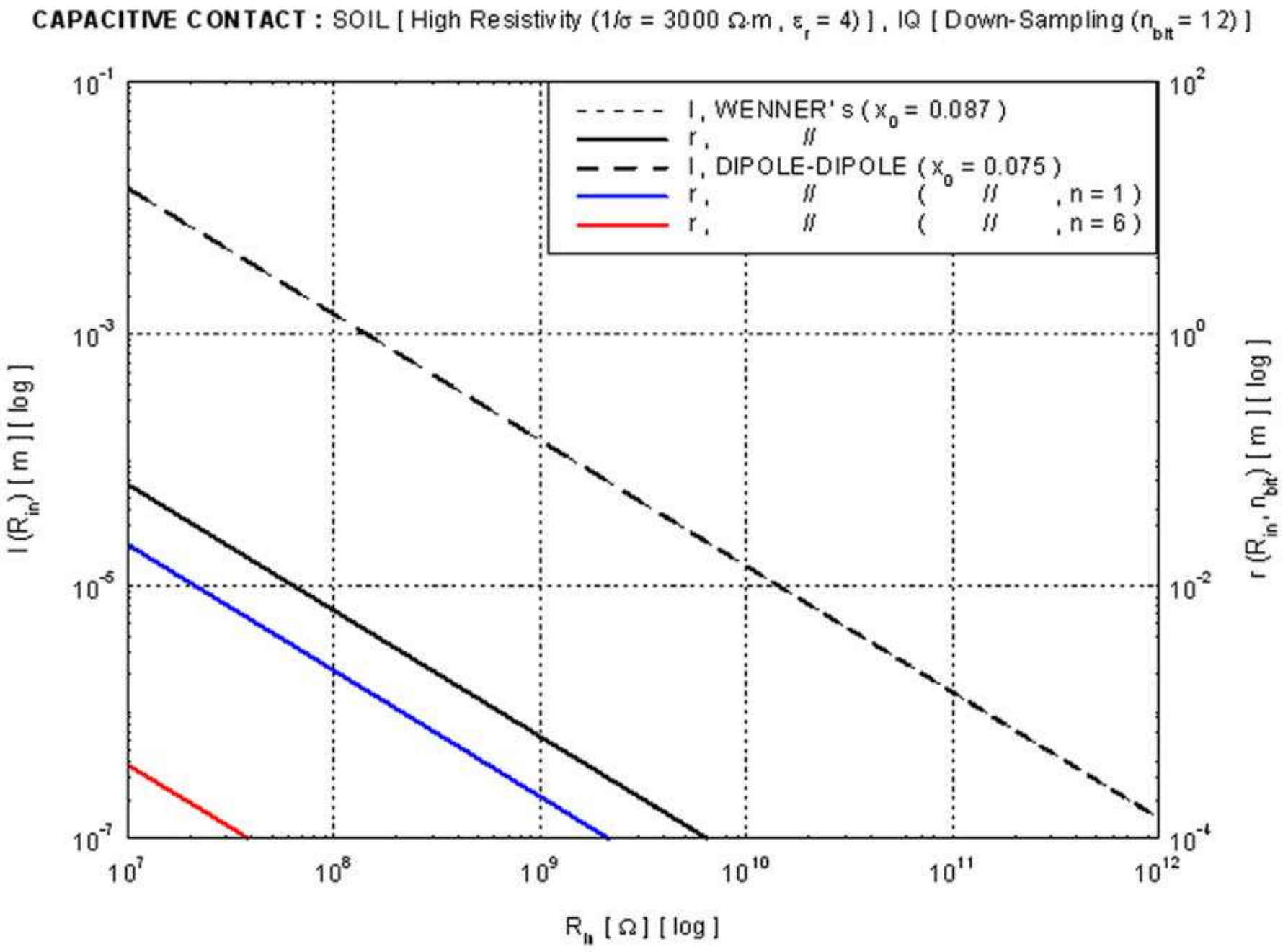
CAPACITIVE CONTACT : SOIL [ High Resistivity (1/σ = 3000 Ω·m , ε_r = 4) ] , IQ [ Down-Sampling (n_bit = 12) ]
l , WENNER' s ( x_0 = 0.087 )
r , //
l , DIPOLE-DIPOLE ( x_0 = 0.075 )
r , // ( // , n = 1 )
r , // ( // , n = 6 )
l (R_in) [ m ] [ log ]
r (R_in, n_bit) [ m ] [ log ]
R_in [ Ω ] [ log ]

**(c)**

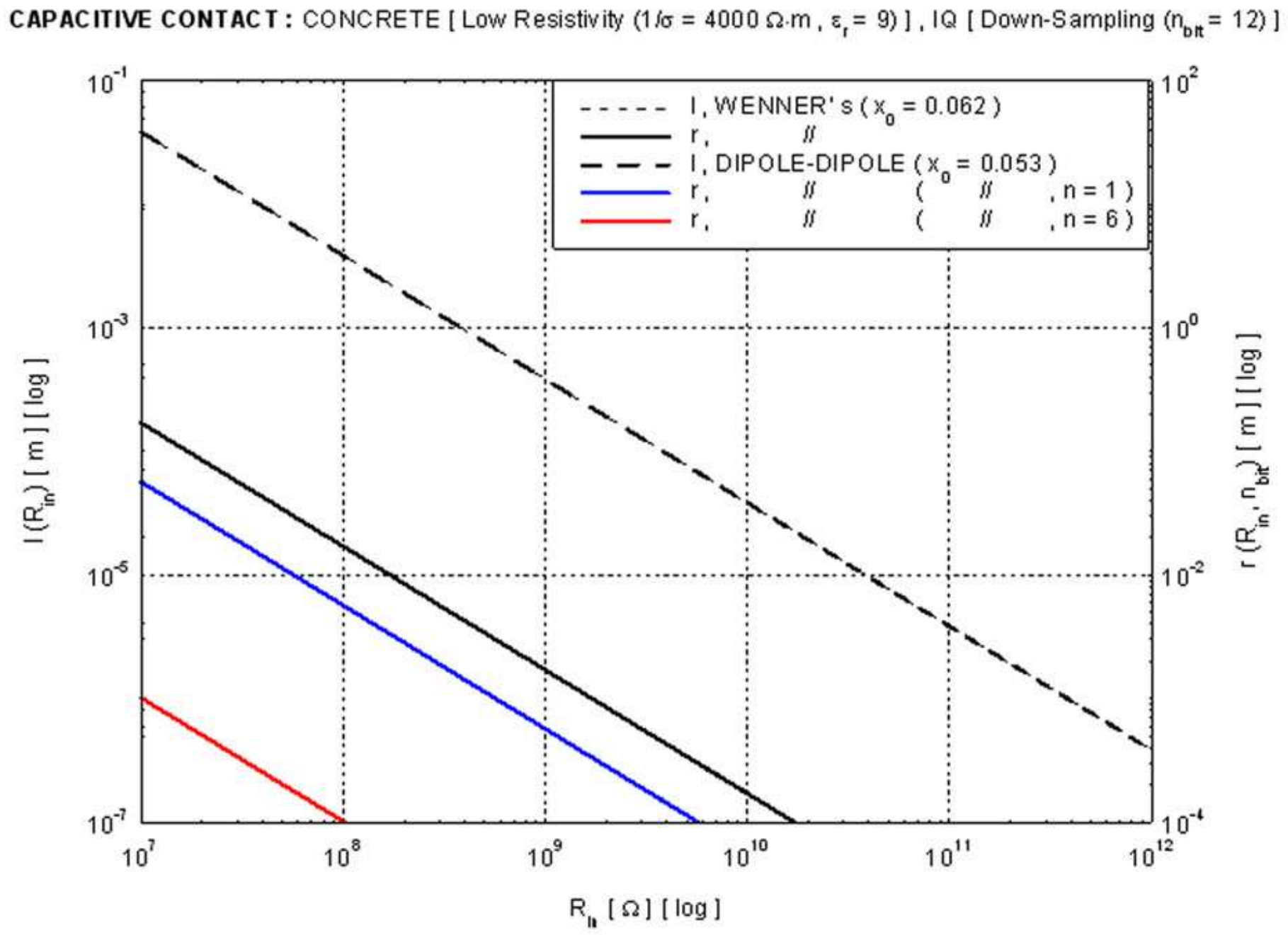

**(d)**

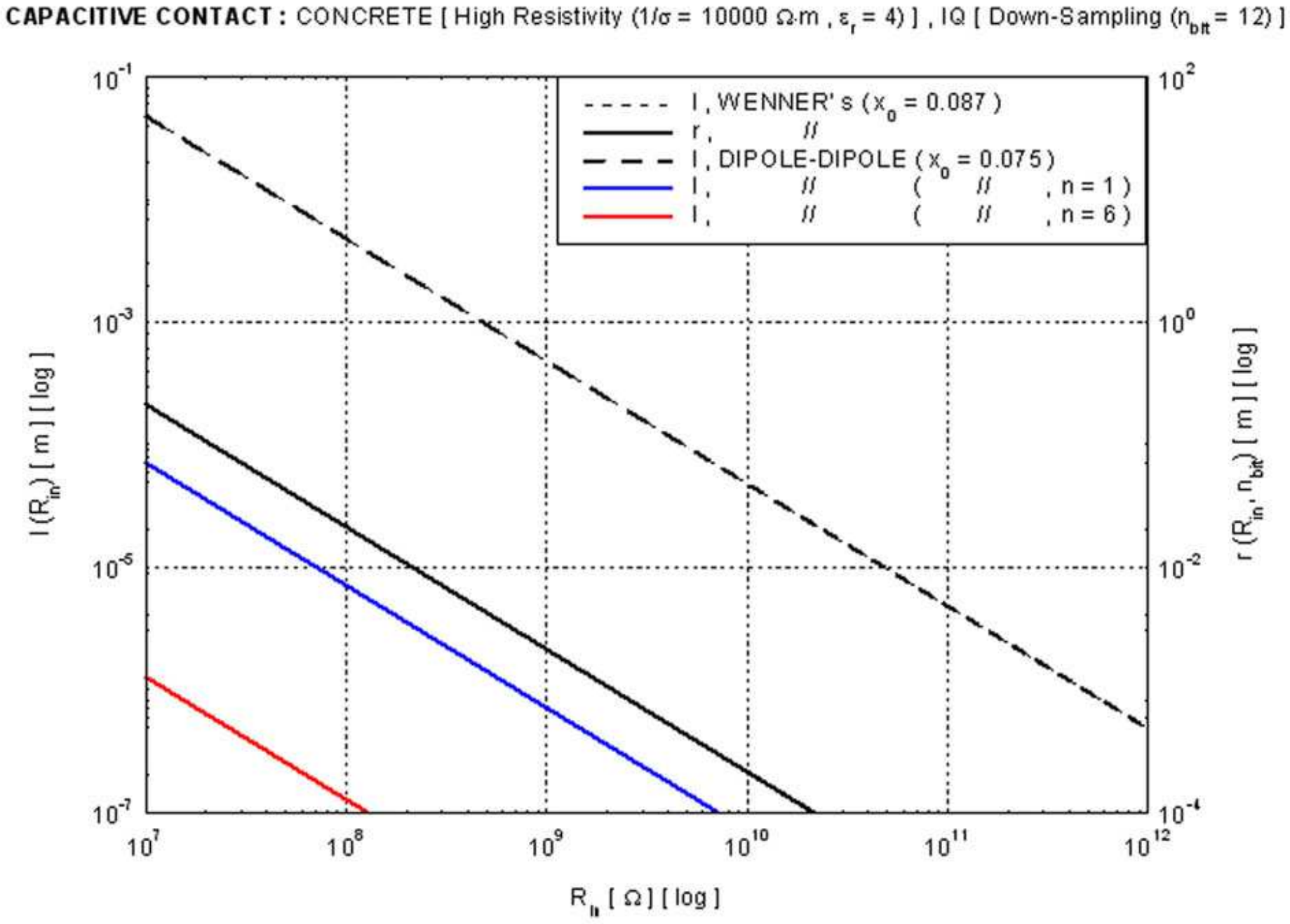

**Figure 15.** Refer to the captions of Tab. 4. The probe show only a capacitive contact both on both the soils (a)-(b) and concretes (c)-(d) with low or high resistivity, it is connected to an IQ ADC and is configured in the Wenner's (W) or dipole-dipole (DD) array ($n=1,6$). Like-Bode's diagrams showing the height $l(R_{in})$ and radius $r(R_{in})$ of the cylindrical electrodes for W or DD array, with $n=1$ and $n=6$, plotted as functions of the input resistance $R_{in}$ for amplifier stage.

## 5. Technical description of the RESPER probe implementation

The RESPER probe should be designed as a *multi* dipole-dipole array, defined by the integer parameters $n=1,2,3,4,5,6$. The RESPER is dimensionally sized by laboratory bench with a total length $L_{TOT} = 30\ cm$, so that the $n$-configuration could be implemented by a characteristic geometrical dimension $L(n) = L_{TOT}/(n+2)$, being $n=1,...6$. Moreover, the probe is designed by cylindrical electrodes, and, according to the Mathcad simulations, each configurations would be implemented by the electrode height $l(n)$ shown in Tab. 5.

| DIPOLE-DIPOLE | L(n) | l(n) |
|---|---|---|
| n = 1 | 10 cm | 0.7 mm |
| n = 2 | 7.5 cm | 2 mm |
| n = 3 | 6 cm | 4 mm |
| n = 4 | 5 cm | 6.75 mm |
| n = 5 | 4.29 cm | 1 cm |
| n = 6 | 3.75 cm | 1.4 cm |

**Table 5.** The RESPER probe would be designed as a *multi* dipole-dipole array, defined by the integer parameters $n=1,2,3,4,5,6$. Each of the *six* configurations is implemented by a characteristic geometrical dimension $L(n)$ and an electrode height $l(n)$ corresponding to $n=1,...,6$.

Finally, although each of the six configurations is characterized by a proper electrode radius, anyway, each configuration is implemented by an electrode radius $\overline{r}_{DD}$ which, for obvious practical reasons, is designed as the arithmetic mean between the values corresponding to the extreme configurations $n=1,\ 6$, i.e. (see Tabs. 4a, b)

$$\overline{r}_{DD} = \frac{r_{DD}(n=1) + r_{DD}(n=6)}{2} \cong 0.7mm\,. \qquad (5.1)$$

Four coaxial cables are connected to the electrodes. The electrical cables should be specified by a very low resistivity, around $\rho_C \approx 1.69 \cdot 10^{-8}\ \Omega \cdot m$ (copper), comparing to the resistivity of both terrestrial soils and concretes, in the range of $130\ \Omega \cdot m \le \rho \le 10^4\ \Omega \cdot m$. The RESPER could feel the physical presence of cables, thus invalidating the measurement of complex impedance on the subjacent medium. A technical remedy has been adopted, so that the coaxial cables are kept at a fixed distance by four guides lying on a plane transversal to the measure, which is parallel to the probe and perpendicular to the medium. The electrical cables would be sized with a section area $S_C = \pi \cdot (r_C)^2 = 4\ mm^2$ and a length around $l_C = 1m$, so providing a very low resistance in the range of $R_C = \rho_C \cdot l_C / S_C \approx 0.0169 - 0.1\ \Omega$.

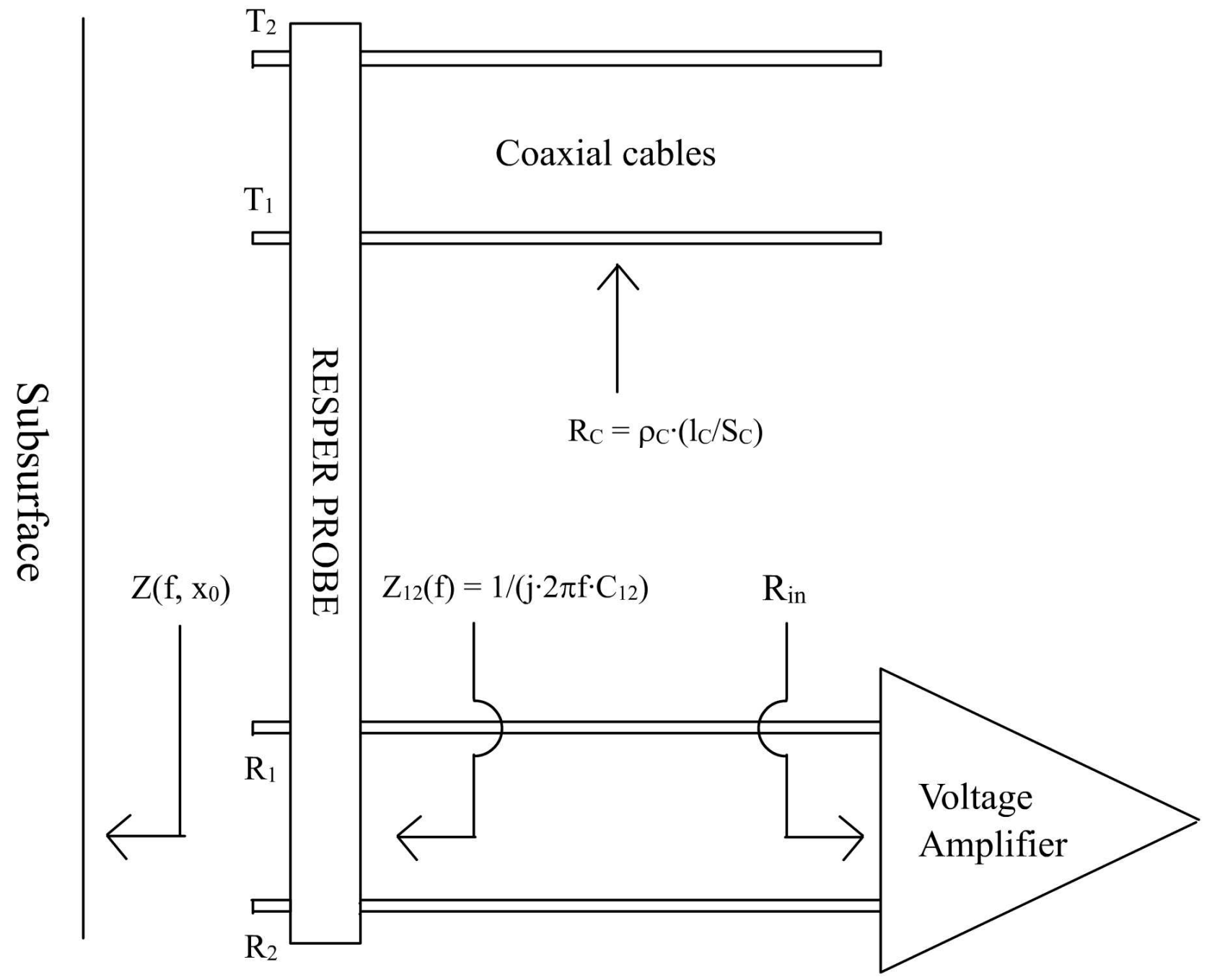

**Figure 16.** Voltage amplifier stage downstream of the RESPER, in galvanic or capacitive contact on a subsurface. Note that the electrical resistance $R_C$ of the coaxial cables is characterized by a very low value, then the input resistance $R_{in}$ of the voltage amplifier, the capacitance $C_{12}$ between the reading electrodes $R_1$, $R_2$ and the complex impedance $|Z|(f,x_0)$ measured by the probe are arranged in a parallel scheme.

The signal processing is well balanced in both the reading and voltage amplifier stages (Fig. 16). Indeed, the RESPER is designed to perform measurements from a suitable minimum frequency $f_{min}$ in the LF band. Thus, the cylindrical electrodes and coaxial cables are naturally sized such that the capacitance $C_{12}$ between the reading electrodes $R_1$ and $R_2$ leads to a maximal impedance modulus $|Z_{12}|_{\max} = 1/(2\pi f_{min} \cdot C_{12})$, a value around some hundreds of $M\Omega$ much higher than the electrical resistance $R_C$ due to the cables, i.e.

$$|Z_{12}|_{\max} = \frac{1}{2\pi f_{\min} \cdot C_{12}} >> R_C . \tag{5.2}$$

Finally, the maximal impedance modulus $|Z_{12}|_{\max}$ is even three magnitude orders higher than the modulus values of complex impedance $|Z|(f,x_0)$ measured by the probe in galvanic or capacitive contact on the subsurface, i.e. terrestrial soil and concretes characterized by values not exceeding $100k\Omega$. Then, $|Z_{12}|_{\max}$ is almost one order lower than the input resistance $R_{in}$ of the voltage amplifier, specified by a value around $1G\Omega$, i.e.

$$|Z|(f, x_0) << \frac{1}{2\pi f_{\min} \cdot C_{12}} < R_{in} . \tag{5.3}$$

**(a)**

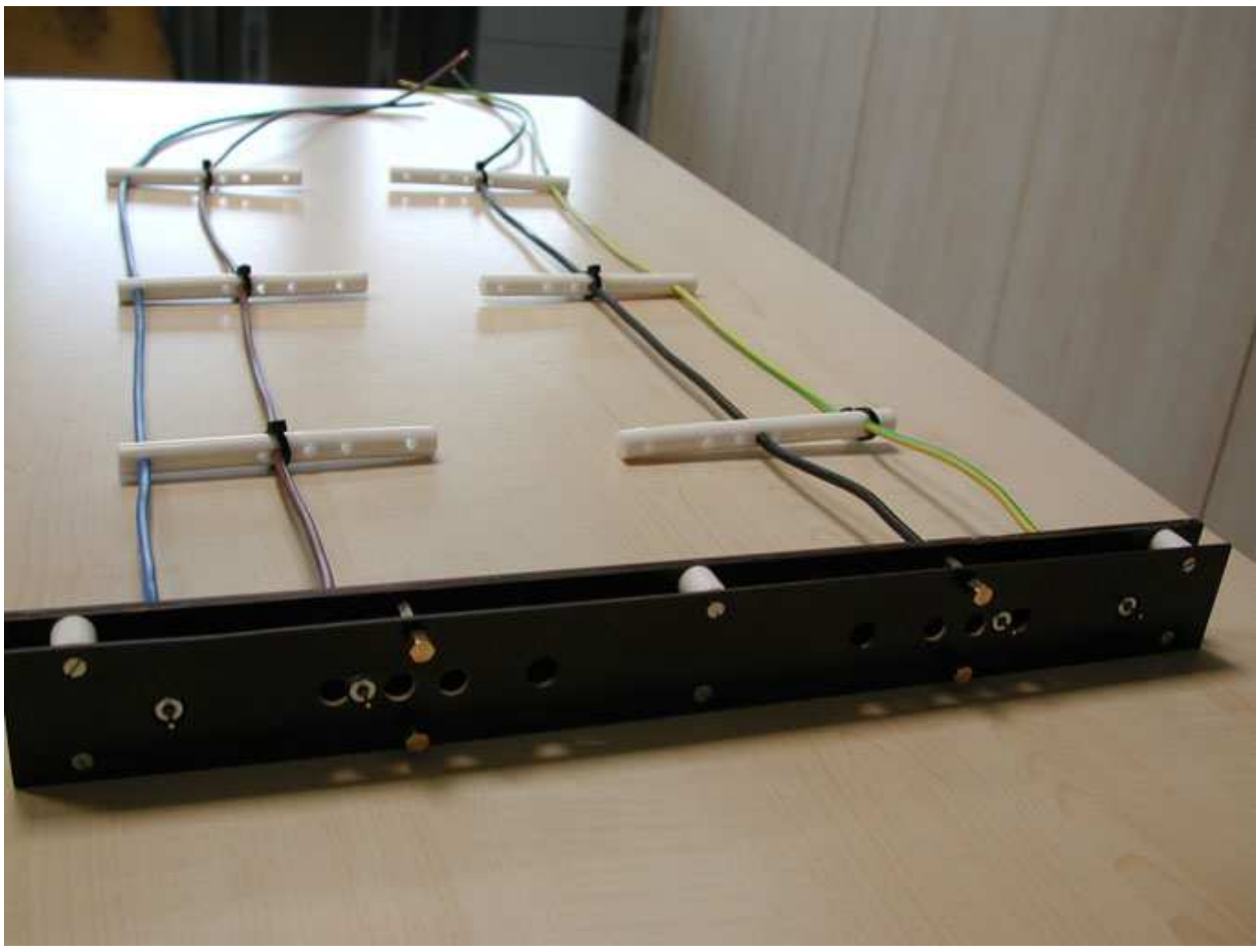

**(b)**

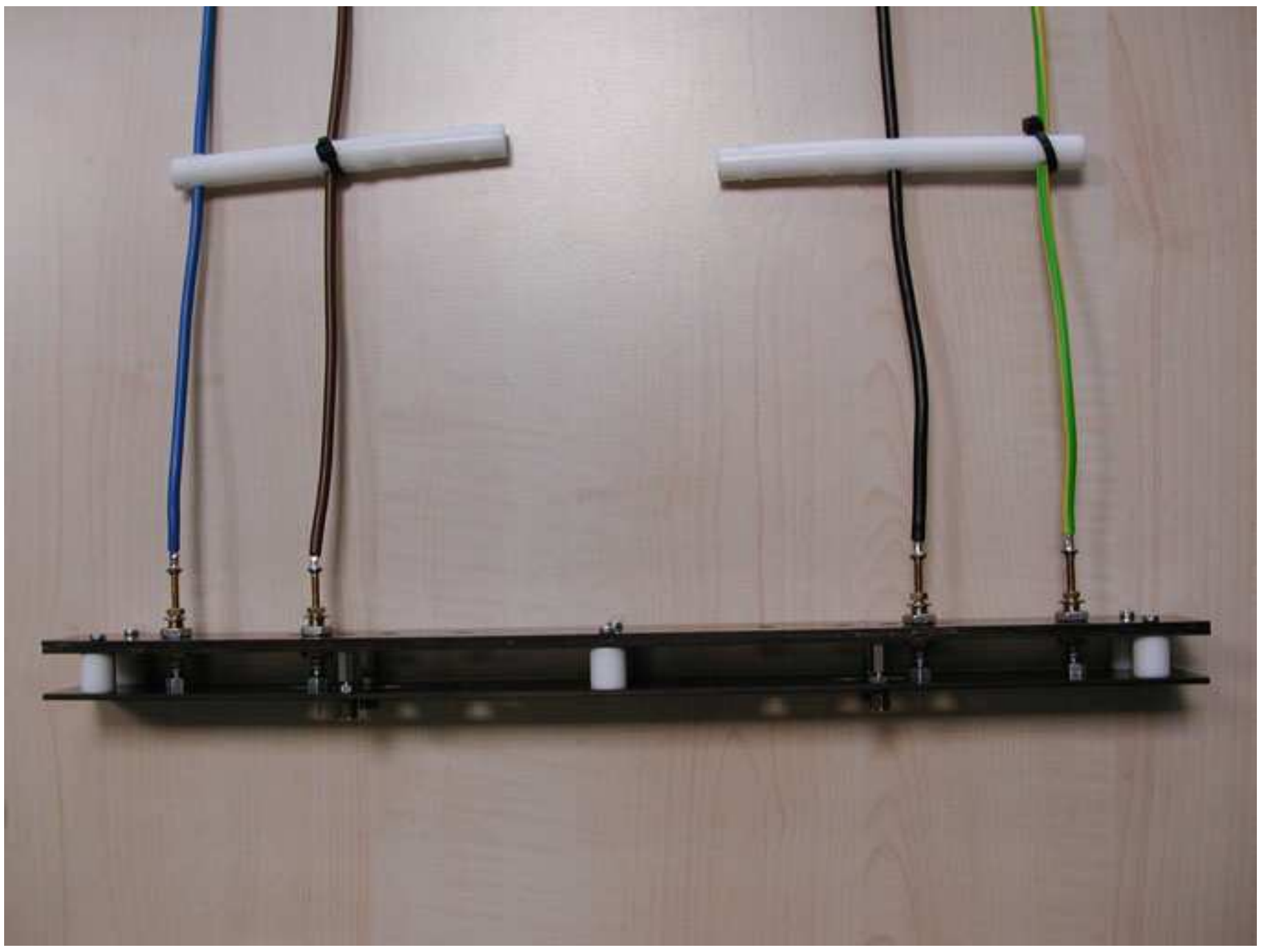

**(c)**

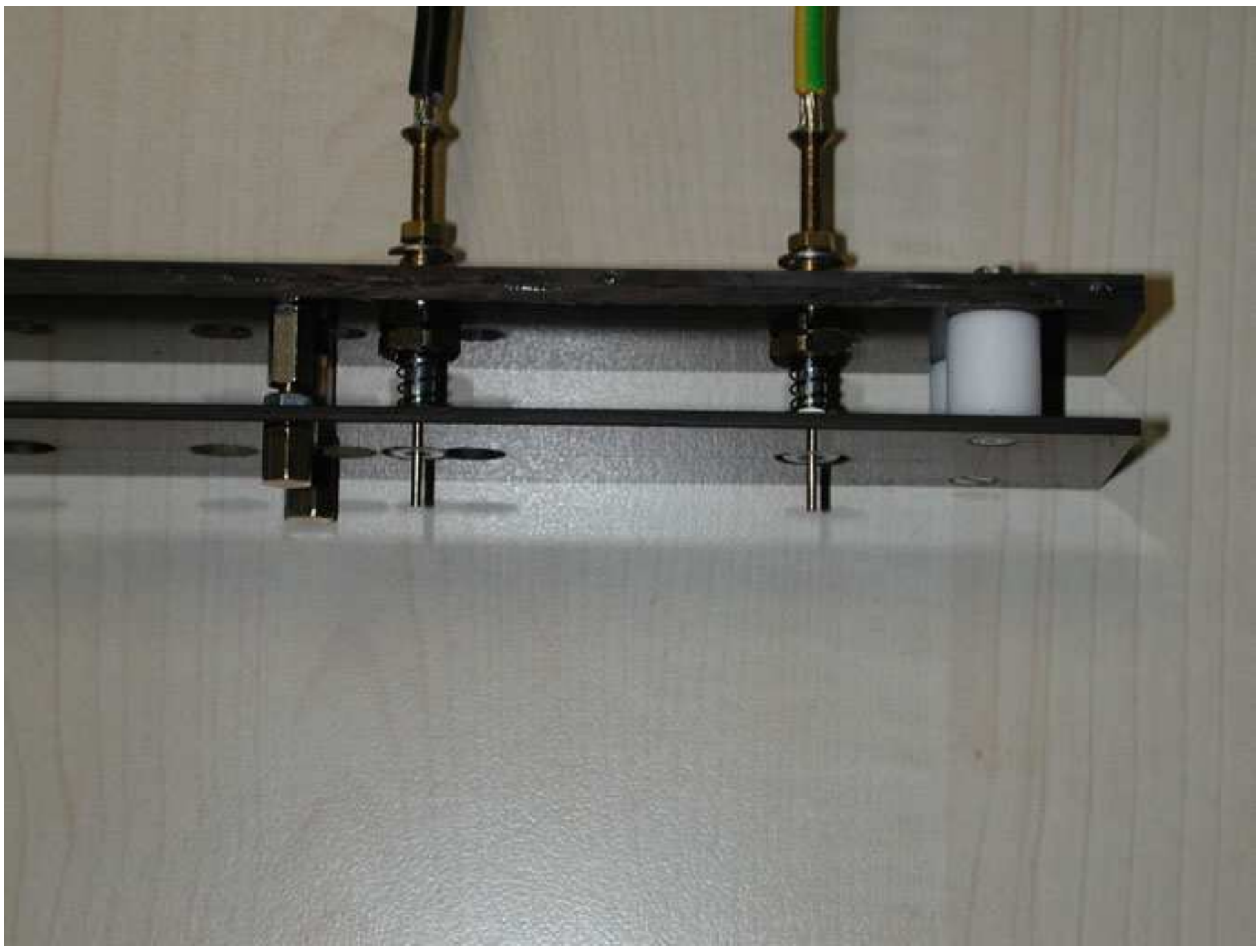

**(d)**

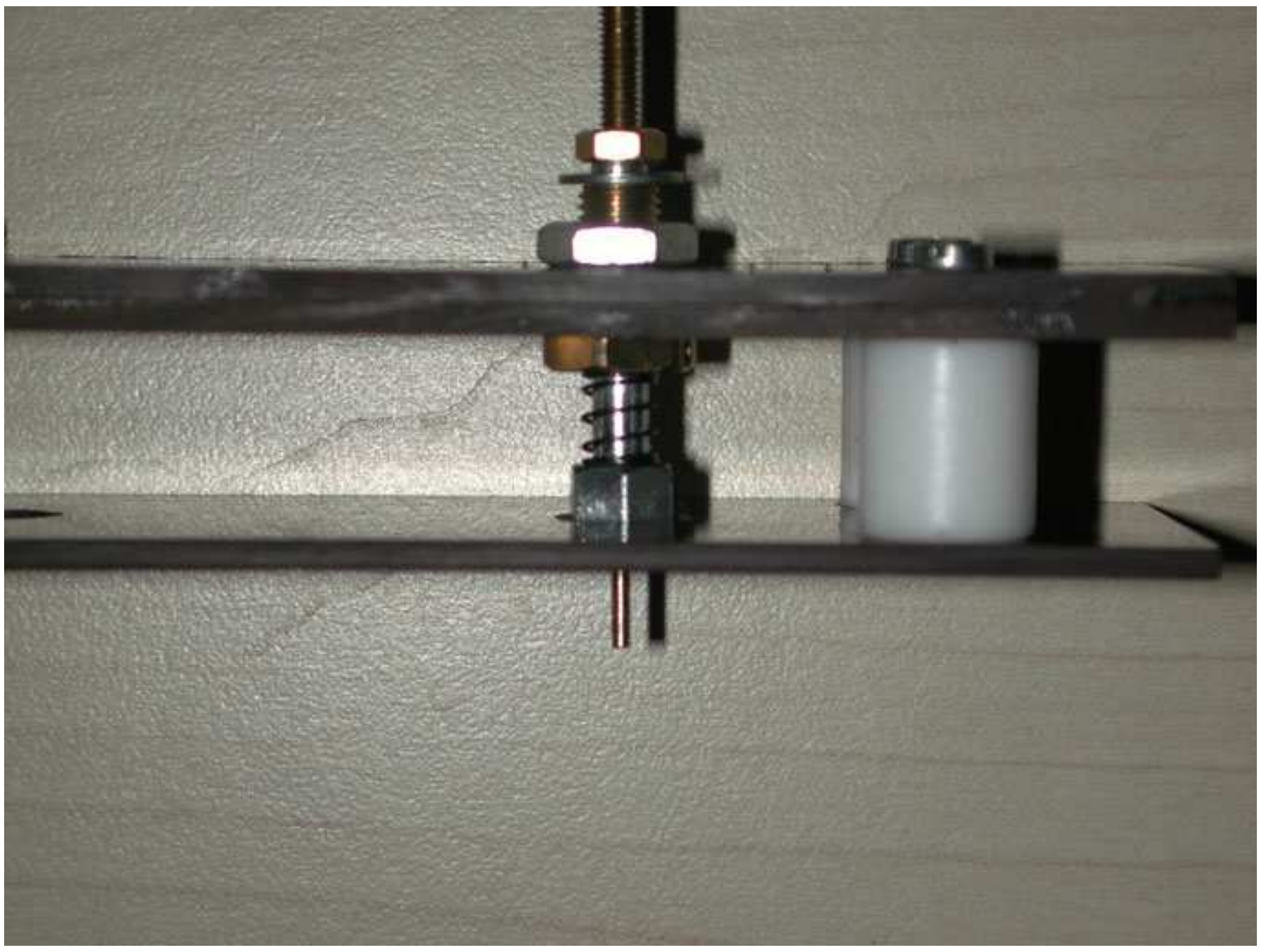

**Figure 17.** Photos of the RESPER probe configured as a *multi* dipole-dipole array from different perspectives (a) - (b) and details of its “spring” poles (c) - (d).

The RESPER probe has been assembled, except the metallic points that must ensure the contact, by insulating materials and more specifically in Tufnol, as regards the support plates, and Teflon, as regards the standoffs (Fig. 17).

A series of holes has been drilled on the surface of support plates in order to allow approaching each other of the two central electrodes to external ones, from a minimum of *4.29 cm* to a maximum of *10 cm*, as shown in the table of configurations (Tab. 6). The dipole-dipole array defined by the integer parameter $n = 6$ could not be implemented, as the positioning of suitable “spring” poles requires *6 mm* holes and an adequate space could not be available to carry out the drilling.

The presence of these “spring” poles allows reaching a right prominence of the tip from the base and, at the same time, a certain amount of pressure which ensures the proper adherence to the artifact that must be tested. There is a brass screw within each pole, which edge has been turned to the measurement of *1.4 mm* (Fig. 17c, d).

Four metal spacers are replaced of time in time depending on the configuration that is to be adopted. The achievement of height $l(n)$ from the plates is ensured by the presence of metallic spacers, which dimensions are reported in Tab. 6.

To complete the description, a copper cable, *1 m* long and with a $4.0\ mm^2$ section area, has been welded to the head of each electrode. The two transmitting or reading cables are kept at a fixed distance $L(n)$ between them, as from Tab. 6, by means of Teflon rods in which has been applied the same series of holes existing on the Tufnol plates.

## Appendix A (Geometrical factor)

All electrical and electromagnetic methods require some form of coupling between a sensor and the ground. The coupling mechanism can have predominantly galvanic, inductive or capacitive character, depending on the nature of the source field, and the frequencies and type of sensors employed (coils, electrodes etc.). The capacitive resistivity (CR) technique exploits the fact that, for electric sources, the quasi-static mode allows for capacitive coupling between sensors and the ground by virtue of the time-varying electric source field. Two poles of a quadrupolar array carry electrostatic charges of opposite sign each of which create an electrostatic potential in the surrounding space. The difference in potential can then be measured at the two remaining poles of the array, its magnitude being linked directly to the permittivity of the medium. The key proposition was that this relation remains valid for time-varying charges, as long as quasi-static conditions are maintained.

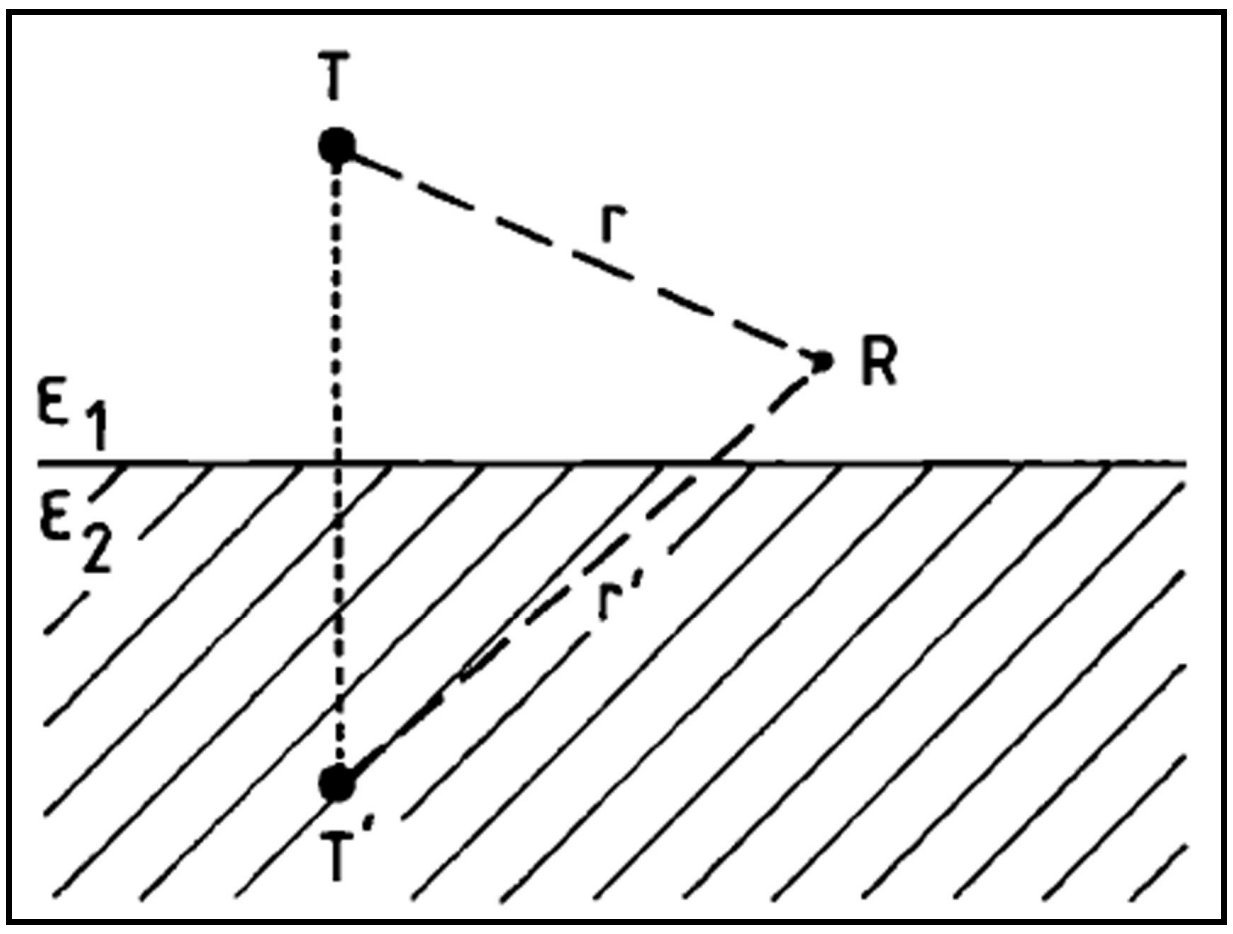

**Figure A.1.** The geometry to determine the potential distribution at R around a point charge at T near the interface of two dielectrics. The quantities *r* and *r'* are the distances from the point R to the point T and to its image T' with respect to the interface, respectively, in the medium of permittivity $\varepsilon_2$.

Grard (1990a) considers electrostatic point charges near a planar interface between two homogeneous and isotropic half-spaces representing electrically different media. A charge located in the vicinity of the interface will create an electrostatic potential in its surroundings, which is a function of the dielectric properties of both media. The electrostatic potential can be readily calculated using the theory of images.

With reference to Fig. A1, if the wavelength of an electromagnetic wave with the given angular frequency $\omega$ is much larger than $r$ and $r'$ in the media of permittivity $\varepsilon_1$ and $\varepsilon_2$, the quasi-static approximation applies [Kaiser, 1962].

The transfer impedance function of a point source at the interface of two half-spaces is calculated as [Grard and Tabbagh, 1991]

$$Z_{RT} = [\frac{1}{r} + \frac{1}{r'}(\varepsilon_1' - \varepsilon_2')/(\varepsilon_1' + \varepsilon_2')] \Big/ (4\pi j \varepsilon_1 \omega), \tag{A.1}$$

being the quantities $\varepsilon'_k$ assumed to be complex and of the form

$$\varepsilon_k' = \varepsilon_k - j/\rho_k \omega, \tag{A.2}$$

where $\varepsilon_k$ and $\rho_k$ are the permittivities and resistivities of the two mediums ($k = 1, 2$).

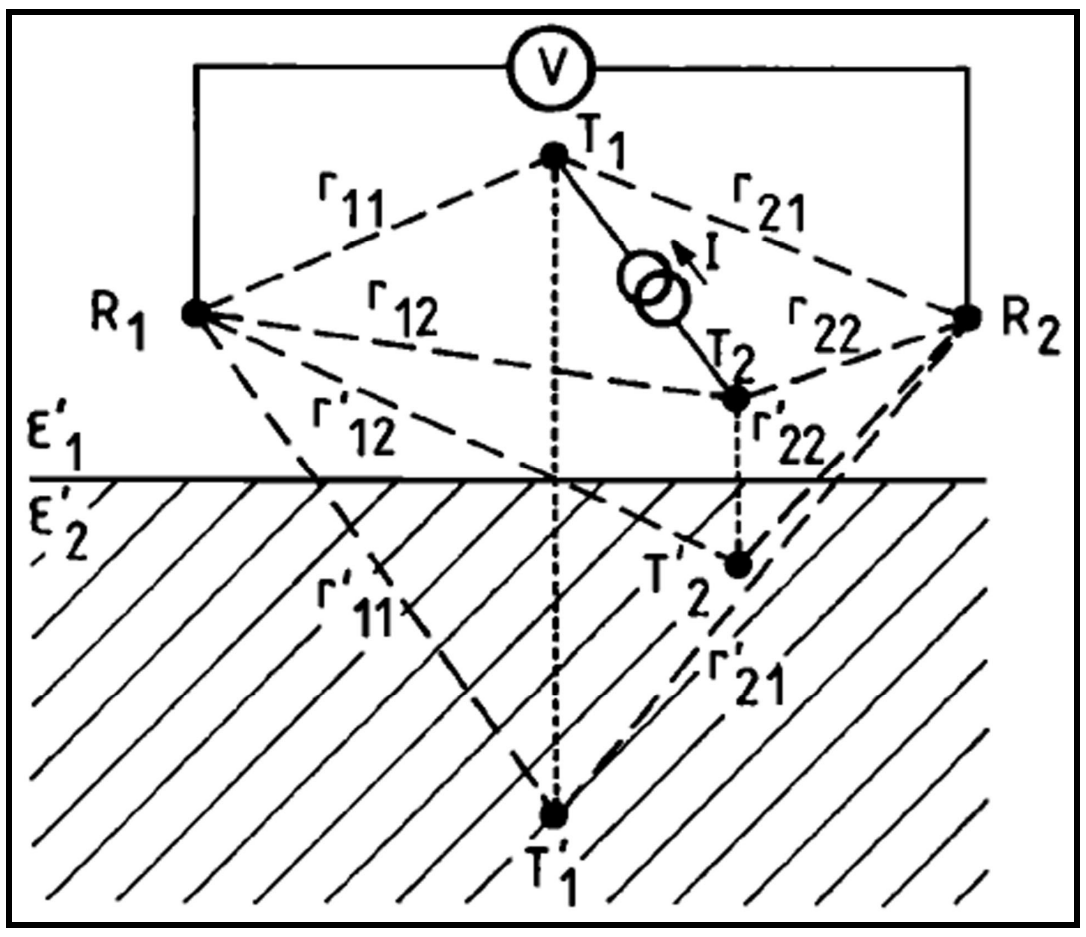

**Figure A.2.** The geometry to determine the transfer impedance of a quadrupolar probe. A current $I = I_1 = -I_2$ is flowing through the medium from a terminal $T_1$ to a terminal $T_2$ and that a voltage $V = V_1 - V_2$ is measured between two other terminals, $R_1$ and $R_2$. The four points $T_1$, $T_2$, $R_1$, and $R_2$ lie in the medium of permittivity $\varepsilon_1$; their positions are arbitrary and not necessarily coplanar.

It is now straightforward to extend this result to a situation with four poles. In practice, an electric current source has two poles which carry opposite electric charges at any moment in time. Equally, potentials can only be measured in respect to a reference potential. This leads to the concept of an *electrostatic quadrupole* where two poles $T_1$, $T_2$ carrying charges $+Q$ and $-Q$, respectively, act as a current source, while the potential difference $\Delta V = V_1 - V_2$ is measured between the two other poles $R_1$, $R_2$ (Fig. A.2).

Referring to Fig. A2, the transfer impedance of a quadrupolar probe is:

$$Z = \frac{1}{jC_0\omega}\frac{\varepsilon_0}{\varepsilon_1'}\frac{2\varepsilon_1' + \delta(\varepsilon_2' - \varepsilon_1')}{\varepsilon_2' + \varepsilon_1'}, \tag{A.3}$$

where

$$C_0 = 4\pi\varepsilon_0 \Big/ \left(\frac{1}{r_{11}} + \frac{1}{r_{22}} - \frac{1}{r_{12}} - \frac{1}{r_{21}}\right) \tag{A.4}$$

is the transfer capacitance of the probe in a vacuum ($\varepsilon_1 = \varepsilon_2 = \varepsilon_0$), and

$$K = 1 - \delta = \left(\frac{1}{r_{11}'} + \frac{1}{r_{22}'} - \frac{1}{r_{12}'} - \frac{1}{r_{21}'}\right) \Big/ \left(\frac{1}{r_{11}} + \frac{1}{r_{22}} - \frac{1}{r_{12}} - \frac{1}{r_{21}}\right) \tag{A.5}$$

is a geometrical factor entirely defined by the configuration of the system.

The above results are derived from electrostatic equations by inserting a time varying charge function associated with the injected current. This procedure ignores the electro-dynamic framework of Maxwell's equations, according to which the time-varying current is associated with an electromagnetic field. It is therefore crucial to establish a condition under which electromagnetic effects can be neglected. Grard and Tabbagh (1991) argue that the *quasi-static approximation* applies and Eqs. (A.3)-(A.5) remain valid if the wavelengths $\lambda_1$, $\lambda_2$ of an electromagnetic wave of angular frequency $\omega$ in media of permittivity $\varepsilon_1$ and $\varepsilon_2$ are much greater than the characteristic distances $r$ and $r'$ in the respective medium. In electromagnetic terms, this corresponds to the well-known conditions of a low induction number regime, i.e. the characteristic distances used must be small compared with the electromagnetic skin depth. Based on experience with applying the electrostatic technique to long-offset resistivity soundings, Benderitter et al. (1994) have proposed the condition

$$\frac{\mu_0 \omega L^2}{\rho} < 1\,, \tag{A.6}$$

where $L$ is a characteristic dimension of the electrostatic array. If this condition is violated, Eqs. (A.3)-(A.5) are no longer valid as the quasi-static approximation breaks down and inductive effects become relevant.

*Wenner's array*

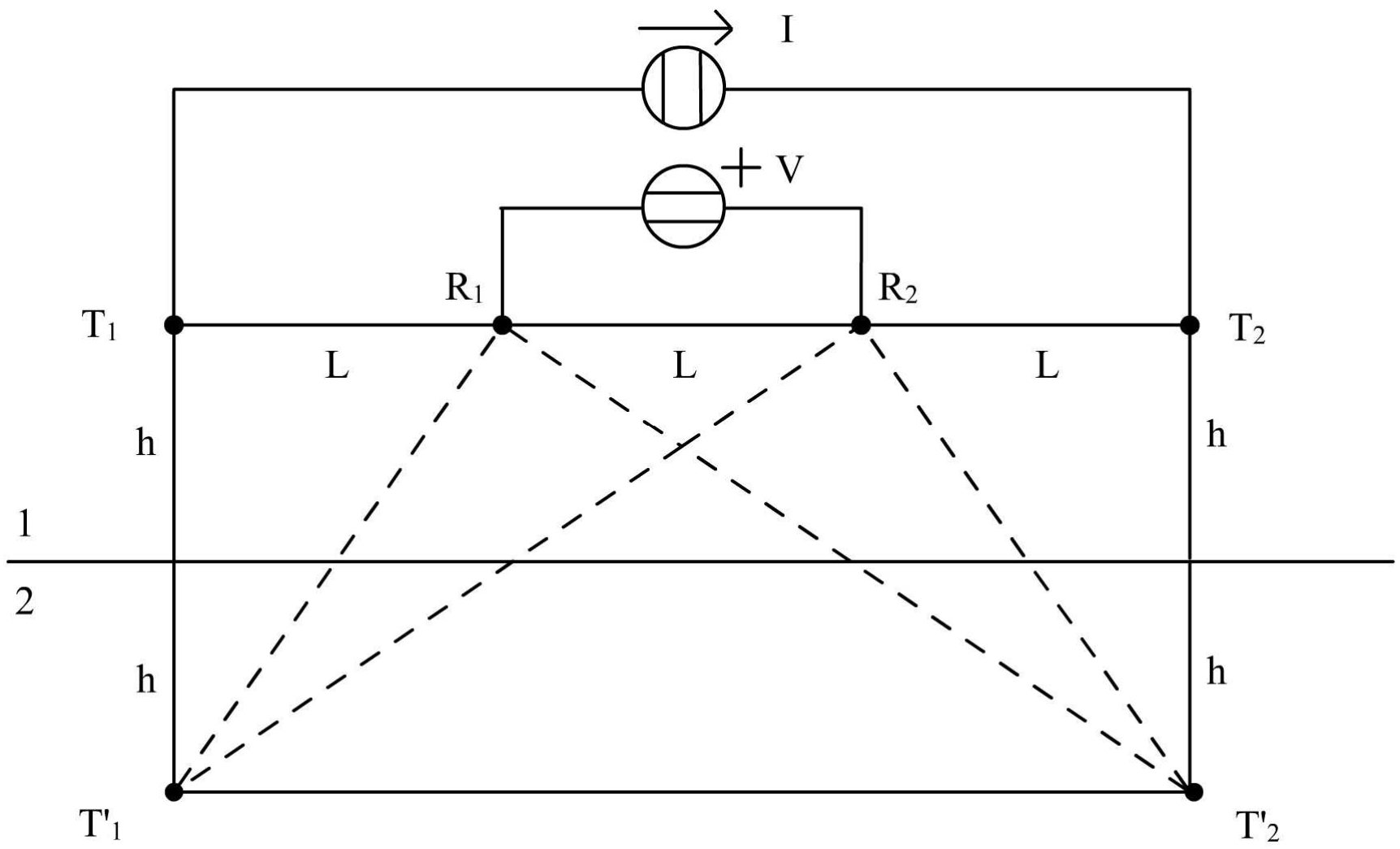

**Figure A.3.** The geometry to determine the transfer impedance of a Wenner's array.

With reference to Fig. A.3:

$$\overline{R_1T_1} = r_{11} = \overline{R_2T_2} = r_{22} = L\,, \tag{A.7}$$

$$\overline{R_1T_2} = r_{12} = \overline{R_2T_1} = r_{21} = 2L\,, \tag{A.8}$$

$$\overline{R_1T_1'} = r_{11}' = \overline{R_2T_2'} = r_{22}' = \sqrt{L^2 + (2h)^2}\,, \tag{A.9}$$

$$\overline{R_1T_2'} = r_{12}' = \overline{R_2T_1'} = r_{21}' = 2\sqrt{L^2 + h^2}\,. \tag{A.10}$$

Once defined the ratio $x$ between the height above ground $h$ and the geometrical characteristic dimension $L$,

$$x = \frac{h}{L}\,, \tag{A.11}$$

the Eqs. (A.9)-(A.10) can be reduced to:

$$r_{11}' = r_{22}' = \sqrt{L^2 + (2xL)^2} = L\sqrt{1+4x^2}\,, \tag{A.12}$$

$$r_{12}' = r_{21}' = 2\sqrt{L^2 + (xL)^2} = 2L\sqrt{1+x^2}\,. \tag{A.13}$$

The Wenner's array measures a capacitance in a vacuum $C_0^{(W)}(L)$, which can be calculated inserting Eqs. (A.7) and (A.8) in Eq. (A.4),

$$C_0^{(W)}(L) = 2\pi\varepsilon_0 \Big/ \left(\frac{1}{r_{11}} - \frac{1}{r_{12}}\right) = 4\pi\varepsilon_0 r_{11} = 4\pi\varepsilon_0 L\,, \tag{A.14}$$

and it can be entirely specified by a geometrical factor $K^{(W)}(x)$ [inserting Eqs. (A.7)-(A.8) and (A.12)-(A.13) in Eq. (A.5)],

$$K^{(W)}(x)=\left(\frac{1}{r'_{11}}-\frac{1}{r'_{12}}\right)\Bigg/\left(\frac{1}{r_{11}}-\frac{1}{r_{12}}\right)=2\left(\frac{1}{r'_{11}}-\frac{1}{r'_{12}}\right)\Bigg/\frac{1}{r_{11}}=2\left(\frac{1}{\sqrt{1+4x^2}}-\frac{1}{2\sqrt{1+x^2}}\right), \tag{A.15}$$

such that: $K^{(W)}(x=0)=1$.

*Dipole-dipole array*

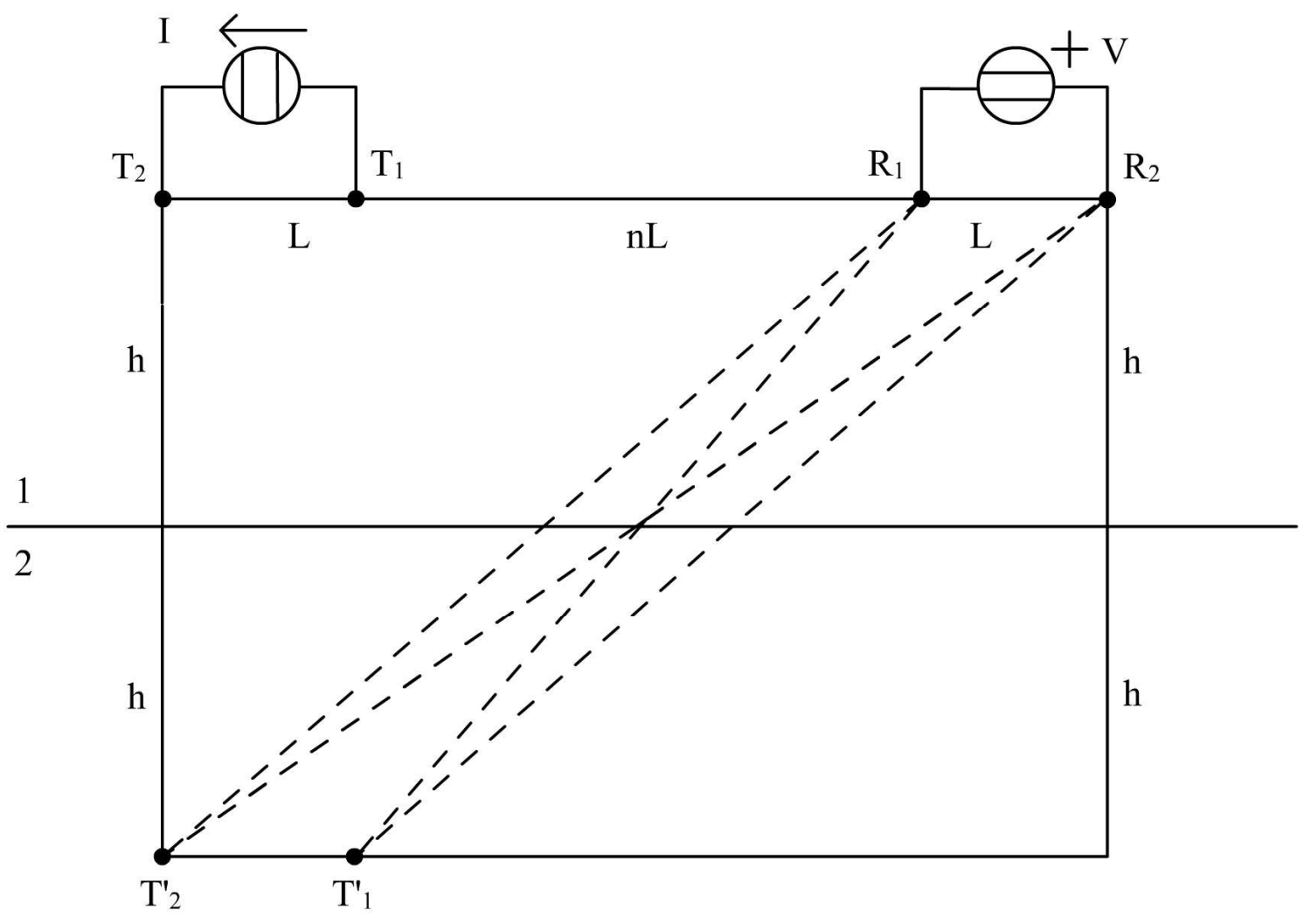

**Figure A.4.** The geometry to determine the transfer impedance of a dipole-dipole array.

With reference to Fig. A.4:

$$\overline{R_1T_1}=r_{11}=nL\,, \tag{A.16}$$

$$\overline{R_2T_2}=r_{22}=(n+2)L\,, \tag{A.17}$$

$$\overline{R_1T_2}=r_{12}=\overline{R_2T_1}=r_{21}=(n+1)L\,, \tag{A.18}$$

$$\overline{R_1T_1'}=r'_{11}=\sqrt{r_{11}{}^2+(2h)^2}=\sqrt{(nL)^2+(2h)^2}\,, \tag{A.19}$$

$$\overline{R_2T_2'}=r'_{22}=\sqrt{r_{22}{}^2+(2h)^2}=\sqrt{(n+2)^2L^2+(2h)^2}\,, \tag{A.20}$$

$$\overline{R_1T_2'}=r'_{12}=\sqrt{r_{12}{}^2+(2h)^2}=\overline{R_2T_1'}=r'_{21}=\sqrt{r_{21}{}^2+(2h)^2}=\sqrt{(n+1)^2L^2+(2h)^2}\,. \tag{A.21}$$

Eqs. (A.16)-(A.18) can be arranged as:

$$L=\frac{r_{11}}{n}, \tag{A.22}$$

$$r_{22} = (n+2)L = \frac{n+2}{n} r_{11}, \tag{A.23}$$

$$r_{12} = r_{21} = (n+1)L = \frac{n+1}{n} r_{11}. \tag{A.24}$$

Inserting Eq. (A.11), i.e. $x=h/L$, the Eqs. (A.19)-(A.21) can be reduced to:

$$r'_{11} = \sqrt{(nL)^2 + (2xL)^2} = L\sqrt{n^2 + 4x^2}, \tag{A.25}$$

$$r'_{22} = \sqrt{(n+2)^2 L^2 + (2xL)^2} = L\sqrt{(n+2)^2 + 4x^2}, \tag{A.26}$$

$$r'_{12} = r'_{21} = \sqrt{(n+1)^2 L^2 + (2xL)^2} = L\sqrt{(n+1)^2 + 4x^2}. \tag{A.27}$$

The dipole-dipole array measures a capacitance in a vacuum $C_0^{(DD)}(L,n)$, which can be calculated inserting Eqs. (A.22)-(A.24) in Eq. (A.4),

$$\begin{aligned} C_0^{(DD)}(L,n) &= \frac{4\pi\varepsilon_0}{\frac{1}{r_{11}} + \frac{1}{r_{22}} - \frac{2}{r_{12}}} = \frac{4\pi\varepsilon_0}{\frac{1}{r_{11}}\left(1 + \frac{n}{n+2} - \frac{2n}{n+1}\right)} = \frac{4\pi\varepsilon_0}{\frac{1}{nL}\left(1 + \frac{n}{n+2} - \frac{2n}{n+1}\right)}, \\ &= 2\pi\varepsilon_0 L n(n+1)(n+2) = C_0^{(W)}(L) \cdot \frac{n(n+1)(n+2)}{2} \end{aligned} \tag{A.28}$$

and it can be entirely specified by a geometrical factor $K^{(DD)}(x,n)$ [inserting Eqs. (A.22)-(A24) and (A.25)-(A.27) in Eq. (A.5)],

$$\begin{aligned} K^{(DD)}(x,n) &= \frac{\frac{1}{r'_{11}} + \frac{1}{r'_{22}} - \frac{2}{r'_{12}}}{\frac{1}{r_{11}} + \frac{1}{r_{22}} - \frac{2}{r_{12}}} = \frac{\frac{1}{r'_{11}} + \frac{1}{r'_{22}} - \frac{2}{r'_{12}}}{\frac{1}{r_{11}}\left(1 + \frac{n}{n+2} - \frac{2n}{n+1}\right)} = \\ &= \frac{\frac{1}{L\sqrt{n^2+4x^2}} + \frac{1}{L\sqrt{(n+2)^2+4x^2}} - \frac{2}{L\sqrt{(n+1)^2+4x^2}}}{\frac{1}{nL}\left(1 + \frac{n}{n+2} - \frac{2n}{n+1}\right)} = \\ &= \frac{1}{2} n(n+1)(n+2)\left(\frac{1}{\sqrt{n^2+4x^2}} + \frac{1}{\sqrt{(n+2)^2+4x^2}} - \frac{2}{\sqrt{(n+1)^2+4x^2}}\right) \end{aligned}, \tag{A.29}$$

being: $K^{(DD)}(x=0,n) = \frac{1}{2} n(n+1)(n+2)\left(\frac{1}{n} + \frac{1}{n+2} - \frac{2}{n+1}\right) = 1,\ \forall n \in \mathbb{N} = \{1,2,3,\ldots\}$ .

## Appendix B (Cylindrical electrodes)

A quadrupolar probe, configured in the Wenner's or dipole-dipole array, is specified by the pairs of transmitting electrodes $T_1$ and $T_2$ at the ends of the quadrupole and the reading electrodes $R_1$ and $R_2$ in the middle of the probe. Suppose that the quadrupole is characterized by a capacitance that is almost invariant for the pairs of electrodes $T_{1,2}$ and $R_{1,2}$,

$$C_{T_1,T_2} \cong C_{R_1,R_2} \cong C_{12} \quad . \tag{B.1}$$

Since the charge $Q$ of the electrodes are equal, the electrical voltage across the pair $T_1$ and $T_2$ approximates the voltage between $R_1$ and $R_2$,

$$\Delta V_{T_1,T_2} \cong \Delta V_{R_1,R_2} \cong \frac{Q}{C_{12}} . \tag{B.2}$$

As first requirement of designing, the voltage amplifier stage downstream of the probe (Fig. 13) must be specified by an input resistance $R_{in}$ that is larger than the reactance associated with the capacitance $C_{12}$, which is characterized by a maximum value in the minimum of frequency $f_{min}$, i.e.

$$R_{in} > \frac{1}{\omega_{\min} C_{12}} = \frac{1}{2\pi f_{\min} C_{12}} \quad \Rightarrow \quad C_{12} > \frac{1}{2\pi f_{\min} R_{in}} . \tag{B.3}$$

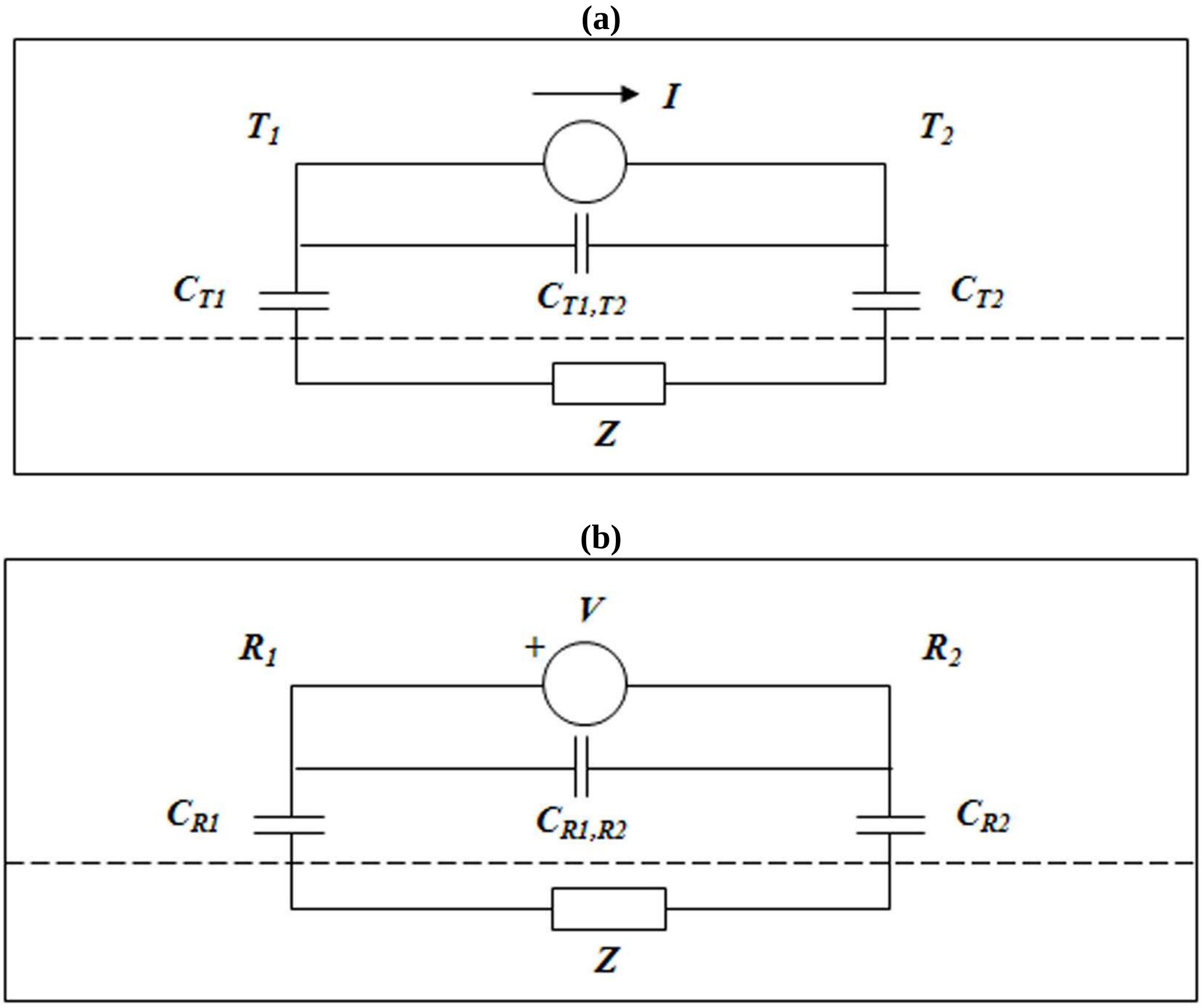

**Figure B.1.** Equivalent capacitance circuits of the quadrupolar probe that schematize the transmission (a) and reception (b) stages.

For the equivalent capacitance circuit that schematizes the transmission stage of the quadrupole (Fig. B.1a), if the effect of the capacitance $C_{12}$ across the electrodes is predominant relative to the shunted capacitances $C_{T1}$ and $C_{T2}$, describing the electrical coupling between the transmitting electrodes and the subjacent medium as,

$$C_{12} << C_{T_1} = C_{T_2} \,, \tag{B.4}$$

then, working in the frequency $f$, the probe injects into the medium a minimum bound for the modulus of the current $|I|_{min}$,

$$\left|I\right|_{\min} \cong \omega \cdot C_{12} \cdot \Delta V_{T_1 T_2} = 2\pi f \cdot C_{12} \cdot \Delta V_{T_1,T_2} \,, \tag{B.5}$$

with $|I|_{min}$ increasing linearly with $f$.

For the equivalent capacitance circuit that represents the reading stage of the quadrupole (Fig. B.1b), the effect of the capacitance $C_{12}$ across the electrodes is predominant even relative to the shunted capacitances $C_{R1}$ and $C_{R2}$, describing the coupling between the reading electrodes and the subjacent medium as,

$$C_{12} << C_{R_1} = C_{R_2} \,. \tag{B.6}$$

If the probe with characteristic geometrical dimension $L$ is immersed in a vacuum, then it measures the vacuum capacitance:

$$C_0 = C_0(L) \,, \tag{B.7}$$

and, in the frequency $f$, measures a minimum limit $|Z|_{min}$ for the transfer impedance in modulus,

$$\left|Z\right|_{\min} = \frac{1}{\omega \cdot C_0} = \frac{1}{2\pi f \cdot C_0(L)} \,, \tag{B.8}$$

which gives rises to a minimum for the electrical voltage $\Delta V_{min,R1,R2}$, flowing the current $|I|_{min}$,

$$\Delta V_{R_1,R_2}^{\min} = \left|Z\right|_{\min} \left|I\right|_{\min} \cong \frac{1}{2\pi f \cdot C_0(L)} \cdot 2\pi f \cdot C_{12} \cdot \Delta V_{T_1,T_2} = \frac{C_{12}}{C_0(L)} \Delta V_{T_1,T_2} \,. \tag{B.9}$$

Note that $\Delta V_{min,R1,R2}$ is independent of $f$, as $|I|_{min}$ is directly and $|Z|_{min}$ inversely proportional to $f$.

As second requirement of designing, the downstream analogical digital converter (ADC) must be specified by a number of bits $n_{bit}$, such that:

$$\frac{\Delta V_{R_1,R_2}^{\min}}{\Delta V_{T_1,T_2}} \cong \frac{C_{12}}{C_0(L)} \simeq \frac{\Delta V_{R_1,R_2}^{\min}}{\Delta V_{R_1,R_2}} = \frac{1}{2^{n_{bit}+1}} \quad \Rightarrow \quad \frac{C_{12}}{C_0(L)} \simeq \frac{1}{2^{n_{bit}+1}} \,. \tag{B.10}$$

*Cylindrical electrodes*

A two cylindrical capacitors series of height $l$, radius $r$ and with spacing distance $L >> r$ is characterized by the electrical capacitance:

$$C = \frac{\pi\varepsilon_0 l}{\ln\frac{L-r}{r}} \cong \frac{\pi\varepsilon_0 l}{\ln(L/r)} \quad , \quad \text{for} \quad L >> r\,, \tag{B.11}$$

being $\varepsilon_0$ the dielectric constant in a vacuum.

Once defined the coefficient of proportionality $\alpha = L/r >> 1$, the radius $r$ depends on the length $L$,

$$r = \frac{L}{\alpha}\,, \tag{B.12}$$

and the electrical capacitance can be expressed as:

$$C \cong \frac{\pi\varepsilon_0 l}{\ln\alpha} \quad , \quad \text{for} \quad \alpha >> 1\,. \tag{B.13}$$

If the Wenner's (W) or dipole-dipole (DD) arrays are implemented by cylindrical electrodes, i.e.

$$C_{T_1,T_2}^{(W)} = \frac{\pi\varepsilon_0 l_W}{\ln\frac{3L-r}{r}} \cong C_{R_1,R_2}^{(W)} = \frac{\pi\varepsilon_0 l_W}{\ln\frac{L-r}{r}} \cong C_{12}^{(W)} = \frac{\pi\varepsilon_0 l_W}{\ln\alpha}\,, \tag{B.14}$$

$$C_{T_1,T_2}^{(DD)} = C_{R_1,R_2}^{(DD)} = \frac{\pi\varepsilon_0 l_{DD}}{\ln\frac{L-r}{r}} \cong C_{12}^{(DD)} = \frac{\pi\varepsilon_0 l_{DD}}{\ln\alpha} \quad , \quad \text{for} \quad \alpha >> 1\,, \tag{B.15}$$

then Eq. (B.3) leads to the first design requirement, relative to the height $l_{W,DD}$:

$$C_{12}^{(W,DD)} = \frac{\pi\varepsilon_0 l_{W,DD}}{\ln\alpha} > \frac{1}{2\pi f_{\min} R_{in}} \quad \Rightarrow \quad l_{W,DD} > \frac{\ln\alpha}{2\pi^2 \varepsilon_0 f_{\min} R_{in}} \quad , \quad \text{for} \quad \alpha >> 1\,. \tag{B.16}$$

The functional trend of the height $l_{W,DD}(R_{in},f)$ with resistance $R_{in}$ is invariant whether the quadrupole is configured in the Wenner's or dipole-dipole array.

Note that the minimum frequency $f_{min}$ is defined by the performed measurement, being specified by the bit number $n_{bit}$ of ADC and the subsurface, i.e. terrestrial soils or concretes.

*Wenner's and dipole-dipole arrays*

Once fixed the total length $L_{TOT}$ of the probe, the characteristic geometrical dimension is $L = L_{TOT}/3$ for the Wenner's (W) (Fig. A.3) and $L = L_{TOT}/(n+2)$ for the dipole-dipole (DD) (Fig. A.4) array defined by the integer parameter $n \in \mathbb{N} = \{1, 2, 3, \ldots\}$, i.e.:

$$L = \begin{cases} L_{TOT}/3 & \text{for Wenner's} \\ L_{TOT}/(n+2) \quad , \quad n \in \mathbb{N} & \text{for dipole-dipole} \end{cases}. \tag{B.17}$$

A Wenner's array with characteristic geometrical dimension $L$ measures a capacitance in a vacuum

$$C_0^{(W)}(L) = 4\pi\varepsilon_0 L\,, \tag{B.18}$$

so that Eq. (B.10) can be specified as:

$$\left.\frac{C_{12}}{C_0(L)}\right|_W = \frac{1}{4\pi\varepsilon_0}\left.\frac{C_{12}}{L}\right|_W \simeq \frac{1}{2^{n_{bit}+1}} \quad \Rightarrow \quad \left.\frac{C_{12}}{L}\right|_W \simeq \frac{\pi\varepsilon_0}{2^{n_{bit}-1}}\,. \tag{B.19}$$

Under the further assumption that the W array is implemented by cylindrical electrodes, then Eq. (B.14) can be combined into Eq. (B.19) leading to the second design requirement, relative to the electrode-electrode distance $L_W$:

$$\left.\frac{C_{12}}{L}\right|_W = \frac{\pi\varepsilon_0}{\ln\alpha}\left.\frac{l}{L}\right|_W \simeq \frac{\pi\varepsilon_0}{2^{n_{bit}-1}} \quad \Rightarrow \quad L_W \simeq l_W\,\frac{2^{n_{bit}-1}}{\ln\alpha} \quad , \quad \text{for} \quad \alpha >> 1\,. \tag{B.20}$$

A dipole-dipole array with characteristic geometrical dimension $L$ measures a capacitance in a vacuum

$$C_0^{(DD)}(L,n) = 2\pi\varepsilon_0 L n(n+1)(n+2) = C_0^{(W)}(L)\cdot\frac{n(n+1)(n+2)}{2} \quad , \quad \forall n \in \mathbb{N} = \{1,2,3,\ldots\}\,, \tag{B.21}$$

so that Eq. (B.10) can be specified as:

$$\left.\frac{C_{12}}{L}\right|_{DD} = \frac{n(n+1)(n+2)}{2}\left.\frac{C_{12}}{L}\right|_W \quad , \quad \forall n \in \mathbb{N}\,. \tag{B.22}$$

Under the further assumption that the DD array is implemented by cylindrical electrodes, then Eq. (B.15) can be combined into Eq. (B.22) leading to the second design requirement, relative to the electrode-electrode distance $L_{DD}$:

$$\left.\frac{l}{L}\right|_{DD} = \frac{n(n+1)(n+2)}{2}\left.\frac{l}{L}\right|_W\,. \tag{B.23}$$

Once fixed the total length of W or DD array $L_{TOT}=3L_W(R_{in})=(n+1)L_{DD}(R_{in})=const$, then are determined the input resistance $R_{in}$ and so the characteristic geometrical dimensions $l_{W,DD}(R_{in})$, $r_{W,DD}(R_{in})$ [Tabs. 4]. Indeed, the dimension $l_{W,DD}(R_{in})$, $L_{W,DD}(R_{in})$ and so $r_{W,DD}(R_{in})$ result as functions depending only on the resistance $R_{in}$ [Figs. 14, 15].

## Acknowledgments

The authors would like to thank Dr. M. Marchetti for interesting discussions on the RESPER probe and for useful pointers on literature regarding Wenner's and dipole-dipole arrays.